\documentclass[ba,preprint]{imsart}% use this for supplement article
\pubyear{2023}
\volume{TBA}
\issue{TBA}

\firstpage{1}
\lastpage{1}

\usepackage[utf8]{inputenc}
\usepackage{amsmath}
\usepackage{amsthm}
\usepackage{amsfonts,epsfig,breakurl}
\usepackage{listings}
\usepackage{lstbayes}
\usepackage{MnSymbol}

\usepackage{mathtools}
\usepackage{url}
\usepackage{graphicx}
\usepackage{subfig}
\usepackage{xcolor}
\usepackage{tabularx}
\usepackage{float}
\usepackage{tikz}
\usepackage{booktabs}
\usepackage{multirow}
\usepackage{verbatim}
\usepackage{threeparttable}
\usepackage{natbib}
\usepackage[colorlinks,citecolor=blue,urlcolor=blue,filecolor=blue,backref=page]{hyperref}

\graphicspath{ {./images/} }
\newcommand{\rsq}{R^2}

\newcommand{\var}{\text{var}}

\newcommand{\normal}{\text{Normal}}
\newcommand{\dirichlet}{\text{Dirichlet}}

\newcommand{\betadist}{\text{Beta}}
\newcommand{\betaprime}{\text{BetaPrime}}
\newcommand{\gammadist}{\text{Gamma}}
\newcommand{\expdist}{\text{Exponential}}
\newcommand{\gigdist}{\text{GenInvGaussian}}
\newcommand{\invgammadist}{\text{InvGamma}}

\newcommand{\rmse}{\text{RMSE}}

\newcommand{\meff}{m_{\text{eff} }}

\newcommand{\elpd}{\text{elpd}}
\newcommand{\lpdhat}{\widehat{\text{elpd}}}

\newcommand{\myosfresults}{\url{https://osf.io/wgsth/}}

\newtheorem{prop}{Proposition}[section]
\newtheorem{lemma}{Lemma}[section]
\usetikzlibrary{shapes,fit} %use shapes library if you need ellipse
\usetikzlibrary{bayesnet}
\usetikzlibrary{shapes,decorations,arrows,calc,arrows.meta,fit,positioning}
\tikzset{
    -Latex,auto,node distance =1 cm and 1 cm,semithick,
    state/.style ={circle, draw, minimum width = 1.0 cm},
    point/.style = {circle, draw, inner sep=0.04cm,fill,node contents={}}, 
    vmissing/.style={
    draw=none, 
    scale=1,
    text height=0.111cm,
    execute at begin node=\color{black}$\vdots$
    },
    hmissing/.style={
    draw=none, 
    scale=1,
    text height=0.111cm,
    execute at begin node=\color{black}$...$
    },
    bidirected/.style={Latex-Latex,dashed},
    el/.style = {inner sep=2pt, align=left, sloped}
}

\begin{document} 

\begin{frontmatter}
%\title{\support{}}

\title{Intuitive Joint Priors for Bayesian Linear Multilevel Models: The R2D2M2 prior}
\runtitle{The R2D2M2 prior}

\date{14 August, 2022}

\begin{aug}
\author{\fnms{Javier Enrique} \snm{Aguilar}\thanksref{addr1}\ead[label=e1]{javier.aguilar-romero@simtech.uni-stuttgart.de}%
\ead[label=u1,url]{https://jear2412.github.io}},
\and
\author{\fnms{Paul-Christian}\snm{ Bürkner}\thanksref{addr2}\ead[label=e2]{paul.buerkner@gmail.com}%
\ead[label=u2,url]{https://paul-buerkner.github.io}}

\runauthor{J. Aguilar and P. Bürkner}

\address[addr1]{Cluster of Excellence SimTech, University of Stuttgart \\
    \printead{e1} \\
    \printead{u1}
}

\address[addr2]{Cluster of Excellence SimTech, University of Stuttgart \\
    \printead{e2} \\
    \printead{u2}
}

%\thankstext{t1}{Some comment}
%\thankstext{t2}{First supporter of the project}
%\thankstext{t3}{Second supporter of the project}

\end{aug}

\begin{abstract}
		\noindent
 
 	   The training of high-dimensional regression models on comparably sparse data is an important yet complicated topic, especially when there are many more model parameters than observations in the data. From a Bayesian perspective, inference in such cases can be achieved with the help of shrinkage prior distributions, at least for generalized linear models. However, real-world data usually possess multilevel structures, such as repeated measurements or natural groupings of individuals, which existing shrinkage priors are not built to deal with.
 	   
        We generalize and extend one of these priors, the R2D2 prior by Zhang et al. (2020), to linear multilevel models leading to what we call the R2D2M2 prior. The proposed prior enables both local and global shrinkage of the model parameters. It comes with interpretable hyperparameters, which we show to be intrinsically related to vital properties of the prior, such as rates of concentration around the origin, tail behavior, and amount of shrinkage the prior exerts. 
    
        We offer guidelines on how to select the prior's hyperparameters by deriving shrinkage factors and measuring the effective number of non-zero model coefficients. Hence, the user can readily evaluate and interpret the amount of shrinkage implied by a specific choice of hyperparameters.
      
        Finally, we perform extensive experiments on simulated and real data, showing that our inference procedure for the prior is well calibrated, has desirable global and local regularization properties and enables the reliable and interpretable estimation of much more complex Bayesian multilevel models than was previously possible. 
	\end{abstract}

%% ** Keywords **

%\tableofcontents

\end{frontmatter}

\section{Introduction}
Regression models are ubiquitous in the quantitative sciences and industry, making up a big part of all statistical data analyses. Their success can be explained by a combination of several factors, involving the ease of interpretation of their additive structure, rich mathematical theory, and (relative) simplicity of their estimation, to name a few key aspects \citep{gelmanregstories2020, harrell2013regression}. Despite these advantages, the vast amount of modeling options and required analyst choices render building interpretable, robust, and well-predicting regression models a highly difficult task.

As the complexity of the analyzed data increases, so does the required complexity of the applied regression models and we are bound to encounter multilevel structures which contain overall (population-level or fixed effects) and varying (group-specific, group-level, or random effects) terms. 	Overall effects influence every member of the population equally, whereas varying effects do so per group. For example, multilevel structures can be found in Psychology, Medicine, or Biology, due to the natural groupings of individuals, or repeated measurement of the same individuals in experiments or longitudinal studies. Multilevel models are designed specifically to account for the nested structure in multilevel data and are a widely applied class of regression models \citep{lme4, gelman_hill_2006, brmsJSSitems}. Their key statistical idea is to assume that a set of regression coefficients, defined according to the multilevel structure, originate from the same underlying distribution whose hyperparameters are subsequently estimated from the data. The implied partial-pooling property  increases the robustness of the parameter estimates, helps with uncertainty calibration, and improves out-of- sample predictive performance \citep{gelman2013bda, gelman_hill_2006}. While multilevel models can be estimated in both frequentist and Bayesian frameworks \citep{lme4, brmsJSS}, we will focus on the latter framework in this work, since it offers considerable flexibility in the specification and regularization of multilevel models \citep{gelman2013bda}.
    
From a Bayesian perspective, the above mentioned distributions of the regression coefficients constitute prior distributions (priors) that describe the uncertainty in the model parameters before seeing the data. The desirable properties of multilevel models can then be explained by the fact that these form joint priors over a set of parameters with shared hyperparameters, rather than separate independent priors for each parameter \citep{gelman2013bda}. Joint priors are a Bayesian success story more generally. For example, joint priors can help to improve the predictive performance of individual additive terms parameterized by more than one parameter, for example, splines, spatial, or monotonic effects \citep{buerknermonotonic, wood2017generalized, Morris2019BayesianHS}. With very few exceptions \citep{Fulgstad2019,rstanarm,Yanchenko}, the development of joint priors for Bayesian multilevel models has been limited to individual additive terms. In contrast, different terms, corresponding to different parameter sets, still receive mutually independent priors, for example, independent (inverse-)gamma, uniform, or half Student-$t$ priors for variance or standard deviation parameters  \citep{BrowneDraper,brmsJSS, pcpriors,Depaoli2015ABA}. As more and more terms are being added to the model while the number of observations remains constant, such models will overfit the data, leading to unreliable or uninterpretable estimates as well as bad out-of-sample predictions \citep{BayesPenalizedRegSara}. Also, overall Type I error rates may be severely inflated if many terms with mutually independent priors are tested \citep{gelman2013bda}.

Without advanced regularization methods such as joint priors, reliable and interpretable estimation of the required highly parameterized models is very hard to achieve. This is particularly obvious in the high dimensional case regime (more covariates than observations: $p>N$), where the number of model parameters $p$ becomes larger than the number of observations $N$ in the data and unregularized regression models would not be estimable \citep{Ridge}; but even if $p<N$, regression models often benefit strongly from regularization, for example, to prevent overfitting and Type I error inflation \citep{Ridge, gelman2013bda, BayesPenalizedRegSara}. While all of these difficulties apply to standard regression models already, they become even worse for multilevel models as the number of additive terms to explore further increases when considering all valid combinations of predictor coefficients and grouping factors which are permitted by the given data structure \citep{catalina2020projection, Barr2013RandomES, Paananen2020GroupHA}. 

In the context of \textit{single-level} linear models, that is, linear models without multilevel structure, a wide range of joint shrinkage priors have been developed \citep{bayesianlasso, Horseshoe, PiironenHorseshoe, DirichletLaplace}. These models enable the estimation, interpretation, and selection of predictor terms and prevent overfitting even when the number of predictors is large and/or the data is small \citep{BayesPenalizedRegSara, vanDP2021theoretical}. On an abstract level, these priors can all be described as Global-Local (GL) priors because they have both, global parameters controlling the shrinkage of all terms jointly and local parameters controlling the (relative) shrinkage of individual terms \citep{vanDP2021theoretical}. Usually, they are combined with a location-scale distribution, such as normal or a double exponential, where the location is fixed to zero and the scale is computed as a product of global and local parameters. While these priors have been very successful within their supported model class \citep{BayesPenalizedRegSara, vanDP2021theoretical}, they remain limited in several key aspects which we aim to address in our work:

\begin{enumerate}
    \item Most existing joint priors for regression models require the manual specification of hyperparameters with potentially strong impact on the obtained inference \citep{BayesPenalizedRegSara}. These hyperparameters are often unintuitive and abstract especially for non-statistical experts, making these priors’ practical application and communication much more difficult.
    \item High dimensionality in multilevel models has so far been studied in cases where the number of overall coefficients $p$ is big while the amount of varying terms $q$ is very small, $p > N, q \ll N$. For example, in their papers, \cite{BuehlmannBio} and \cite{LinLina} carried out theoretical research and simulation studies of frequentist multilevel models with large $p$ yet $q \leq 2$.
    \item Existing joint priors on multilevel models focus only on the group-specific coefficients \citep{Fulgstad2019, rstanarm},  without at the same time considering the overall coefficients.
    Further, it remains unclear how these priors scale with the number of predictors when their coefficients vary across an increasing number of grouping factors. However, in order to ensure consistent and robust regularization in multilevel models, we need to define a single global joint prior on all coefficients, and ensure it scales well to large numbers of additive terms and complex multilevel structure.
\end{enumerate}

An initial step towards addressing these issues, at least for single-level linear models, is the R2D2 prior developed by \cite{r2d2zhang}. The main idea is to specify a prior on the coefficient of determination $R^2$ – also known as \textit{proportion of explained variance} – and decompose the \textit{explained variance} into individual variance components via Dirichlet Decomposition (abbreviated as D2), thus propagating uncertainty from $R^2$ on to the regression coefficients. We further motivate the use of priors over $R^2$ in Section \ref{sec:r2d2m2prior}. The R2D2 prior shares several desirable properties with other global-local shrinkage priors for single-level linear models  \citep{PolsonDefault, Horseshoe, r2d2zhang} and is easy to understand even for non-expert analysts since the only hyperparameters it requires are the shape parameters of the prior on $R^2$ and a single concentration parameter for the Dirichlet distribution. However, its scope is limited to single-level linear models. These simple models do not live up to the requirements of model-based research, where multilevel data structures are omnipresent, so a much more general solution is needed. 

Therefore, the aim of our paper is to fill this gap by generalizing the R2D2 prior to models with increased complexity in terms of multilevel structures.  We name our new prior the \textit{R2D2M2 prior}, where M2 stands for \emph{M}ultilevel \emph{M}odels.
In parallel to our work, related ideas were put forward by \cite{Yanchenko} although with a different focus: they concentrate on defining useful $R^2$ measures for non normal likelihoods without studying the theoretical properties on the resulting priors as we do here.  
\subsection{Main contributions}

Our main contributions are as follows:

\begin{itemize}
    \item We propose a prior over a global $R^2$ that jointly regularizes both the overall and varying coefficients at the same time. Our new prior's hyperparameters are easier to interpret, which facilitates their specification by the user.
    \item We expand upon the theoretical properties of the original R2D2 prior by considering normal base for distributions for the coefficients. We study concentration properties around the origin and behavior in the tails and show how both are dependent on the hyperparameters selected by the user. 
    \item We propose the use of shrinkage factors to evaluate how the chosen hyperparameters determine the amount of shrinkage for each individual coefficient. We also propose a global measure of shrinkage, which quantifies the effective total amount of non-zero coefficients in the model. Both of these quantities were not considered for the R2D2 prior before.
    \item We implement both the R2D2 prior and the R2D2M2 prior in the probabilistic programming language Stan \citep{stan2022,StanJSS} and demonstrate that inference for our implementations is well calibrated. We also provide an implementation in the \texttt{brms} (version 2.19.2) \citep{brmsJSS} R package, which provides a high level interface to fit Bayesian generalized (non-)linear multivariate multilevel models using Stan. We further perform extensive simulations to test the capabilities of the R2D2M2 prior and discuss its shrinkage behavior, out-of-sample predictive performance, as well as the relationship of the latter with the amount of shrinkage. We also show that frequentist error metrics such as Type I and Type II errors can be controlled with the help of our prior even in complex scenarios. 
    \item Our joint prior offers the possibility to study cases in which both the number of overall coefficients $p$ and varying terms $q$ are large and potentially both greater than $N$, that is, high dimensionality in both the overall and varying coefficients. It is to the best of our knowledge the first method that provides this. To demonstrate the prior's applicability and usefulness in such high-dimensional cases not only in simulations but also in the real world, we reanalyse the riboflavin dataset provided by \cite{BuehlmannBio}.
 
\end{itemize} 

The remainder of the paper is organized as follows: Section \ref{sec:r2d2m2prior} starts by motivating the use of priors over $R^2$ (and thus the use of the R2D2M2 prior) by showing how weakly informative priors on the coefficients can translate into very informative priors for $R^2$. The construction of the R2D2M2 prior is presented along with insights that lead to better understanding of the prior. Section \ref{sec:Properties} presents theoretical properties of both the R2D2 and R2D2M2 prior such as marginal distributions and concentration properties as well as guidelines for hyperparameter selection. We introduce the concept of shrinkage factors for both the R2D2 and R2D2M2 prior as well as the relationship the prior's hyperparameters have with the effective number of nonzero coefficients in the model, thus extending the idea of \cite{PiironenHorseshoe} to multilevel models. This provides an easy and intuitive way of setting up the prior, since it is possible to visualize how much sparsity will be induced. Section \ref{sec:examples} shows the results of testing the R2D2M2 prior in intensive simulations. We make use of simulation based calibration \citep{taltssbc} and provide evidence that that our implementation and estimation of both the R2D2 and the R2D2M2 prior models are well calibrated. What is more, we simulate data from sparse multilevel models and provide detailed discussions about how the prior behaves with respect to estimation error, out-of-sample predictive performance, credible interval coverage and Type I and Type II errors. Finally, we show how the R2D2M2 prior performs in real life data by testing it on the Riboflavin production data set made publicly available by \cite{BuehlmannBio}. 
The paper ends with a discussion in Section \ref{sec:discussion} where we summarize our findings and suggest future lines of research. All mathematical proofs can be found in  Appendix \ref{appendixA}.

\section{The R2D2M2 prior}

\label{sec:r2d2m2prior}

In the following, we denote the response variable as $y$, its $n$th observed value as $y_n$ where $n \in \{1,...,N\}$, and the predictor variables (features or covariates) as $x_i$ with $i \in \{1,...p\}$. Additionally we assume conditional independence of the $N$ observations in the data. We define the overall coefficients as those that affect every member of the general population and the group or varying coefficients as those that do so per group.  In the statistical literature, the terms overall and varying coefficients have been known as fixed and random effects, respectively, nevertheless we believe that this nomenclature is problematic in the Bayesian setting, since all of the parameters of interest are treated as (random) variables whose uncertainty is represented via probability distributions \citep{gelman_hill_2006}. We denote the overall coefficients by $b_i,  i \in \{0,...,p\}$ and the group specific or varying coefficients by $u_{i g_j}$. Here $g$ denotes a categorical grouping variable, encoding the individual groups $g_j$ across which the coefficient of the $i$th predictor $x_i$ is expected to vary. If we denote $G_i$ as the index set of all grouping variables across which the coefficient of $x_i$ is expected to vary, we can write the observation model of a linear multilevel model as
%--------------------
\begin{align}
\label{eq:lklhd}
    y_n &\sim \normal \left( \mu_n, \sigma^2 \right) \\
    \mu_n &= b_{0}+ \sum_{i=1}^p x_{ni} b_{i}+  \sum_{g\in G_0} u_{0 g_{j[n]}  }  + \sum_{i=1}^p x_{ni} \left( \sum_{g \in G_i} u_{i g_{j[n]}} \right),
\end{align}
%--------------------
for $n \in \{1,...,N\}, i \in \{0,...,p\},g\in \{1,..., G_i\} $ and $j\in J_g$, where $J_g$ is the index set over the levels of grouping variable $g$. Here, $b_0$ and $u_{0 g_j}$ denote the overall intercept and varying intercepts, respectively, whereas the former will not be subject to extra prior regularization (see below). \\	

When doing Bayesian inference in multilevel models, it is a common practice to set independent normal priors on all regression coefficients: 
\begin{align}
\label{eq:priorsoncoefs}
	b_i \sim  \normal \left( 0,  \tilde{\lambda}_i^2   \right) , \ \ 
	u_{ig_j} \sim  \normal \left( 0,  \tilde{\lambda}_{ig}^2  \right),
\end{align}
where, in current practice, $\tilde{\lambda}^2_i, \tilde{\lambda}^2_{ig} $ are either fixed or are assigned independent priors. The literature shows that either fixing the values of $\tilde{\lambda}$ or assigning independent priors for $\tilde{\lambda}$ is suboptimal, as the total prior variance of the model scales with the number of included terms, which is implausible in reality \citep{Fulgstad2019}. Beyond that, it remains unclear how these priors scale with the number of predictors when their coefficients vary across an increasing number of grouping factors. 

\subsection{Implied priors on R2}

The $\rsq$ measure expresses the proportion of variance explained by the model in relation to the total variance $\sigma_y^2$ of the response $y$ \citep{gelmanregstories2020}. For Gaussian models with residual standard deviation $\sigma$ we define 
%---------------------
\begin{align}
\label{r2def}
	\rsq &\coloneqq \text{corr}^2(y, \mu)=  1-\frac{\sigma^2}{\sigma_y^2},
\end{align}
%---------------------
as a \textit{global} measure of proportion of variance explained \citep{r2mlmssterba}. Global means that it jointly comprises all overall and group specific coefficients at the same time. This stands in contrast to some common definitions of $R^2$ in multilevel models, where $R^2$ is defined for each level of the model separately, rather than jointly across all levels and terms \citep{r2mlmNakagawa, r2mlmssterba}. Our work emphasizes the use of a global $R^2$ as our aim is a joint regularization of all regression coefficients. In that sense, we use the above $R^2$ metric simply as a starting point for a joint regularizing prior for multilevel models, independently of which metric one would prefer to measure explained variances after estimating the model. 

 Given that $\rsq$ is invariant under changes to the mean of the response $y$ or any of the predictors $x_i$, we can assume $\mathbb{E}(y)=\mathbb{E}(x_i)=0, \forall i \in \{ 1,..., p\}$ without loss of generality in the following. This implies that the model's overall intercept $b_{00}$ is zero and we can rewrite $\rsq$ as
 %---------------------
\begin{align}
\label{eq:r2fvarmu} %r2 as a function of var mu
	\rsq&= \frac{\var(\mu) }{\var(\mu)+\sigma^2}.
\end{align}
%---------------------

 Assume a-priori independence of the regression coefficients and consider a prior for $b_i$ and $u_{ig_j}$ that satisfies  $\mathbb{E}(b_i)=0, \mathbb{E}(u_{ig_{j}})=0$. Additionally, denote by $\sigma^2_{x_i}$ the variance of the $i$th covariate, then we can write
%---------------------
\begin{align}
\label{eq:vardecomposition}
	\var(\mu)&= \sum_{g\in G_0} \tilde{\lambda}_{0g}^2   + \sum_{i=1}^p \sigma_{x_i}^2 \left(   \tilde{\lambda}^2_i+  \sum_{g\in G_i} \tilde{\lambda}_{ig}^2  \right) .
\end{align}
%---------------------

Equation (\ref{eq:vardecomposition}) shows that $\var(\mu)$ can be decomposed into the variance components belonging to the overall and varying effects, respectively. Furthermore, the variance corresponding to the varying coefficients can be sub-decomposed into the variance associated to the varying intercepts and varying slopes. 
%---------------------
 \begin{figure}[t!]%
    \centering
    \includegraphics[keepaspectratio , width=\textwidth]{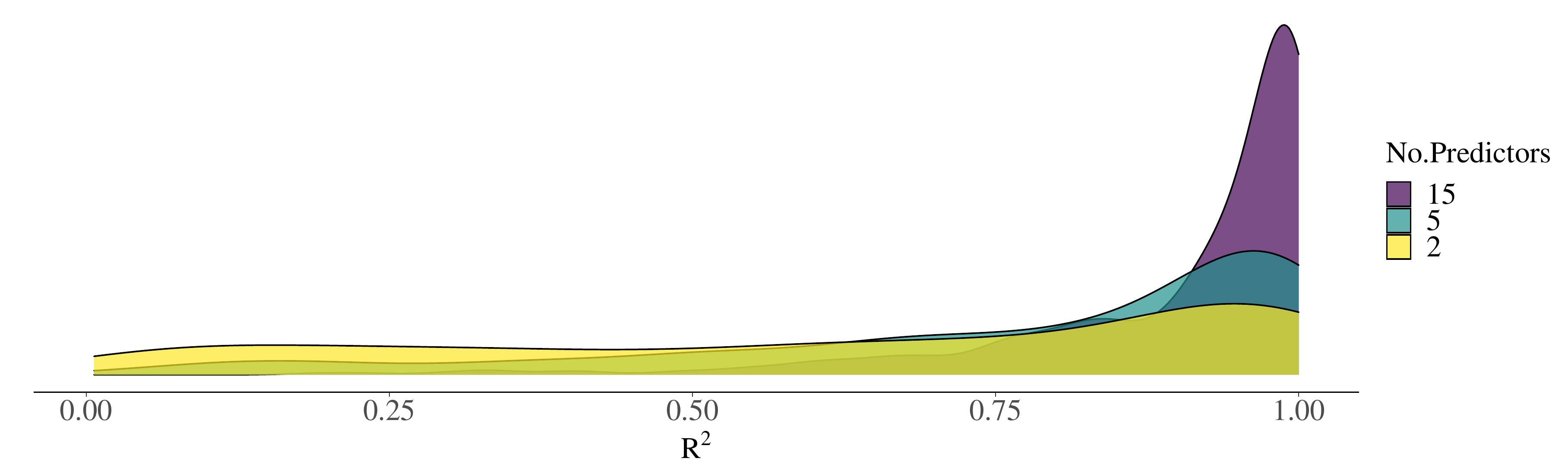}
    \caption{Implied prior on $R^2$ varying number of predictors in a single level linear model with $\normal(0,1)$ priors on the coefficients and an $\expdist(1)$ prior on the error $\sigma$. Notice how weakly informative priors implicate a highly informative prior on $R^2$.}
    \label{fig:r2}
\end{figure}
%--------------------- 

The decomposition of $\var(\mu)$ shown in Equation (\ref{eq:vardecomposition}) allows us to investigate the effect the standard deviations $\tilde{\lambda}$ (or variances $\tilde{\lambda}^2$ ) have on the implied prior for $R^2$. As an example, consider a simple linear model without grouping variables $y_n \sim \normal(\mu_n, \sigma^2)$ where $\mu_n= \sum_{i=1}^p x_{ni} b_i$, with independent $\normal(0,\tilde{\lambda}_i^2)$ priors on $b_i$ with $\tilde{\lambda}_i=1$, and an $\expdist(1)$ prior on $\sigma$. These type of priors are usually considered weakly informative when the variables are standardized, however Figure \ref{fig:r2} shows that the implied prior on $R^2$ is actually very informative. The implied prior on $R^2$ becomes increasingly narrow with a peak close to $R^2=1$ as the number of predictors increases. This behavior becomes even stronger for larger values of $\lambda_i$ and when adding more terms. Therefore, it is also present in the context of multilevel models, where including additional levels per grouping factor and/or additional grouping factors increases the amount of parameters substantially. For instance, when considering one grouping factor $g$ with $L$ different levels and all $p$ coefficients varying across the $L$ levels, one would have $(p+1)L$ additional coefficients (the $1$ is for the varying intercept). 

An $R^2 \approx 1$ represents almost perfect in-sample predictions. This is usually an indicator of overfitting when relevant amounts of noise are present in the data, such that $R^2 \approx 1$ can only be achieved by partially fitting on noise. Any prior implying such a large $R^2$ is highly unrealistic and would not be able to regularize sufficiently. Accordingly, it is more sensible to specify a prior on the overall variance explained by the model and then decompose this variance into components assigned to the individual terms, thereby inducing a joint prior over the variances of all terms. For this purpose, we propose (as have others \citep{rstanarm,r2d2zhang}) to specify a prior directly on $R^2$ as it provides an intuitive and widely understood measure of model fit.  

As mentioned before, our main intention is to generalize the R2D2 prior proposed by \cite{r2d2zhang} to models with increased complexity in terms of multilevel structures. On account of the inclusion of \emph{M}ultilevel \emph{M}odels (M2) in the R2D2 prior, we name our new prior the \emph{R2D2M2 prior}. \\

\subsection{Derivation of the R2D2M2 prior}

In the following we present the derivation of our new prior. For this purpose, we begin by decomposing the priors' variances $\tilde{\lambda}^2_i$ and $\tilde{\lambda}_{ig}^2$ as
 %---------------------
 \begin{align}
 \label{tlambdas}
 \tilde{\lambda}_i^2= \frac{\sigma^2}{  \sigma_{x_i}^2 } \lambda_i^2 , \ \ 
  \tilde{\lambda}_{ig}^2  =  \frac{\sigma^2}{  \sigma_{x_i}^2 } \lambda_{ig}^2.	
 \end{align}
%---------------------
The factor $\frac{\sigma^2}{ \sigma_{x_i}^2}$ accounts for the scales of $y$ and $x_i$ and ensures that $\lambda_i$ represents prior variances on standardized variables, which are comparable across terms and $\lambda_i$ are appropriate for any scale of $y$ and $x$. In practice, $\sigma_{x_i}$ is unknown and it could be  replaced by the sample variance $ \hat{\sigma}_{x_i}^2 $ obtained from the data. Plugging (\ref{tlambdas}) into (\ref{eq:vardecomposition}) results in 
%--
\[  \var(\mu)= \sigma^2  \left( \sum_{g\in G_0} \lambda_{0g}^2   + \sum_{i=1}^p  \left(   \lambda^2_i+  \sum_{g\in G_i} \lambda_{ig}^2  \right) \right) = \sigma^2 \tau^2,    \]
%--
where we have taken $\tau^2$ as the sum of standardized prior variances given  by
%---------------------
\begin{align}
\label{totvariance}
\tau^2 &= \sum_{g\in G_0} \lambda_{0g}^2   + \sum_{i=1}^p  \left(   \lambda^2_i+  \sum_{g\in G_i} \lambda_{ig}^2  \right) .
\end{align}
%---------------------
$\tau^2$ is the (standardized) explained variance in correspondence to the interpretation of $R^2$ as proportion of explained variance. Using $\var(\mu)=\sigma^2 \tau^2$, $R^2$ can be written as
%---------------------
\begin{align}
\label{eq:r2ft2}
	R^2&= \frac{\var(\mu)}{\var(\mu)+\sigma^2 }=   \frac{  \sigma^2 \tau^2  }{  \sigma^2 \tau^2 + \sigma^2} = \frac{ \tau^2 }{\tau^2+1}.
\end{align}
%---------------------

Following \cite{r2d2zhang}, we set a Beta distribution on $R^2$ and we write $R^2\sim \betadist (\cdot, \cdot )$. The Beta distribution offers a wide number of different parametrizations \citep{meanprecbeta}. Here we choose to parametrize the Beta distribution on $R^2$ in terms of the prior mean $\mu_{R^2}$ and prior "precision" $\varphi_{R^2}$ (also known as mean-sample size parameterization), since it provides users with an intuitive way to incorporate prior knowledge into the $R^2$ prior; Both hyperparameters are understandable and expressible in terms of domain knowledge that represent the existing relationship between the included covariates and the response variable. 

\begin{figure}[t!]%
    \centering
    \includegraphics[keepaspectratio, width=\textwidth ]{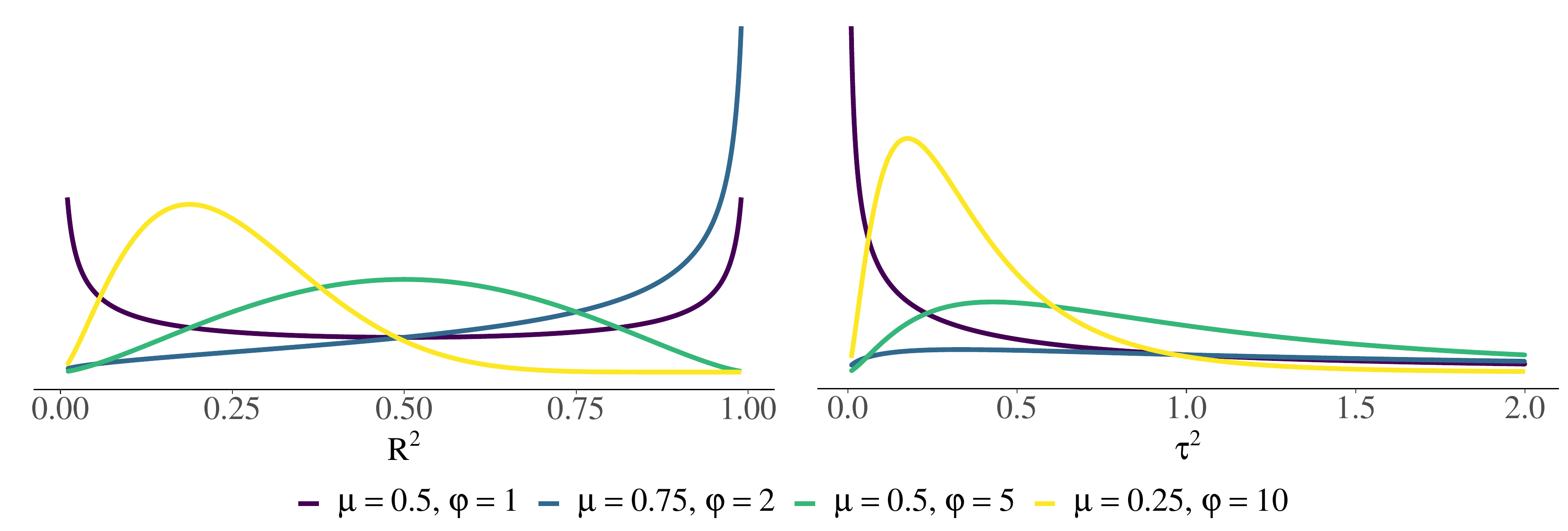}
    \caption{Exemplary densities of the Beta prior for $R^2$ (left) and corresponding Beta-Prime prior for $\tau^2$ (right) with varying mean and precision parameters $\mu, \varphi$ respectively.}
    \label{fig:betapriors}
\end{figure}

Equation (\ref{eq:r2ft2}) implies that, by definition, $\tau^2$ has a Beta-Prime distribution with parameters $\mu_{R^2}, \varphi_{R^2}$, represented by $\tau^2 \sim \betaprime(\mu_{R^2}, \varphi_{R^2})$. Figure \ref{fig:betapriors} illustrates the flexibility that the Beta distribution can offer in expressing prior knowledge about $R^2$. The corresponding Beta-Prime prior for $\tau^2$ is also shown. For example, for values of $(\mu_{R^2},\varphi_{R^2})=(0.5,1)$, we get a bathtub-shaped prior on $R^2$ that places most mass near the two extremes $R^2=0$ and $R^2=1$. A priori, this indicates that the user expects the model to contain a substantial amount of either noise or signal. 

%---------------------
%---------------------
In a consecutive decomposition step, we follow a Global-Local (GL) prior framework and set
%---------------------
\begin{align}
	\label{lambdaphi}
	\lambda_i^2=\phi_i\tau^2,\ \ \lambda_{ig}^2=  \phi_{ig}\tau^2  , \ \ 
\end{align}
%---------------------
with a vector $\phi$ containing all elements $ \phi_i, \phi_{ig} \geq 0, \forall i \in \{1,...,p\},g \in G_i$ such that $\sum_{i}\phi_i+\sum_{i}\sum_{g \in G_i}  \phi_{ig}  =1$, thus rendering $\phi$ as a simplex. $\phi_i$ and $\phi_{ig}$ represent the proportion of standardized explained variance attributed to the $i$th and $ig$th term respectively. Decomposition (\ref{lambdaphi}) describes the proportion of standardized explained variance attributable to the $ig$th term, therefore controlling the variability allocated to each individual variance $\lambda_{ig}^2$.

A natural prior for $\phi$ is a Dirichlet distribution with hyperparameter $\alpha$ chosen by the user, where $\alpha$ is a vector of length equal to the number of variance components that are included in the model and it governs the shape of the Dirichlet distribution. For instance, if the model has $p$ overall coefficients, one grouping factor $g$ and $q$ varying coefficients in the grouping factor, then $\alpha$ is of length $p+q+1$, due to the inclusion of the varying intercept. The amount of levels $L$ does not affect the length of $\alpha$ according to our prior specification (e.g., see Figure \ref{fig:r2d2prop2} below).We denote the elements of $\alpha$ as $\alpha_i$ and $\alpha_{ig}$, which represent the a priori expected relevance on each overall and varying regression term, respectively. In our context we will call $\alpha$ a concentration vector, since $\alpha_0= \sum_{i}\alpha_i+\sum_{ig}\alpha_{ig}$ determines the narrowness of the distribution, that is, how peaked it is \citep{OnTheDirichlet}. 

Naturally, the question arises of how $\alpha$ is to be specified. When $\alpha=(1,...,1)'$ the prior is flat over all possible simplexes, which renders this value a good choice in the absence of additional prior knowledge. More generally, setting $\alpha=(a_\pi,...,a_\pi)'$ with a concentration parameter $a_\pi > 0$ (a symmetric Dirichlet distribution) can also be sensible since it drastically reduces the number of hyperparameters to specify, providing us with the ability to globally control the shape of the Dirichlet distribution with a single value. Additionally, the user can control the induced sparsity of the model by use of this single value as we show in Section \ref{subsection:tailcon}. A symmetric Dirichlet distribution is useful when there is no prior preference that favours one component over another. Values for which $a_\pi<1$ produce simplexes that concentrate their mass on the edges ("bathtub" distribution) and most coefficient values within a single sample will be close to 0. Instead, setting $a_\pi>1$ results in concentration around the center of the simplex, favouring an even distribution among the different components \citep{OnTheDirichlet}. Finally, the user can also specify asymmetric Dirichlet distributions by designating different values for the components of $\alpha$, which represent the differently a priori expected importance of the corresponding regression term.

The values of $\tau^2$ and $\phi_{ig}$ play different roles in the prior. The explained variance $\tau^2$ controls the amount of global shrinkage and is therefore deemed as a global scale, whereas the $ig$th attributed variance $\phi_{ig}$ serves as a local scale that controls the individual shrinkage the prior induces over each specific term. The components of $\phi$ compete to increase and to acquire a higher proportion from the total variance $\tau$. This can be seen when examining the covariance between two different components, which is given by $\text{cov}(\phi_i, \phi_j)= \frac{-\alpha_i \alpha_j}{ \alpha_0^2 (\alpha_0+1)}, i\neq j$. Thus, for a fixed value of $\tau^2$, as the importance of one term increases the importance of other terms decrease, and vice versa.  

In the GL framework, the next step is to assign a base distribution that is able to express the attributed variance to each regression term $b_i$ and $u_{ig_j}$. In their original R2D2 paper, \cite{r2d2zhang} consider the coefficients to follow double exponential base distributions since they ensure both a higher concentration of mass around a zero and heavier tails than normal distributions. In the present paper, we will instead employ normal base distributions for each regression term for several, not mutually independent reasons:
\begin{enumerate}
    \item It is customary in the multilevel model literature to 
    consider normal base distributions for the overall and varying coefficients \cite{ gelman_hill_2006, wakefield2013bayesian}. What is more, Global-Local priors with normal base distributions are very flexible and have been the gold standard when proposing shrinkage priors, as can be seen in  \citep{Horseshoe,DirichletLaplace, BayesPenalizedRegSara}, among others.
    \item  Posterior approximation via gradient-based MCMC methods encounter issues when sampling from non-differentiable distributions, such as the double exponential \citep{handbookmcmc}. Additionally, it is straighforward to reparametrize the posterior distribution to improve the rates of convergence of either Gibbs or Hamiltonian Monte Carlo based MCMC samplers in the context of conditionally normal multilevel models \citep{brmsJSS, GibbsMultigrid}. 
    \item  The use of normal distributions provides us with closed analytical expressions for several quantities of interest as described in Section \ref{sec:Properties} and shows, for example, good shrinkage and tail behavior.
\end{enumerate}

 To ease prior specification and speed up sampling, we will set a prior on the intercept $\tilde{b}_{00}$ implied when all $\mathbb{E}(x_i)=0$ and then recover the original intercept $b_{00}$ after model fitting using a simple linear transformation \citep{rstanarm, brmsJSS}. Common priors on $b_{00}$ are a normal prior with mean $\mathbb{E}(y)$ and a user chosen scale depending on the scale of $y$ as well as Jeffrey's prior which is improper flat in this case \citep{Jeffreys}. 
 
What remains to be specified is a prior for the residual variance $\sigma^2$ (or equivalently residual standard deviation $\sigma$). We follow the recommendations of \cite{GelmanHalfStudentt} to consider a half Student-$t$ prior on $\sigma$ with $\nu$ degrees of freedom and scale $\eta$. We propose setting $\eta \approx \text{sd}(y)$, since both the prior expected mean and variance are proportional to $\eta$. To be able to implement a Gibbs sampling approach, an inverse Gamma distribution $\sigma^2 \sim \text{InvGamma} (c,d) $, with shape and scale parameters $c,d$ respectively, could also be considered. We show how to implement a Gibbs sampler for the R2D2M2 prior in Appendix \ref{section:appendixB}. 
 
The full R2D2M2 model can summarized as 
%----
\begin{equation}
\label{r2d2m2model}
    \begin{aligned}
y_n & \sim \normal (\mu_n, \sigma^2) \\
\mu_n &= b_{0}+ \sum_{i=1}^p x_{ni} b_{i}+  \sum_{g\in G_0} u_{0 g_{j[n]}  }  + \sum_{i=1}^p x_{ni} \left( \sum_{g \in G_i} u_{i g_{j[n]}} \right) \\
b_{0}& \sim  p(b_{0}) \\
b_i &\sim \normal \left(0, \frac{\sigma^2}{\sigma_{x_i}^2}   \phi_i \tau^2\right), \quad 
    u_{ig_j} \sim \normal \left(0, \frac{\sigma^2}{\sigma_{x_i}^2}   \phi_{ig} \tau^2\right) \\
    \tau^2&= \frac{R^2}{1-R^2}\\
R^2  &\sim \betadist (\mu_{R^2},\varphi_{R^2}), \ \
     \phi \sim \dirichlet (\alpha), \ \ 
     \sigma \sim p(\sigma). \\
\end{aligned}
\end{equation}
%----
\begin{figure}[t!]%
    \centering
    \begin{tikzpicture}
    % x node set with absolute coordinates
    \node[state] (r2) at (0,0) {$R^2$};

    \node[state] (t2) at (2,0) {$\tau^2$};
    
    %\node[state] (phi1) at (4,3)  {$\phi_1$};
    %\node[vmissing] (dots1) at (4,2) {};
    \node[state] (phip) at (4,1) {$\phi_i$};

    \node[state] (phip1) at (4,-1)  {$\phi_{ig}$};
    %\node[vmissing] (dots3) at (4,-2) {};
    %\node[state] (phipq) at (4,-3) {$\phi_{p+q}$};
    
    %\node[state] (lambda1) at (6,3)  {$\lambda_1$};
    %\node[vmissing] (dots5) at (6,2) {};
    \node[state] (lambdap) at (6,1) {$\lambda_i^2$};

    \node[state] (lambdap1) at (6,-1) {$\lambda_{ig}^2$};
    %\node[vmissing] (dots7) at (6,-2) {};
    %\node[state] (lambdapq) at (6,-3) {$\lambda_{p+q}$};

    %\node[state] (beta1) at (8,3)  {$b_1$};
    %\node[vmissing] (dots8) at (8,2) {};
    \node[state] (betap) at (8,1) {$b_i$};
    
    \node[state] (u11) at (8,-1)  {$u_{ig_1}$};
    \node[hmissing] (hdots1) at (9,-1) {};
    %\node[vmissing] (dots8) at (8,-2) {};
    \node[state] (ul1) at (10,-1)  {$u_{ig_l}$};
    %\node[vmissing] (dots9) at (10,-2) {};
    %\node[state] (u1q) at (8,-3) {$u_{1q}$};
    %\node[state] (ulq) at (10,-3) {$u_{lq}$};
    %\node[hmissing] (hdots2) at (9,-3) {};
    % Directed edge
    \path (r2) edge (t2);
    %\path (t2) edge  (phi1);
    \path (t2) edge  (phip);
    \path (t2) edge  (phip1);
    %\path (t2) edge  (phipq);

    %\path (phi1) edge  (lambda1) ;
    \path (phip) edge (lambdap) ;
    \path (phip1) edge (lambdap1) ;
    %\path (phipq) edge (lambdapq) ;
    
    %\path (lambda1) edge  (beta1);
    \path (lambdap) edge (betap);
    
    \path (lambdap1) edge (u11);
    %\path (lambdapq) edge (u1q);

    % \node[draw=black, thick,fit=(phi1)(phipq), inner sep=0.2cm] (blackbox) {};

    %\node[draw=blue,dotted, thick,fit=(beta1)(beta1), inner sep=0.2cm] (bluebox) {};

    \node[draw=red,dotted,thick,fit=(u11)(ul1), inner sep=0.2cm] (redbox) {};
    
\end{tikzpicture}
  \caption{Construction of the R2D2M2 prior schematically. We begin by assigning a distribution to $R^2$, afterwards its uncertainty is propagated to the regression terms via a Dirichlet decomposition of the (a priori) explained variance. Finally, we make use of a distribution that can express the amount of variance that has been allocated to each term. Notice how all $u_{ig_jl}$ share the same local variance $\lambda_{i_g}$ for a given pair $(i, g_j)$ and varying level $l$. \\}
  \label{fig:r2d2prop2}
\end{figure}
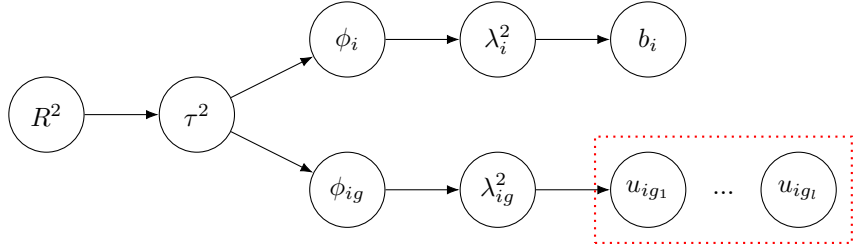%
%----
A schematic construction of the R2D2M2 is shown in Figure \ref{fig:r2d2prop2}. We do not specify distributions in this diagram, thus emphasizing that there is flexibility in the way uncertainty can be propagated from $R^2$ to the coefficients (as long as consistent rules that respect the corresponding domains of each parameter are followed). For example, the user might want to use a different distribution for $R^2$, such as the Kumaraswamy distribution which offers advantages when using quantile-based approach to statistical modelling \citep{Kumaraswamy}, or they might desire to model correlations among the components of $\phi$ explicitly, in which case they could use a distribution that allows it, such as a logistic normal distribution \citep{logisticnormal}. 

We close this section commenting on how the intuitiveness of the R2D2M2 prior assists in eliciting prior knowledge or lack of it. Users can readily communicate a prior expected value for $R^2$ through $\mu_{R^2}$. Given a value for $\mu_{R^2}$, the value selected for $\varphi_{R^2}$ is an indicator of how confident the user is on how much of the response variability the model is explaining. This falls in line with the recommendations given by \cite{prioreli} on how to improve prior elicitation: the hyperparameters on the prior for $R^2$ are easy to understand and the priors for the regression terms regularized jointly (see Figure \ref{fig:r2d2prop2}), hence significantly decreasing the number of hyperparameters to the user has to choose. Thus, the R2D2M2 prior brings us a step forward in terms of prior elicitation for multilevel models and provides a valuable tool in the Bayesian workflow for data analysis more generally \citep{bayesianworkflow}.

\section{Properties of the R2D2M2 prior}
\label{sec:Properties}
In this section, we derive several mathematical properties of the R2D2M2 prior. We begin by presenting the marginal prior distributions of the overall and varying coefficients. We show how the behavior around the origin and the tails of the marginal densities can be controlled by the hyperparameters $\mu_{R^2}, \varphi_{R^2}, a_\pi$. The distinct behaviors induced by hyperparameter choices can be exploited by the user to represent their prior knowledge and desired degree of regularization.

We then present the conditional posterior distributions of the overall and varying coefficients. Specifically, based on their conditional posterior means, we introduce shrinkage factors that allow us to calculate the effective number of coefficients and give further guidelines on how to choose the hyperparameters of the R2D2M2 prior. \\

\subsection{Tail and concentration behaviors}
\label{subsection:tailcon}

The original R2D2 prior proposed by \cite{r2d2zhang} uses a double exponential (Laplace) base distribution for the coefficients $b_i$. However, as mentioned before, in the multilevel model literature it is customary to treat $b_i$ and in particular the varying coefficients $u_{ig_j}$ as normally distributed (see Section \ref{sec:r2d2m2prior} for details). The use of normal base distributions for the coefficients may lead to different marginal prior distributions, different concentration properties around the origin and tail behavior, which we will study in the following. 

Consider the hyperparameters of the R2D2 prior $\mu_{R^2}$, $\varphi_{R^2}$ and let $a_1= \mu_{\rsq} \varphi_{\rsq},  a_2=(1-\mu_{R^2})\varphi_{R^2}$ be the corresponding shape parameters of the Beta distribution on $R^2$ in the standard parameterization. In the following choose $\alpha=(a_\pi,...,a_\pi)'$ with $a_\pi > 0$. \cite{r2d2zhang} showed that, under specific conditions depending on the hyperparameters $a_\pi$ and $a_2$, their prior is able to attain polynomial concentration rates around the origin and polynomial decay rates in the tails, thus allowing for shrinkage of noise towards zero and detection of signals. This shows that the R2D2 prior is able to compete with other well known GL shrinkage priors. We arrive to similar conclusions when using normal base distributions and taking multilevel structure into consideration. In the following propositions, the marginal distributions presented depend on a fixed value of $\sigma$. Other authors \citep{Horseshoe, BaiHypothesisNB, r2d2zhang} usually fix the value of $\sigma=1$ for simplicity, however the marginal priors they present are still conditional on $\sigma$, as are ours. 

\begin{prop}
\label{prop:marginalpriors}
Given the R2D2M2 prior, the marginal prior densities of $b_i, u_{ig_j}$ given $\sigma$ for any $i=1,...,p, g\in \{1,..., G_i\}, j\in J_g$ are
%---------------------
\begin{subequations}
    \begin{equation}
    p(b_i|\sigma)= \frac{1}{ \sqrt{2\pi q_i^2}\text{B}(a_\pi, a_2) }\Gamma(\eta) U( \eta, \nu,z_i)
    \end{equation}
    \begin{equation}
    p(u_{ig_j}|\sigma)=\frac{1}{ \sqrt{2\pi q_i^2}\text{B}(a_\pi, a_2) }\Gamma(\eta) U( \eta, \nu,z_{ig_j}),
    \end{equation}
\end{subequations}
%---------------------
where $q_i^2=\frac{\sigma^2}{\sigma_{x_i}^2},a_2=(1-\mu_{R^2})\varphi_{R^2}, \eta=a_2+1/2$ and $\nu= 3/2-a_\pi$. $\text{B}(\cdot,\cdot)$ and $\Gamma(\cdot)$ represent the Beta and Gamma functions respectively and $U(\eta, \nu, z_i)$ represents the confluent hypergeometric function of the second kind, as well as $z_i=\frac{|b_i|^2}{2q_i^2}$ and $z_{ig_j}=\frac{|u_{ig_j}|^2}{2q_i^2}$. If there are no varying coefficients, the marginal prior densities of $b_i$ remain unaltered. 
\end{prop}
%---------------------
\begin{prop}
\label{prop:originprior}
As $|b_i|\to 0, |u_{ig_j}| \to 0$, $0<a_\pi \leq 1/2$, and $a_2>0$, the marginal prior densities are unbounded with a singularity (i.e., undefined) at zero.  Moreover, the marginal priors satisfy
\begin{align*}
    p(b_i|\sigma)&\sim
    \begin{cases}
        c_1 b_i^{2a_\pi-1}+\mathcal{O}\left( |b_i|^{2a_\pi+1} \right), \  &a_\pi<1/2 \\
         -c_2 \ln (b_i^2/ 2q_i^2)+\mathcal{O}\left(  b_i^2 \ln (b_i^2/ 2q_i^2) \right), \ &a_\pi=1/2
    \end{cases} \\ 
    p(u_{ig_j}|\sigma)&\sim
    \begin{cases}
        c_1 u_{ig_j}^{2a_\pi-1}+\mathcal{O}\left( |u_{ig_j}|^{2a_\pi+1} \right), \  &a_\pi<1/2 \\
         -c_2 \ln (u_{ig_j}^2/2q_i^2)+\mathcal{O}\left(  u_{ig_j}^2 \ln (u_{ig_j}^2/ 2q_i^2) \right), \ &a_\pi=1/2
    \end{cases} \\ 
\end{align*} 
%---------------------
where $c_1= (2q_i^2)^{1/2-a_\pi} \Gamma(1/2-a_\pi)/\Gamma(a_2+1/2), c_2=1/\Gamma(a_2+1/2)$. When instead $a_\pi>1/2$, the marginal prior densities are bounded and continuous at zero. The marginal prior densities are differentiable at $t=0$ for all values of $a_\pi\geq 1$. 
\end{prop}
%---------------------
Proposition \ref{prop:originprior} implies that the value of $a_\pi$ controls the boundedness of the marginal priors near the origin, which in turn influences the level of shrinkage that is enforced upon the coefficients. Sufficient mass near zero allows the prior to shrink small signals sufficiently \citep{BayesPenalizedRegSara}. As $a_\pi$ decreases from $1/2$ to zero, the R2D2M2 prior shifts its mass even stronger to the origin and we can expect the posterior distribution to be concentrated near zero. This leads to prior concentration around sparse coefficient vectors and favors sparse estimation of the coefficients. On the other hand, an increasing $a_\pi$ from $1/2$ produces a bounded prior and indicates that we are in favor of less sparse coefficient vectors. 

We can further understand the behavior of the marginal priors by considering that, when assuming $\phi \sim \dirichlet(\alpha)$ where $\alpha= (a_\pi,..., a_\pi)'$, the value of $a_\pi$ determines where the density places its mass \citep{OnTheDirichlet}. When $0< a_\pi< 1$ the density congregates at the edges of the simplex and prefers sparse distributions, i.e., most of the values within a single sample will be close to 0, and the vast majority of the mass will be concentrated in a few of the $\phi$ values. As $a_\pi$ increases to and exceeds 1, the density concentrates near the center of the simplex, with a mode appearing in $\left(\frac{1}{\text{dim}(\alpha)},..., \frac{1}{\text{dim}(\alpha)}\right)'$, where $\text{dim}(\alpha)$ denotes the length of $\alpha$. Values of $a_\pi>1$ prefer variates that are dense and evenly distributed, i.e., all the values within a single sample are similar to each other. This property is inherited by the R2D2M2 prior as Proposition \ref{prop:originprior} shows, since selecting $a_\pi \leq 1/2$ will attempt to concentrate the total variance into few regression terms, whereas $a_\pi>1/2$ will distribute it evenly among the terms. Translation of prior beliefs over sparsity or lack thereof is thus possible by selection of the parameter $a_\pi$. 
%---------------------
 \begin{figure}[t!]
	\centering
	\includegraphics[keepaspectratio,width=\textwidth]{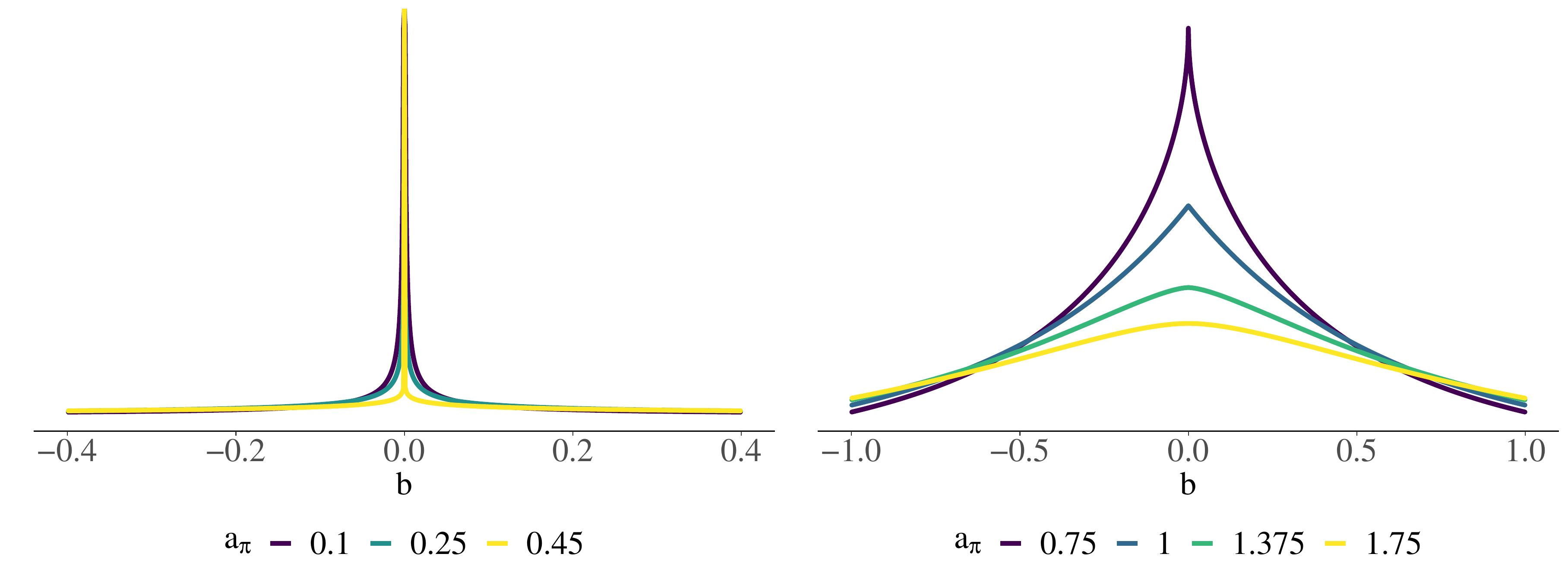}
	\caption{Marginal prior densities of the coefficients with $\sigma=1$. The value of $a_\pi$ controls the behavior around the origin. If $a_\pi \leq 1/2$ the marginal density increases towards infinity for values close to zero and is unbounded and non-differentiable at the origin. In contrast, if $a_{\pi}>1/2$ the marginal density is bounded also at the origin. When $a_\pi \geq 1$ the densities become differentiable at the origin. }
	\label{fig:priorb}
\end{figure}
%---------------------

Figure \ref{fig:priorb} shows the marginal density of $p(b_i|\sigma)$ for different values of $a_\pi$ with fixed $q_i^2=1$, the latter implying an equal scale of predictors and response variable without loss of generality. For the value of $a_\pi=0.25$, the prior distribution shows high concentration of mass near the origin and is undefined (singular) at zero, which will lead to stronger shrinkage of weak signals towards the origin. On the other hand, when $a_\pi=0.75$, the singularity is not present anymore and the prior becomes bounded at zero. A case of interest that presents bounded marginal priors is that of the (uniform) flat Dirichlet distribution over the simplex, that is when $a_\pi=1$, meaning that the use of the flat Dirichlet prior over $\phi$ enforces less sparsity. Both types of behavior can be beneficial, since they can play in favor of the prior beliefs about sparsity the user might have. 

The tails of the prior dictate whether or not it is possible that large signals can be detected (even when specific levels of sparsity are assumed) and how much these are shrunken towards zero \citep{Horseshoe, DirichletLaplace, vanDP2021theoretical}. It is in principle desirable that heavy tails are present in order to prevent over-shrinkage of true signals \citep{Horseshoe, PiironenHorseshoe}, which in turn can result in priors that are of bounded influence, i.e., sufficiently strong signals are left unshrunken by the prior. This behavior, known as tail robustness \citep{Horseshoe, vanDP2021theoretical}, is vital in sparse settings to be able to shrink coefficients near zero much more forcefully than those far from it. However, we should keep in mind that in practice, this behavior does not always plays in our favour, especially when parameters are weakly identified by the data. To circumvent this issue, authors such as \cite{PiironenHorseshoe, ShrinkingShoulders} have proposed \textit{regularized} versions of shrinkage priors that also regularize very strong signals to some degree.

Our proposed R2D2M2 prior is able to attain heavier tails than the Cauchy distribution and, when no varying terms are present (the R2D2 prior with normal base distributions), it is of bounded influence, as is shown in the following propositions.
%---------------------
\begin{prop}
\label{prop:tailprior}
Given $|b_i|\to \infty, |u_{ig_j}| \to \infty$ for any $a_\pi>0$ and $a_2>0$ the marginal prior densities satisfy 
\begin{align*}
 p(b_i|\sigma ) &\sim \mathcal{O}(1/|b_i|^{2a_2+1}), \ \ 
p(u_{ig_j}) \sim  \mathcal{O}(1/|u_{ig_j}|^{2a_2+1}).
\end{align*}
Moreover, when $0<a_2 \leq 1/2$ the R2D2M2 marginal priors have heavier tails than the Cauchy distribution. 
\end{prop}
%---------------------
%---------------------
\begin{prop}
\label{prop:r2d2bounded}
Assume there are no varying coefficients $u_{ig_j}$ in the model, then the R2D2M2 prior is of bounded influence, i.e., sufficiently strong signals are not shrunken.
\end{prop}
%---------------------

While we were not able to prove that bounded influence also holds for the R2D2M2 prior, we have no reason to believe that it does not hold. Indeed, the results from the simulations presented in Section \ref{sec:examples} show the bounded influence for the R2D2M2 prior empirically, at least for a range of different hyperparameter and data generating setups.

Propositions \ref{prop:originprior} and \ref{prop:tailprior} show the adaptability of the R2D2M2 prior, in the sense that the user can tune the hyperparameters according to their prior beliefs on the degree of sparsity and signal strength. However, one should keep in mind that the complexity of the problem rapidly increases when considering multilevel structures and care should be taken when selecting the hyperparameters as to neither overfit (too little sparsity) nor over-shrink (too much sparsity).  \\

\subsection{Shrinkage factors}
\label{subsection:shrinkagefactors}

 When performing Bayesian inference, the prior can be seen as a regularizer. This is clearly exemplified when using the Bayesian LASSO or other Bayesian shrinkage priors \citep{bayesianlasso,BayesPenalizedRegSara}, since the posterior modes under these priors correspond to certain frequentist regularizers \citep{DirichletLaplace}. However, usual frequentist properties are not precisely equivalent; for instance, the posterior distribution for a continuous random variable is not able to produce exact sparse results, as the latter would require a point mass at zero. When using shrinkage priors, some authors have been able to express the posterior means as shrunken versions of (unshrunken) frequentist estimators \citep{PiironenHorseshoe}, which in practice leads to a better out-of-sample predictive performance and a decrease in the variance of the estimators \citep{Ridge}. Several authors have used the concept of \emph{shrinkage factors} to study the amount of shrinkage the prior induces and the effect the different values of the hyperparameters \citep{Horseshoe,PiironenHorseshoe, BaiHypothesisNB, PolsonHalfCauchy}. Here, we present the theory of shrinkage factors for the R2D2M2 prior as well as for the original R2D2 prior, since it has not been done before. 
 
It can be shown that the conditional posterior distributions of the coefficients $b_i$ and $u_{ig_j}$ given $b_{-i}, u_{-ig_j}, \phi, \sigma, \tau, y$ in (\ref{r2d2m2model}) are normal with means given by 
%------------------------
\begin{equation}
\label{postbmeancond}
    \begin{aligned}
     \mathbb{E}(b_i|  b_{-i}, u,\sigma, \phi, \tau^2, y )&= \left( \frac{\sigma^2_{x_i}}{\phi_i 
    \tau^2}+ SS_i  \right)^{-1}  \sum_{n=1}^N e_{n}^{(-i)} x_{ni}     \\
    \mathbb{E}(u_{ig_j}| b, u_{-ig_j} ,\sigma, \phi, \tau^2, y )&= \left(  \frac{\sigma^2_{x_i}}{\phi_{ig}\tau^2} + SS_{ig_j}  \right)^{-1} \sum_{n \in L_{g_j}}  e_{n}^{(-ig_j)} x_{ni}, \\
    \end{aligned}
\end{equation}
%---------------------
where $b_{-i}$ and $u_{-ig_j}$ represent the exclusion of the $i$th overall coefficient and the $ig_j$ varying coefficient, respectively, $SS_i=\sum_{n} x^2_{ni}$ and $SS_{ig_j}=\sum_{n \in L_{g_j}} x_{ni}^2$. We use $e_{n}^{(-i)}$ and $e_{n}^{(-ig_j)}$ to represent the errors that would be obtained when excluding $b_{i}$ and $u_{ig_j}$ when trying to predict the $n$th observation. 
%--------------------
\begin{align*}
e_{n}^{(-i)} &=  y_n    -\sum_{i'\neq i} x_{ni'} b_{i'} -\sum_{i=0}^p  x_{ni} \left( \sum_{g \in G_i} u_{i g_{j[n]}} \right)   \\     
e_{n}^{(-ig_j)}&=  y_n - \sum_{i=1}^p x_{ni} b_i - \sum_{i'\neq i}\sum_{g \in G_{i'}} x_{ni'} u_{i'g_{j[n]}}- x_{ni} \sum_{g \in G_i, g'\neq g} u_{ig'_{j[n]}}.
\end{align*}

If we assume there are no varying terms in the model, i.e., we are considering the original R2D2 prior, and $\sum_{n} x_{ni} x_{nj} = 0$ for $i\neq j$ then the conditional posterior mean of $b_i$ is given by 
%-------------------------
\begin{align}
    \label{postbmle}
    \mathbb{E}(b_i|  \sigma, \phi, \tau^2, y )= \left( \frac{\sigma^2_{x_i}}{\phi_i 
    \tau^2}+ SS_i  \right)^{-1} SS_i \, \hat{b}_i, 
\end{align}
%-------------------------
where $\hat{b}_i$ is the maximum likelihood estimate (MLE) of $b_i$. Thus, the posterior mean under the R2D2 prior is a shrunken version of the MLE, a common behavior that is encountered when regularizing \citep{Ridge} or performing Bayesian inference with shrinkage priors such as the regularized horseshoe \citep{PiironenHorseshoe}. We are able to quantify the amount of shrinkage of $b_i$ from their MLE towards zero by introducing shrinkage factors denoted as $\kappa_i$, where $0 \leq \kappa_i \leq 1$. Let 
%---------------------
\begin{align}
    \label{defkappa}
    \kappa_i= \frac{1}{1+\frac{SS_i}{\sigma_{x_i}^2} \phi_i\tau^2}
\end{align}
%---------------------
then Equation \eqref{postbmle} becomes
%-------------------------
\begin{align}
    \label{postbmleshrinkage}
    \mathbb{E}(b_i| \sigma,\kappa_i, y )= (1-\kappa_i) \hat{b}_i.
\end{align}
%-------------------------
The shrinkage factor $\kappa_i$ quantifies the amount of shrinkage that is exerted by the use of the R2D2 prior, relative to the non-regularized MLE. Shrinkage values close to 1 indicate full shrinkage from the MLE towards 0 and vice-versa, values close to 0 indicate no or only minor shrinkage. 

When considering the presence of varying terms in the model, frequentist inference for $b_i$ is typically performed by first integrating out $u_{ig_j}$ \citep{lme4}. From a practical perspective, this is necessary as it is not possible to fully identify the overall and varying coefficients simultaneously when solving the optimization problem related to maximum likelihood. Thus, classical inference chooses to predict, rather than estimate,  the varying terms $u_{ig_j}$ after having obtained estimates $\hat{b}_i$ of the overall coefficients $b_i$ \citep{lme4,wakefield2013bayesian}. In Bayesian inference on the other hand, both unknown quantities $b_i$ and $u_{ig_j}$ are considered random variables and are present in the full posterior distribution. Considering this, the generalization of shrinkage factors from an MLE when including varying terms is not immediate, because the joint MLE of the $b_i$ and $u_{ig_j}$ does not exist in this case. However, Equations \eqref{postbmeancond} still show shrinkage is being carried out from some fixed quantities that involve the conditioned parameters. The approach to define shrinkage from quantities different than the MLE has been considered by other authors too \citep{BaiHypothesisNB,PolsonHalfCauchy}, although not in a multilevel context. Doing so here opens up the way to define shrinkage factors $\kappa_i$ and $\kappa_{ig_j}$ for overall and varying coefficients respectively by
%-------------------
\begin{equation}
    \label{def:kappa}
\begin{aligned}
\kappa_i= \frac{ 1 }{1 + r_i \phi_i\tau^2  }, \ \ \ \
\kappa_{ig_j}= \frac{ 1 }{1+r_{igj}\phi_{ig}\tau^2 },\\
\end{aligned}
\end{equation}
%------------------- 
where $r_i=\frac{ SS_i}{ \sigma^2_{x_i}}$ and $r_{ig_j} = \frac{ SS_{ig_j}}{ \sigma^2_{x_i} }$. Substituting in Equations \eqref{postbmeancond}  results in
%---------------------
\begin{equation}
    \label{postcoeffmeankappa}
    \begin{aligned}
    \mathbb{E}(b_i| b_{-i}, \sigma, \kappa_i, y ) &= (1-\kappa_i)  (SS_i)^{-1} \sum_{n=1}^N e_{n}^{(-i)} x_{ni} \\
      \mathbb{E}(u_{ig_j}| b, u_{-ig_j} ,\sigma, \kappa_{igj}, y ) &=  (1-\kappa_{ig_j})   (SS_{ig_j})^{-1}   \sum_{n \in L_{g_j}}  e_{n}^{(-ig_j)} x_{ni}. \\
    \end{aligned}
\end{equation}
%---------------------
Shrinkage factors are useful in providing insight on the influence of the prior hyperparameters $\mu_{R^2}, \varphi_{R^2}$, $\alpha$, and $a_\pi$ on the posterior distribution of each term. For instance, notice that in the extreme cases that $\alpha$ is chosen such that for a fixed $i$, $\phi_i \to 0$, then $\kappa_i \to 1$.  If $\phi_i \to 1$ we have $\kappa_i \to (1+ r_i \tau^2)^{-1}$, so regularization takes place even if a priori it is believed that all the explainable variance is explained by a single component $b_i$. Additionally, by comparing $\kappa_i$ and $\kappa_{igj}$ in Equation \eqref{def:kappa} we can see that if $\phi_i \approx \phi_{ig}$ (which can be expected a-priori if $\alpha= a_\pi(1,...,1)'$ when $a_\pi>1$  \citep{OnTheDirichlet}), then the varying coefficients will exhibited stronger shrinkage since $SS_{ig_j}\leq SS_i$ naturally, due to observations being partitioned into different levels. It is also worth mentioning that the shrinkage phenomenon present in the posterior of $u_{ig_j}$ is two-fold: (1) resulting from the usual shrinkage of varying coefficients that hierarchical models exhibit (i.e., partial pooling, \cite{gelman_hill_2006}) and (2) resulting from the local and global scales that are part of the R2D2M2 prior. 

\begin{figure}[t!]%
	\centering
  	\includegraphics[keepaspectratio, width=0.99\textwidth, height=0.4\textheight]{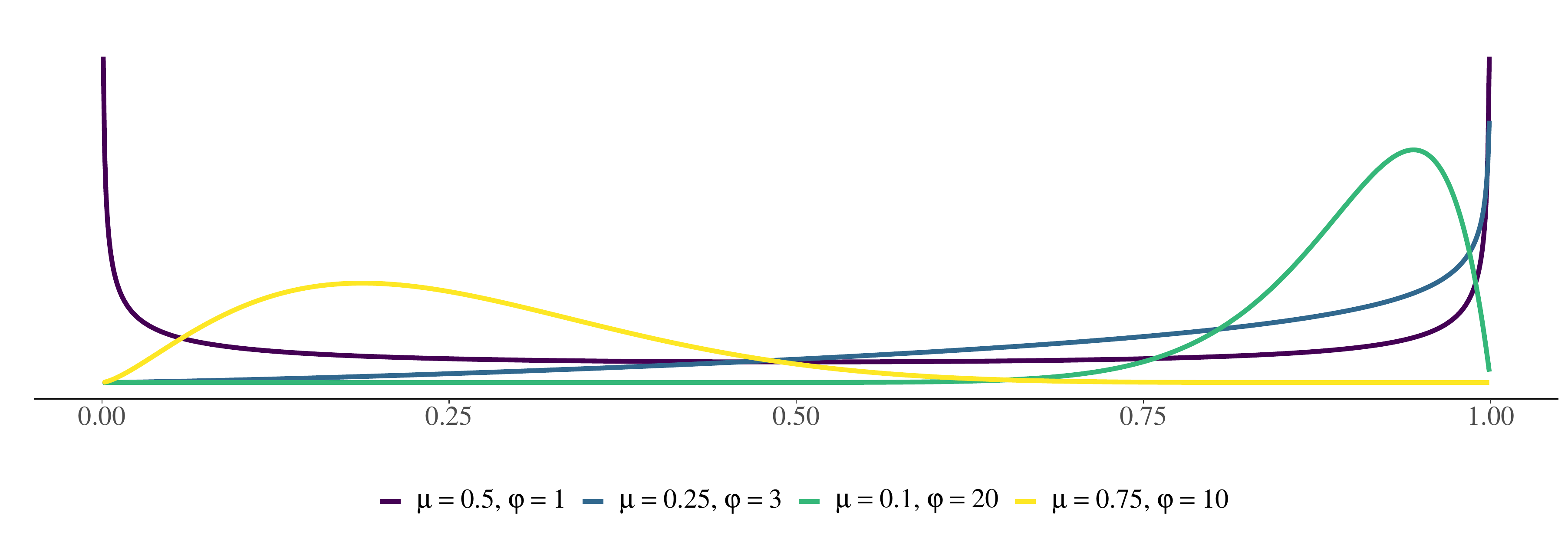}
	\caption{Densities of shrinkage factors $\kappa$ for several hyperparameter choices. The distribution offers  flexibility to express different shrinkage behaviors. Notice that when $\mu=0.5$ and $\varphi=1$ the distribution is symmetric, categorizing the coefficient as either complete noise ($k\approx 1$) or as a strong signal ($k\approx 0$).  }
	\label{fig:kappa_dist2}
\end{figure}

It is possible to find a closed expression for the densities of $\kappa_i$ and $\kappa_{ig_j}$, as well as for their moments. Consider Equations \eqref{def:kappa} with $\phi$ fixed and $\tau^2$ random. Using the fact that $\tau^2 \sim \betaprime ( \mu_{R^2}, \varphi_{R^2})$, the prior densities of the shrinkage factors, conditional on $\phi$ but integrated over $\tau$, are given by 
%------------------------
\begin{equation}
\label{distkappa}
    \begin{aligned}
	p \left( \kappa_i | \phi_i \right) &= \frac{ ( r_i \phi_i)^{a_2} }{\text{B}(a_1, a_2)}   (1-\kappa_i)^{a_1-1} \kappa_i^{a_2-1} \left(  ( r_i\phi_i-1)\kappa_i+1  \right)^{-a_1-a_2}, \\
	p \left( \kappa_{ig_j} | \phi_{ig} \right)&= \frac{ ( r_{ig_j} \phi_{ig})^{a_2} }{\text{B}(a_1, a_2)}   (1-\kappa_{ig_j})^{a_1-1} \kappa_{ig_j}^{a_2-1} \left(  ( r_{ig_j}\phi_{ig}-1)\kappa_{ig_j}+1  \right)^{-a_1-a_2} . \ \ 
    \end{aligned}
\end{equation}
%------------------------
The $m$-th moments of $\kappa_{i} | \phi_{i}$ and $\kappa_{ig_j} | \phi_{ig}$ with $m \in \mathbb{Z}^+$ are given respectively by
%------------------------
\begin{align*}
	\mathbb{E}\left(  \kappa_i^m | \phi_i \right) 
	&=  \frac{ (r_i \phi_i)^{a_2} }{\text{B}(a_1, a_2)} \text{B}(a_2+m, a_1)  \prescript{}{2}{F}_1   \left(   \xi , \beta, \gamma, z_i  \right)   , \\ 
	\mathbb{E}\left(  \kappa_i^m | \phi_i \right) 
	&=  \frac{ (r_{ig_j} \phi_{ig})^{a_2} }{\text{B}(a_1, a_2)} \text{B}(a_2+m, a_1)  \prescript{}{2}{F}_1   \left(   \xi , \beta, \gamma, z_{ig_j}  \right)   , \ \ \ 
\end{align*}
%------------------------
where $\prescript{}{2}{F}_1   \left(   \xi, \beta, \gamma, z  \right)$ is the hypergeometric function \citep{Zwillinger} with $\xi= a_1+a_2, \beta=a_2+m , \gamma= a_1+a_2+m$, $z_i= 1-r_i \phi_i$ and $z_{ig_j}=1-r_{ig_j}\phi_{ig}$ . In the important case of $m=1$, we have
%------------------------
\begin{equation}
    \label{expkappab}
\begin{aligned}
	\mathbb{E}\left(  \kappa_i | \phi_i \right)&=(r_i \phi_i)^{a_2} \left(  1-\mu_{R^2} \right)  \prescript{}{2}{F}_1   \left(   \xi , \beta, \gamma, z_i  \right) , \\
	\mathbb{E}\left(  \kappa_{ig_j} | \phi_{ig} \right)&=(r_{ig_j} \phi_{ig})^{a_2} \left(  1-\mu_{R^2} \right)  \prescript{}{2}{F}_1   \left(   \xi , \beta, \gamma, z_{ig_j}  \right).
\end{aligned}
\end{equation}
%---------
The distribution of the shrinkage factors $\kappa$ depends directly on the hyperparameters $\mu_{R^2}, \varphi_{R^2}$ specified for the prior on $R^2$ and is sufficiently flexible to attain different forms that represent a diversity of a priori beliefs as we show in Figure \ref{fig:kappa_dist2}. For example, the prior density of $\kappa$ can become unbounded near 0 or 1 (or both at the same time), favoring limited or full shrinkage respectively. It is also possible to obtain bounded densities, which can represent mild shrinkage. These differences influence how noise and large signals are treated, which is reflected in the values of posterior means \citep{Horseshoe}. Equations (\ref{expkappab}) show the effect of $\mu_{R^2}$ on the amount of shrinkage and how the prior is able to relate prior belief on $R^2$ with expected shrinkage on the coefficients. The expected shrinkage of each individual coefficient in the model is proportional to $1-\mu_{R^2}$, implying that $\mu_{R^2}$ acts in a global manner. This is expected due to the relationship between the model's global scale $\tau^2$ and $R^2$ established in Equation \eqref{eq:r2ft2}. Prior beliefs about $R^2$ are straightforwardly inherited to the amount of shrinkage the prior exhibits. If the user considers $R^2$ is low and represents this via a low value of $\mu_{R^2}$, then a noticeable amount of shrinkage will take place. Analogously, a high value of $\mu_{R^2}$ results in less shrinkage. 
%-----------

%-----------
Figure \ref{fig:kappa_dist2} shows that the density of $\kappa$ is able to reflect a horseshoe like prior for the amount of shrinkage when $(\mu_{R^2}, \varphi_{R^2})=(0.5,1)$. This type of prior can be considered fairly non-informative on the $\kappa$ scale, since it places 1/3 of its mass on $1/4 \leq \kappa \leq 3/4$, with  the rest of the density equally distributed to the left of 1/4 and to the right of 3/4. This shape indicates that we expect to see both relevant strong signals ($\kappa \approx 0$) and irrelevant variables a priori ($\kappa \approx 1$) \citep{Horseshoe, PiironenHorseshoe}.

\subsection{Effective number of non-zero coefficients}
\label{subsection:meff}
\cite{PiironenHorseshoe} proposed that the prior's hyperparameters can be understood intuitively by analyzing the imposed prior on the \textit{effective number of coefficients}. In the case of single level models they define the effective number of overall coefficients as
%--------
\begin{align}
\label{meffcoefoverall}
m_{\text{eff}_O}= \sum_{i=1}^{p}  (1-\kappa_i).
\end{align}
%----------
We generalize Equation (\ref{meffcoefoverall}) to multilevel models (\ref{r2d2m2model}) by also including the varying coefficients:
%--------
\begin{align}
\label{meffcoef}
m_{\text{eff}}= \sum_{i=1}^{p}  (1-\kappa_i)+\sum_{i \in \{0,...,p\}} \sum_{g \in G_i} \sum_{j \in J_g} (1-\kappa_{ig_j}).
\end{align}
%--------
When the shrinkage factors $\kappa$ are close to 0 or 1, resulting in no shrinkage and total shrinkage respectively, Equation \eqref{meffcoef} can help us understand how specific values for $\mu_{R^2}, \varphi_{R^2}$ determine the amount of unshrunken ("active") variables present in the model and can serve as an indicator of the effective model complexity. 

\begin{figure}[t!]%
	\centering
  	\includegraphics[keepaspectratio, width=\textwidth,]{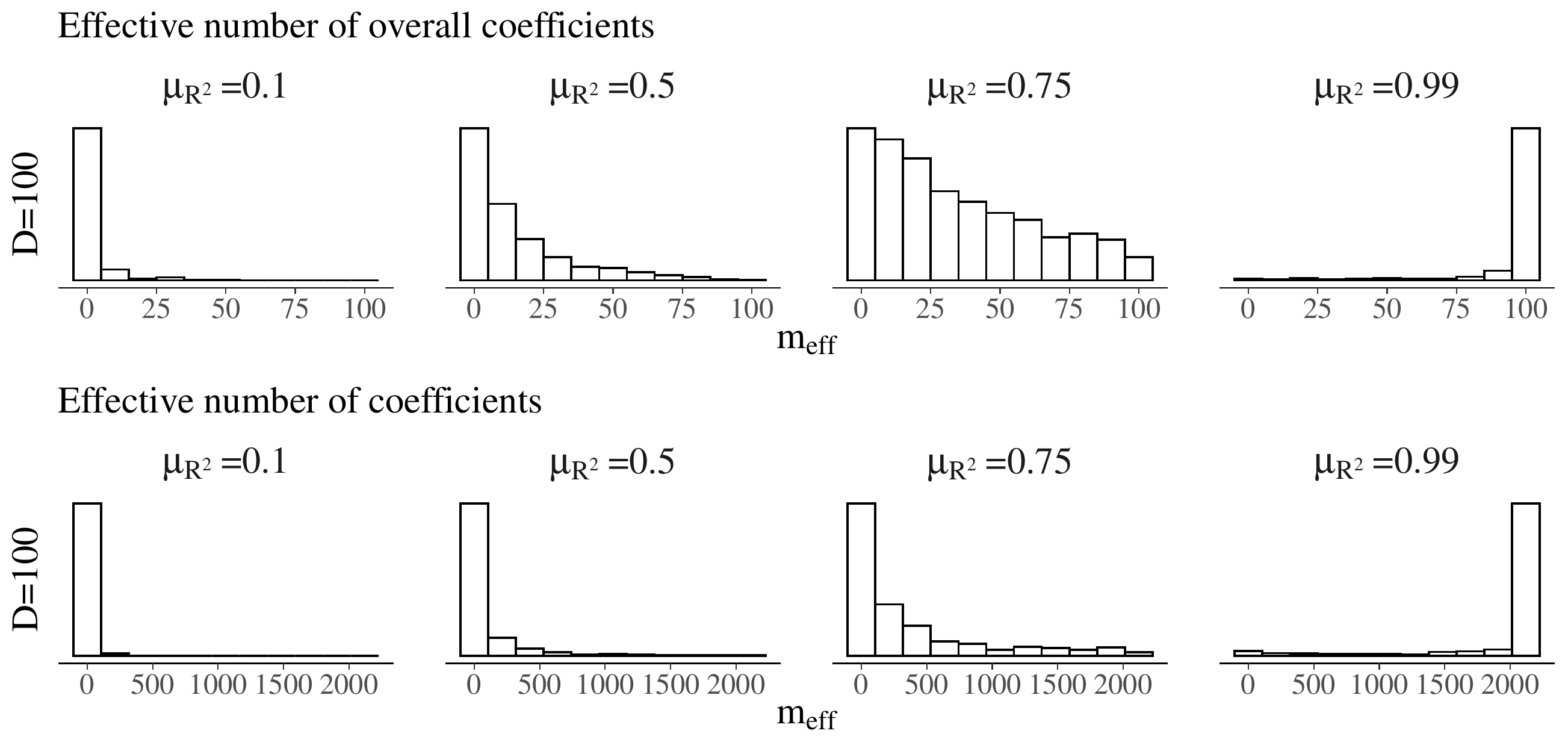}
	\caption{Prior densities for the effective number of coefficients under different conditions for $\mu_{R^2}$ with fixed $\varphi_{R^2} = 1, a_\pi=0.5$ and $p=100$ overall coefficients, one grouping factor and 20 levels.  Decreasing $\mu_{R^2}$ implies less signal expected a priori, which results in increased a priori shrinkage of the coefficients. These plots serve as an intuitive way to understand the effect the hyperparameters have on the amount of shrinkage. }
	\label{fig:meffcombined10}
\end{figure}

For a given fixed dataset, we can visualize the prior imposed on $m_{\text{eff}}$ for different hyperparameter values of $\mu_{R^2}, \varphi_{R^2}$. This can be done by simulating observations from $m_{\text{eff}}$ directly. To do so, we first generate random variates $\tau, \phi$ independently from their prior distributions, then we compute $\kappa_i, \kappa_{ig_j}$ from Equations \eqref{def:kappa} and finally calculate $m_{\text{eff}}$. Figure \ref{fig:meffcombined10} shows the histograms of the effective number of overall coefficients and total effective number of coefficients (which include the overall plus varying coefficients). In both cases, the behavior is as previously described; as $\mu_{R^2}$ increases the effective number of overall coefficients increases too and vice versa. 
%-----------

We recommend that the user visualizes the prior of $m_{\text{eff}}$ for different values of $\mu_{R^2}, \varphi_{R^2}$ to get a better idea of how the global shrinkage is affected by the selection of the hyperparameters; and to verify if this choice represents their prior expectations on the number of non-zero coefficients. The prior plots of $m_{\text{eff}}$ provide an intuitive way of understanding the induced overall shrinkage and can even become a valuable tool of communication between the user and non-specialized audiences.

\section{Numerical experiments and case studies}
\label{sec:examples}

We have implemented the R2D2M2 prior model in the probabilistic programming language Stan \citep{StanJSS, stan2022}. The corresponding code can be found in Appendix B as well as on {\myosfresults}. Based on this implementation, we performed two large-scale simulations studies and analysed a high-dimensional real-world data set using the R2D2M2 prior as detailed below.

\subsection{Simulation Based Calibration}
\label{subsection:SBC} 

First, we demonstrate that inference for our implementation of the R2D2M2 prior is probabilistically well calibrated. For this purpose, we applied simulation based calibration (SBC) \citep{taltssbc}. SBC is a general procedure for validating inferences from Bayesian algorithms capable of generating samples from posterior distributions. The procedure gives us the ability to identify inaccurate computations as well as inconsistencies in model implementations. In works by first sampling true parameter values $\tilde{\theta}$ from the prior, $\tilde{\theta} \sim p(\theta)$ and subsequently sampling data $\tilde{y}$ from the likelihood $\tilde{y} \sim p(y|\tilde{\theta})$. Afterwards, using the algorithm that needs to validated, we obtain $S$ samples $\{\theta_1,...,\theta_S  \} \sim p(\theta | \tilde{y})$ from the (approximated) posterior. Finally, for a quantity of interest $f(\theta)$, we calculate a rank statistic of how many values in $\{ f(\theta_1),..., f(\theta_n) \}$ fall below the true value $f(\tilde{\theta})$. If the algorithm is well calibrated, this rank statistic of the prior sample $f(\tilde{\theta})$ in relation to the posterior draws is uniformly distributed \citep{taltssbc}. We will use the terms posterior draws and samples interchangeably. 

To test uniformity of the rank statistics, we applied a intuitive graphical test proposed by \cite{teemubuerknergraphical}, which provides simultaneous confidence bands of the empirical cumulative distribution function (ECDF) that satisfy a pre-specified joint coverage assuming uniformity. For ease of illustration, the plots show differences between the perfectly uniform CDF (a diagonal line) and the ECDF. In other words, it rotates a regular ECDF plot by 45 degrees to the right to make the uniform CDF a flat line and make the relevant part of the plot stand out more clearly. The graphical test is straightforward to use and is implemented in the R package \texttt{SBC} \citep{sbcmanual}. 

To sample from the posterior distribution we use Stan \citep{StanJSS, stan2022}, a probabilistic programming language that provides the user with a (now substantially extended) implementation of the No-U-Turn Sampler (NUTS) from \cite{nuts}, an adaptive form of Hamiltonian Monte Carlo (HMC) sampling \citep{handbookmcmc}. To avoid sampling issues often encountered in hierarchical models  we use non-centered parametrizations of the model (see {\myosfresults} and \cite{ProbProgrammingGorinova} for details).  

Calibration is a function of both the probabilistic model under consideration and the posterior approximation algorithm. Here we have tested calibration of our Stan implementations of both the R2D2 and the R2D2M2 prior models, since both of them are highly practically relevant. Note that calibration tests for previous implementations of the R2D2 prior model have not been reported beforehand by other authors. 

There are two quantities that play a key role in how computationally intensive SBC can turn out to be: the number of simulation trials per configuration (i.e., the number of fitted models per configuration) and how many posterior draws to extract per simulation (i.e., per fitted model). We have fixed these to $T = 100$ and $S = 3000$, respectively. For MCMC samplers such as random walk Metropolis-Hastings or Gibbs sampling, this amount of posterior draws  might seem as a bit small, however for adaptive Hamiltonian Monte Carlo estimates, a few thousand samples is sufficient for most models \citep{brmsJSS, nuts,stan2022}; and, as we show below, is indeed sufficient for the here-considered models. 

We have tested calibration for a total of 96 different simulation configurations as detailed below. This amount comes from fully crossing all the factors that are varied in the simulations. An overview of the different configurations can be found in Appendix \ref{section:appendixC} in Table \ref{tab:sbchyperparams}. The different conditions are chosen to represent several realistic scenarios that analysts might encounter in a similar fashion in real-life data analysis. 

To sample prior values $\tilde{\theta}$ from the R2D2 or the R2D2M2 prior we need to specify several quantities: the prior mean $\mu_{R^2}$ and prior precision $\varphi_{R^2}$ of the prior on $R^2$ as well as the concentration parameter $a_\pi$ of the Dirichlet prior on $\phi$. We chose $\mu_{R^2} \in \{0.1, 0.5 \}, \varphi_{R^2} \in \{0.5,1 \}$, and $a_\pi \in \{0.5, 1 \}$. The number of grouping factors $K$ was chosen as $K \in \{0, 1\}$, implying either an R2D2 or an R2D2M2 prior. The number of covariates $p$ was varied in $p \in \{10,100,300\}$. When considering the R2D2M2 prior, we used $q=p+1$ varying terms per grouping factor $K$, that is, varying coefficients for each covariate plus a varying intercept. The number of levels per grouping factor was held constant with a value of $L=20$. Finally, the residual standard deviation $\sigma$ was sampled from a half Student-$t$ with 3 degrees of freedom and scale of 1 \citep{GelmanHalfStudentt} and the overall intercept $b_{0}$ was sampled from a centered normal distribution with standard deviation of 5. 

The covariates $x_i, i=1,...,p$ were sampled independently of $\tilde{\theta}$ from a multivariate normal distribution centered at the origin with a covariance matrix $\Sigma_x$, formed from a correlation matrix $\rho_x$ which has  an autoregressive structure $\text{AR}(1)$. The correlations considered here were $\rho \in  \{0,0.5\}$. Setting $\rho=0$ results in $\Sigma_x=I_p$ such that predictor variables are sampled independently. Finally, the outcome data $\tilde{y}$ was sampled as per Equation \eqref{eq:lklhd}. 

The obtained SBC results indicate that inference for our implementations of the R2D2 and R2D2M2 prior models is well calibrated in all considered configurations (see {\myosfresults} for the full results). For brevity, we only showcase a few selected results below. Specifically, we consider $(p,\mu_{R^2}, \varphi_{R^2}, a_{\pi})=( 100, 0.5, 1,0.5)$ and $K \in \{0, 1\}$. The absence of grouping factors, $K=0$, implies the use of the R2D2 prior.  

\begin{figure}[t!]%
	\centering
	\includegraphics[keepaspectratio, width=0.96\textwidth, ]{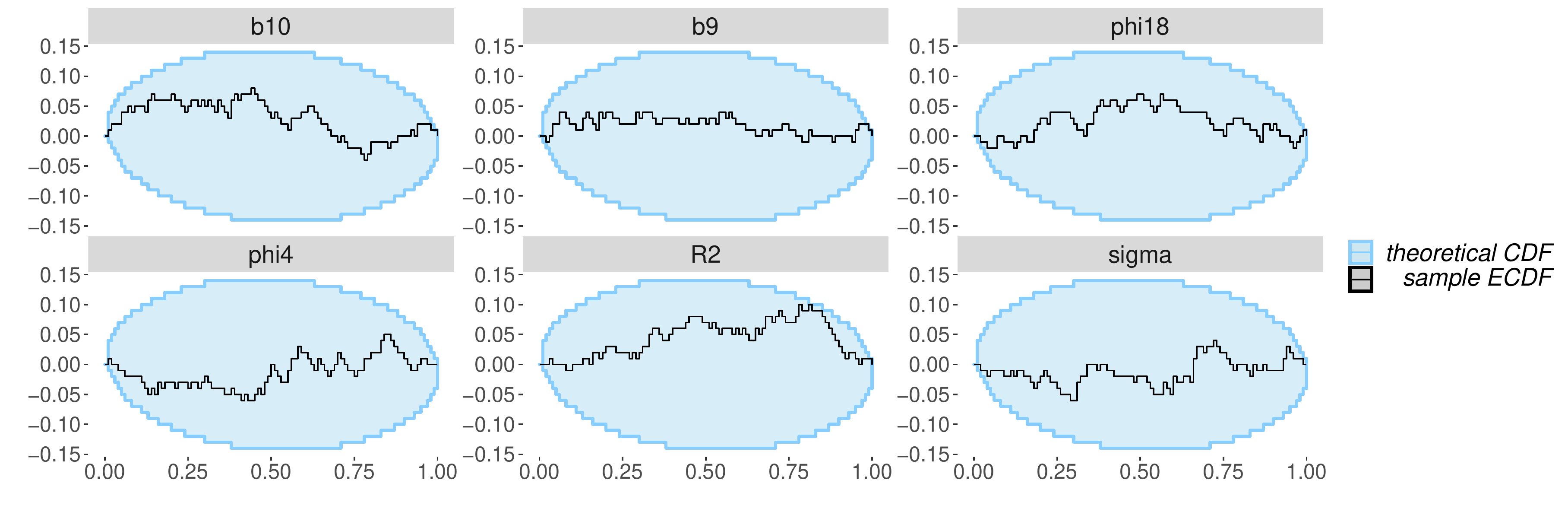}
	\caption{ECDF difference plots for the distributions of randomly chosen parameters in the R2D2 model. The blue areas in the ECDF difference plots indicate 95\%-confidence intervals under the assumptions of uniformity and thus allow for a null-hypothesis significance test of self-consistent calibration. The plots show that proper calibration has been achieved.}
	\label{fig:sbcresults-r2d2}
\end{figure}

\begin{figure}[t!]%
	\centering
	\includegraphics[keepaspectratio, width=0.96\textwidth, height=0.25\textheight]{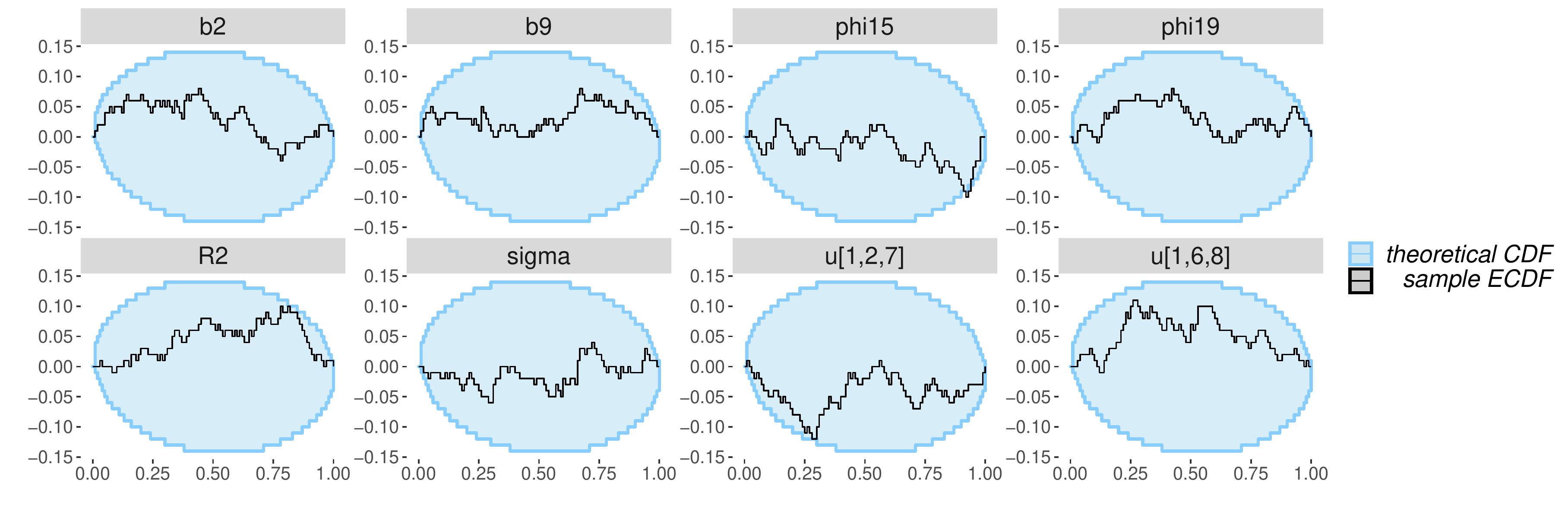}
	\caption{ECDF difference plots for the distributions of randomly chosen parameters in the R2D2M2 model. The plots show that proper calibration has been achieved.}
	\label{fig:sbcresults-r2d2m2}
\end{figure}

Figures \ref{fig:sbcresults-r2d2} and \ref{fig:sbcresults-r2d2m2} show ECDF difference plots of several model parameters for the R2D2 and R2D2M2 models, respectively. For the R2D2 model, we show results for $R^2$, $\sigma$, two overall coefficients and two components of $\phi$ chosen randomly. For the R2D2M2 model, we additionally show two varying coefficients chosen randomly too. Selection of the presented terms can be arbitrary, since the coefficients are exchangeable in the simulation process of all configurations. If ranks were perfectly uniform, the plots would display a horizontal line, but deviations from exact uniformity are to be expected because of the simulation process. All trajectories inside the shaded blue area indicate good calibration. 
These results provide evidence that inference for our implementation is correctly calibrated for the quantities of interest we have investigated (i.e., all model parameters). However, as \cite{sbcmanual} mention, SBC provides only a necessary but not sufficient condition for correct calibration. It is possible the incorrect calibration still occurred for some pushforward quantities of the model, although unlikely given the good calibration of all model parameters \citep{taltssbc}.
%-----------
%-----------
%-----------
%-----------
%------
\subsection{Simulations from sparse multilevel models}
\label{subsection:GeneralSimMLM}
\subsubsection{Data generation of multilevel models}
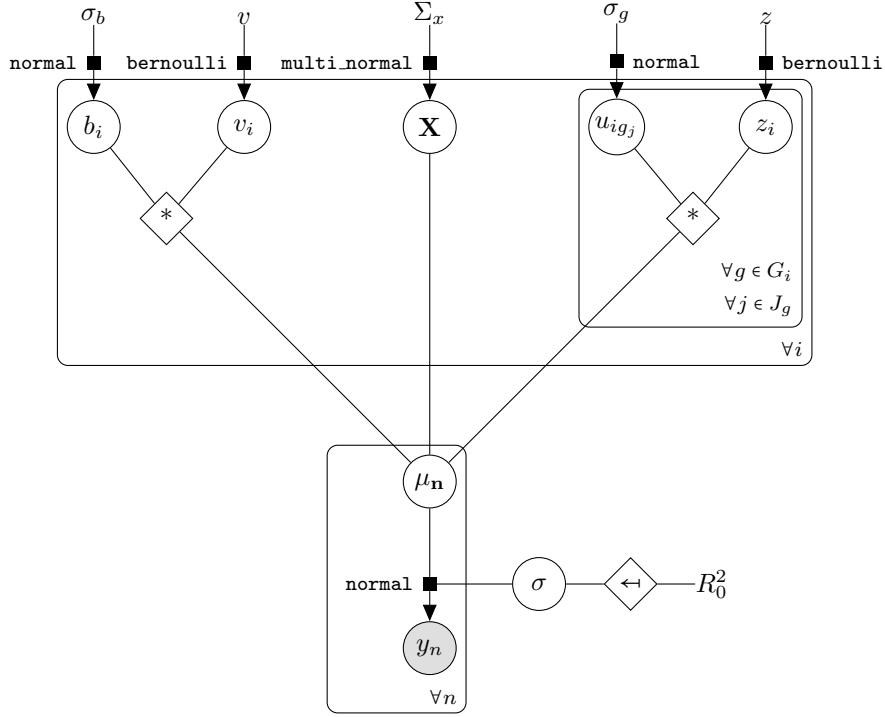
\begin{figure}[t!]
\centering
\begin{tikzpicture}

 %--- Y
 \node[obs]                               (y) {$y_n$};
 
  %---- mu_n
 \node[latent, above=1.5cmof y] (mun) {$\mathbf{\mu_n}$};
 
  %---- X
  \node[latent, above=4cm of mun, xshift=0cm]  (x) {$\mathbf{X}$};
 
%------ dot dot b
  \node[det, below= .5cm of x, xshift=-3.5cm]            (dotb) {*} ; %
  
 %------ dotu dot uigj
  \node[det, below= .5cm of x, xshift=3.5cm]            (dotu) {*} ; % 
 
 %---- b_i, v_i
 \node[latent, left=3.75 of x]   (bi) {$b_i$};
 \node[latent, left=1.75 of x]   (vi) {$v_i$};

 %---- u_igj, z_i
 \node[latent, right=1.75 of x]  (uigj) {$u_{ig_j}$};
 \node[latent, right=3.75 of x]  (zi) {$z_i$};
 
%-------Hyperparams
% bi, vi 
\node[const, above=1  of bi]  (sigmabi) {$\sigma_b$}; %
\node[const, above=1 of vi]  (pvi) {$v$}; %

% uigj, zi 
\node[const, above=1  of uigj]  (sigmauigj) {$\sigma_g$} ; %
\node[const, above=1 of zi]  (pzi) {$z$} ; %

 % X 
 \node[const, above=1  of x, xshift=0cm]  (sigmax) {$\Sigma_x$} ; %

%---- Factors
%--x
\factor[above=of x] {x-f} {left: \texttt{multi\_normal}} {sigmax} {x} ; %

%-- bi
\factor[above=of bi] {bi-f} {left: \texttt{normal}} {sigmabi} {bi} ; 
%-- vi
\factor[above=of vi] {vi-f} {left: \texttt{bernoulli}} {pvi} {vi} ; 

%-- uigj
\factor[above=of uigj] {uigj-f} {right: \texttt{normal}} {sigmauigj} {uigj} ; 
%-- vi
\factor[above=of zi] {zi-f} {right: \texttt{bernoulli}} {pzi} {zi} ;

 %--- y
 
\factor[above=of y] {y-f} {left: \texttt{normal}} {mun} {y} ;

%--- plates
\plate {uplate} { %
    (uigj)
    (zi)
    (dotu)
}{$ \begin{aligned}
    \forall g 
\in G_i\\ \forall j \in J_g
\end{aligned}$ };

\plate {bplate} { %
    (bi)%
    (vi)%
    (dotb) %
    (uplate)
}{$ \forall i$ };

\plate {yplate} { %
    (y)(y-f)(y-f-caption) %
    (mun) %
  } {$\forall n$} ;

 % sigma
\node[latent, right=1cm of y-f] (sigma) {$\sigma$}; %

% dot sigma

 \node[det, right=0.5cmof sigma]            (dotsigma) {$\leftmapsto$} ; % 

 % R2
\node[const, right=0.5cm of dotsigma]         (r2)   {$R^2_0$}; %

 % edges 
 % Connect bi and vi to the dot node
  \edge[-] {bi,vi} {dotb} ;

  \edge[-] {uigj,zi} {dotu} ;
  
  \edge[-] {dotb,dotu,x} {mun} ;
  
  %\edge[-] {r2,mun} {dotsigma} ;
  \edge[-] {r2} {dotsigma} ;
  
  \edge[-]{sigma}{y-f}
  
  \edge[-]{dotsigma}{sigma}
  
\end{tikzpicture}
  \caption{ A graphical model showing the data generating process for the simulations. The $*$ symbol denotes the usual product and $\leftmapsto$ represents the use of Equations (\ref{eq:r2fvarmu}) and (\ref{eq:vardecomposition}) to find the value of $\sigma$ to maintain the pre-specified value of $R_0^2$. The normal distributions considered for the coefficients $b_i,u_{ig_j}, \Sigma_x$ have an expected value of 0. The mean of $y_n$, given by $\mu_n$, is calculated using Equation \eqref{eq:lklhd}.  }
  \label{fig:datagen}
\end{figure}

We conducted simulation studies to analyze the performance of the R2D2M2 prior under different scenarios. To generate multilevel data we follow the approach of \cite{catalina2020projection}. Their simulation approach is sufficiently general to encompass the different forms (sparse or non-sparse) multilevel data that might be encountered in practice. A graphical model that represents the data generating process is shown in Figure \ref{fig:datagen} (see \cite{TikzBayesNet} for more details on this type of diagrams). The different data generating conditions are summarized in Table \ref{tab:sim2datahyperparams} in Appendix \ref{section:appendixC}. 

Again, we included all possible $q=p+1$ varying terms per grouping factor $K \in\{1,3\}$, with $p\in \{10,100,300\}$. The number of levels per grouping factor remains constant with a value of $L=20$. The covariates $x_i, i=1,...,p$ are generated from a multivariate normal distribution centered at the origin with a covariance matrix $\Sigma_x$, formed from a correlation matrix $\rho_x$ which has  an autoregressive structure $\text{AR}(1)$. The correlations considered were $\rho \in  \{0,0.5\}$.  The overall intercept $b_{0}$ and the overall coefficients $b_i, i=1,...,p$ are simulated independently from a normal distribution with mean zero and variances $\sigma_I^2=4, \sigma_b^2=9$, respectively. For a fixed covariate $i$, group $g$ and level $l$, the varying coefficients $u_{ig_j}$ are simulated from a normal distribution with mean zero and variance $\sigma_g^2=4$. 

To induce sparsity in the data generating process we set each coefficient $b_i, u_{ig_j}$ to zero with probabilities $v, z$ respectively. If an overall coefficient is set to zero, then we also set the corresponding varying counterparts to zero (i.e if $b_i=0$ then $u_{ig_j}=0,\, \forall g, j$). The values for $v$ were set to $\{0.5,0.95\}$ and $z$ was always set equal to $v$. The value of the residual standard deviation $\sigma$ is adjusted in each simulation to maintain the prespecified value for the true proportion of explained variance $R^2_0 \in \{0.25, 0.75\} $, to do so we used Equations (\ref{eq:r2fvarmu}) and (\ref{eq:vardecomposition}).
%-----------------------------
%-----------------------------
\subsubsection{Prior Hyperparameters}
\label{subsubsection:priorhyperparams}
In the model, we set the $R^2$ prior hyperparameters to $\mu_{R^2}\in \{0.1,0.5\}$, $\varphi_{R^2} \in \{0.5,1\}$, which leads to $a_2 \in \{ 0.45, 0.25, 0.9, 0.5\}$. Values for which $a_2 \leq 1/2$ allow the marginal prior distributions of the regression terms to have heavier tails than the Cauchy distribution (see Proposition \ref{prop:tailprior}) and vice-versa. The hyperparameter $\alpha$ is set to $\alpha=(a_\pi,...,a_\pi)'$ with $a_\pi \in \{0.5,1\}$ to test the different conditions near the origin, since values of $a_\pi \leq 1/2$ lead to unbounded marginal priors and vice-versa (see Proposition \ref{prop:originprior}). Therefore, after fully crossing all these factors, we have a total of 8 different hyperparameter setups, which are summarized in Table \ref{tab:sim2hyperparams}. For brevity in the upcoming discussion of results, we will only refer to the indexes of these setups rather than to the specific hyperparameter values. 
%--------------------------
\begin{table}[ht]
\caption{Configurations of the prior hyperparameters that were selected to test the R2D2M2 model under a sparse multilevel simulation setup.  }
\label{tab:sim2hyperparams}
\centering
\scalebox{1}{
\begin{tabular}{lrrrrrrrr}
  \hline
Hyperparameter  & 1 & 2 & 3 & 4 & 5 & 6 & 7 & 8 \\ 
  \hline
$a_\pi$ & 0.5 & 1.0 & 0.5 & 1.0 & 0.5 & 1.0 & 0.5 & 1.0 \\ 
  $\mu_{R^2}$ & 0.1 & 0.1 & 0.5 & 0.5 & 0.1 & 0.1 & 0.5 & 0.5 \\ 
 $\varphi_{R^2}$ & 0.5 & 0.5 & 0.5 & 0.5 & 1.0 & 1.0 & 1.0 & 1.0 \\ 
   \hline
\end{tabular}
}
\end{table}
%--------------------------
The selected values of the hyperparameters were chosen to create scenarios in which the  marginal priors of the regression terms exhibit (un)boundedness near the origin and heavier (lighter) tails than the Cauchy distribution. For instance, consider setup (5) in Table \ref{tab:sim2hyperparams}, when $a_\pi= 0.5$ and $(\mu_{R^2}, \varphi_{R^2})=(0.1,1)$ (leading to $a_2=0.9$) we can expect that the prior will exert high shrinkage while presenting lighter tails. In this case, the prior is unbounded near the origin and there is major concentration of mass near the origin. In contrast, we can expect setup (4) to exert comparably little shrinkage, since the combination of boundedness ($a_\pi=1$) at the origin and heavy tails ($a_2=0.25$) of the marginal priors will lead the R2D2M2 prior to shift its mass to the tails, thus focusing even more on detecting relevant signals. These different behaviors show the flexibility of the R2D2M2 prior. Similar analyses can be conducted for the other hyperparameter setups. The prior for the residual standard deviation $\sigma$ was chosen as a half Student-$t$ distribution with scale $\eta = \text{sd}(y)$ and $\nu=3$ degrees of freedom, allowing for a finite variance but still with a heavy right tail.

The simulation setup provides an amount of 48 different data generation configurations, on which each of the different 8 hyperparameter setups will be tested. After fully crossing these conditions we have a total of 384 different simulation configurations. Table \ref{tab:sim2datahyperparams} in Appendix \ref{section:appendixC} summarizes the different conditions we have considered. For each configuration, we generated $T = 40$ datasets consisting of $N = 200$ training observations each. Out-of-sample predictive metrics were computed based on $N_{\rm test} = 200$ test observations independent of the training data. The test set was designed to include all current grouping factors and levels that are already in the training test, i.e., we are not performing prediction of new levels. To study the performance of the Bayesian models with the R2D2M2 prior, we have made use of several metrics. The complete list of metrics recorded for the experiments can be found in our OSF repository \footnote{\myosfresults}. The primarily important ones are explained as they come up below. 

\subsubsection{Analysis of results}
\label{subsubsection:analysis}
We show an analysis for three out of the total 48 different data generation configurations, whose results are representative of the overall results. They describe how the R2D2M2 prior performs in these scenarios with the different hyperparameter setups shown in Table \ref{tab:sim2hyperparams}. The complexity of the different scenarios considered increases in terms of number of coefficients included in the model. For the selected configurations $K=1,3, \rho=0, v=z=0.95, R^2_0= 0.75$ and $p \in \{10,100,300 \}$. For the different values of $p$, the total number of coefficients included in the linear predictor are  $\{230, 2120, 6320\}$ when $K=1$ and $\{670, 6160, 18360\}$ when $K=3$, respectively. The $K=1$ and $K=3$ scenarios paint qualitatively the same picture. Of course, model performance in $K=3$ scenarios was uniformly worse as it is considerably harder, but no new patterns (e.g., in terms of prior hyperparameter choice) emerged. For brevity, the results for $K=3$ are only shown in Appendix \ref{section:appendixC}. The combination of $R^2_0=0.75$ and $p_i=0.95$ implies that  few (strong) signals account for most of the variability of $y$, a scenario which represents that only a handful of covariates have a substantial effect on the response. The R2D2M2 prior performs similarly for the other data generation setups and the results can be found on {\myosfresults}.  

\paragraph{Estimation error}\mbox{}

We present summaries for several metrics in Table \ref{tab:predtab1} considering $K=1$. Similar results can be found when $K=3$ in Table $\ref{tab:predtab3}$ in Appendix \ref{section:appendixC}. The Root Mean Square Error ($\rmse$) of parameter estimates (as represented by the parameter's posteriors) was calculated for all the parameters using the posterior mean as an estimator across the $T$ replications. The $\rmse$ is an indicator of estimation error and balances bias and variance. A low value is desirable with zero indicating the ideal case (perfectly precise and unbiased estimation). In this sense, it can be seen as criterion for proper parameter recovery. As $p$ increases, the task of parameter recovery rises in complexity; however, $\rmse$ values are low even in the most difficult scenarios, when $p=300$. 

%--- K=1
\begin{table}[t!]
\centering
\caption{Predictive Table results with $K=1$. }
\resizebox{\textwidth}{!}{
\begin{tabular}{llrrrrrrrr}
\hline
\multicolumn{10}{c}{Scenario $K=1, \rho=0,  p_i=0.95, R_0^2=0.75, N=200 , \alpha=0.05$} \\
\hline
p & Metric & 1 & 2 & 3 & 4 & 5 & 6 & 7 & 8 \\ 
  \hline
10 & RMSE & 0.02 & 0.03 & 0.02 & 0.02 & 0.02 & 0.02 & 0.01 & 0.02 \\ 
   & elpd & -434 & -427 & -441 & -449 & -430 & -433 & -430 & -438 \\ 
   & $\meff$ & 32 & 38 & 33 & 44 & 31 & 41 & 31 & 45 \\ 
  100 & RMSE & 0.02 & 0.02 & 0.02 & 0.02 & 0.02 & 0.02 & 0.02 & 0.01 \\ 
   & elpd & -1271 & -1712 & -1319 & -1654 & -1029 & -1567 & -1388 & -1292 \\ 
   & $\meff$ & 334 & 485 & 333 & 548 & 289 & 481 & 382 & 440 \\ 
  300 & RMSE & 0.03 & 0.03 & 0.02 & 0.02 & 0.02 & 0.02 & 0.02 & 0.02 \\ 
   & elpd & -3848 & -3572 & -4219 & -3340 & -3234 & -3335 & -3708 & -3367 \\ 
   & $\meff$ & 1412 & 1662 & 1560 & 1756 & 1111 & 1208 & 1330 & 1652 \\ 
   \hline
\end{tabular}
}
\label{tab:predtab1}
 \begin{tablenotes}
      \small
      \item  Results are shown for the Root Mean Squared Error (RMSE), expected log-pointwise predictive density (\elpd) on the test set and the effective number of nonzero coefficients $\meff$. The numbers 1-8 indicate the hyperparameter setup considered.
\end{tablenotes}
\end{table}

\begin{figure}[t!]%
	\centering
	\includegraphics[keepaspectratio, width=0.99\textwidth, height=0.25\textheight]{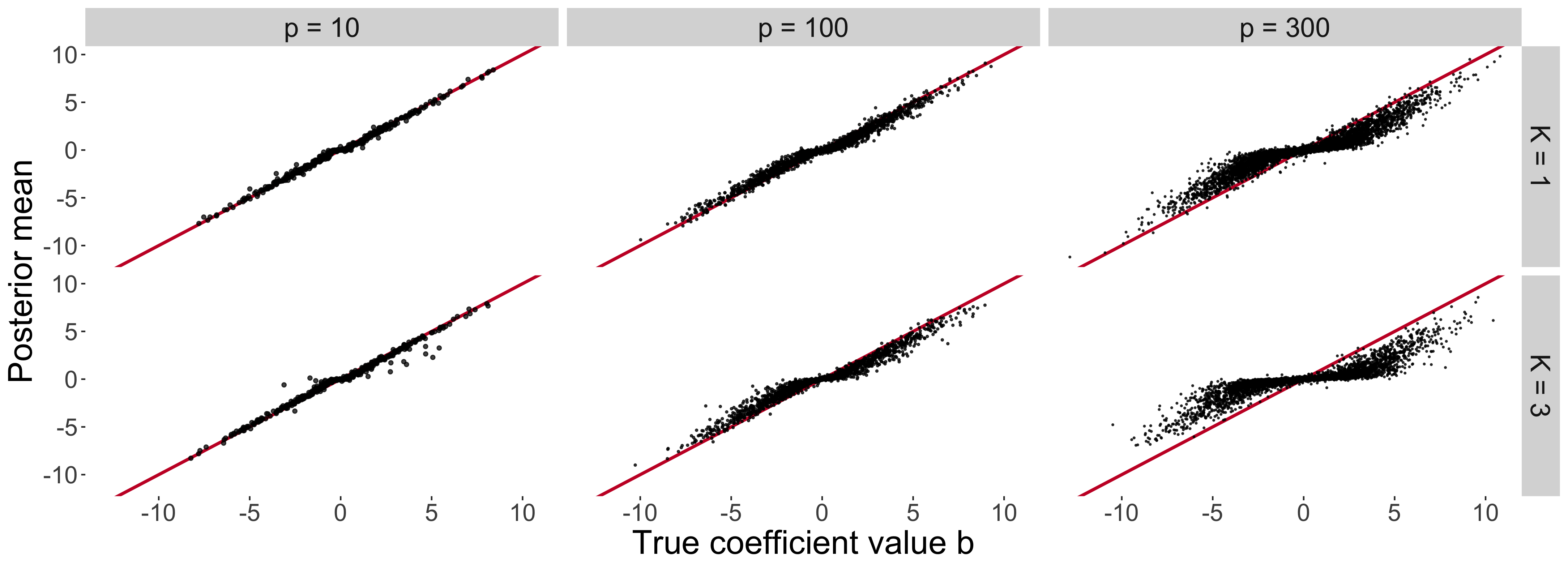}
	\caption{Posterior shrinkage of the overall coefficients pooled over the different hyperparameter setups and simulations. We show the true value versus the posterior mean. The red diagonal line is the identity function. Shrinkage is higher for weaker (in magnitude) signals resulting in ''chair-like" plots. As the complexity increases, the prior increases the threshold above which (in absolute terms) shrinkage is noticeably weaker. }
	\label{fig:posteriorbshrinkage}
\end{figure}

 Figure \ref{fig:posteriorbshrinkage} shows the posterior shrinkage towards zero on the overall coefficients for the case in which $K=1,p=300$ when pooling over the different hyperparameter combinations and all simulations together. To show this, we compare the true value of the coefficient $b$ with its posterior mean. The red diagonal line is the identity function and points that fall exactly on it indicate perfect posterior point estimation and no shrinkage. If the true $b$ is positive (negative) and the corresponding dot is below (above) the diagonal line, then the posterior is shrinking the term towards zero. This 'chair-like' appearance is usually exhibited by shrinkage priors \citep[see][for more details]{Horseshoe, BayesPenalizedRegSara} and is present under the different setups we have considered. This behavior indicates that the prior is exerting more shrinkage on weak signals and is consistent with the bounded influence and tail robustness properties the R2D2M2 prior possesses. Sufficiently strong signals are correctly detected and deemed as important, hence being shrunken less by the prior. 
 
We see that, as the complexity of the problem increases (e.g., going from $K=1$ to $K=3$), the model responds by performing more shrinkage and by increasing the threshold a true coefficient needs to exceed to remain relatively unshrunken. This is especially clear for the $p=300$ case since, as $K$ increases, a flatter and wider region appears in the center of the chair-like plot, thus representing that the threshold for detection and bounded influence has increased and everything below it will be highly shrunken. This is an expected behavior of robust shrinkage priors \citep{Horseshoe}. As complexity increases, they react by shrinking more and thus being more strict as to what they consider as a relevant signal. 
%-----------
%-----------  
%
\paragraph{Out-sample-predictive performance and shrinkage}\mbox{}\\

\begin{figure}[t!]%
	\centering
	\includegraphics[keepaspectratio, width=0.99\textwidth, height=0.23\textheight]{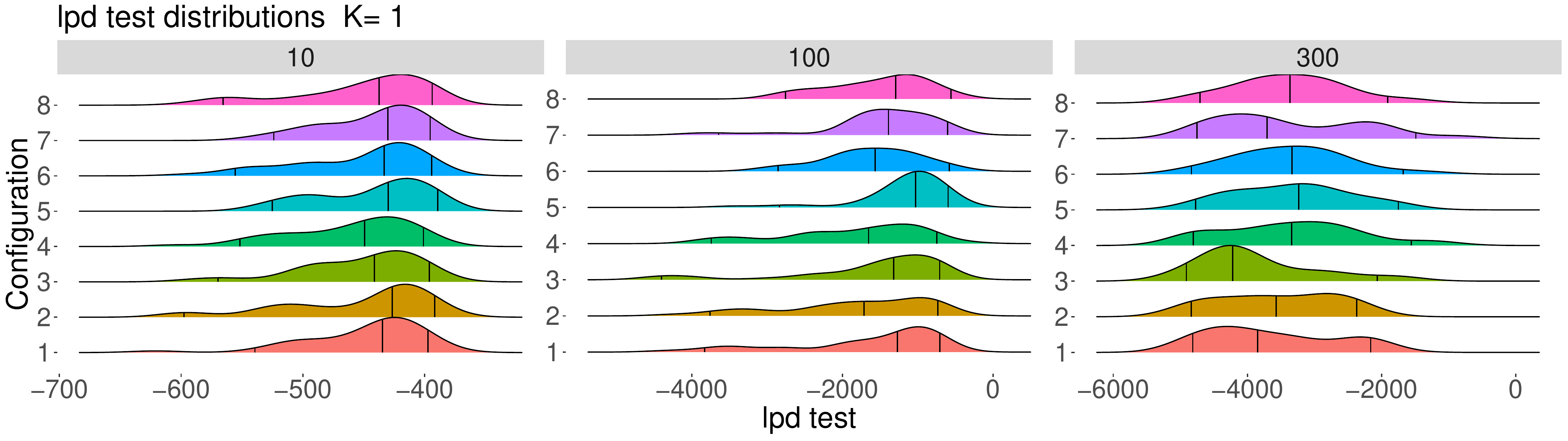}
	\caption{Estimated densities of the $\elpd$ estimators on the test set as $p$ increases and arranged by hyperparameter setups when $K=1$. The variation depicted is across $T=40$ simulation trials. The vertical lines inside each density represent the $5\%, 50\%, 95\%$ quantiles.}
	\label{fig:lpd_test_1_ridges}
\end{figure}

To assess predictive performance we estimate the expected log-pointwise predictive density (\elpd) defined by \cite{AkiSurvey} as 
\[ \elpd=  \sum_{i=1}^N \int p_t(\tilde{y})\log p(\tilde{y}_i|y) d\tilde{y}_i ,  \]
where $p(\tilde{y}_i|y)$ is the posterior predictive distribution and $p_t(\tilde{y}_i)$ is the distribution representing the true data generating process. The $\elpd$ is as measure of predictive accuracy for the $N$ data points taken one at a time and measures the goodness of the whole predictive distribution. We compute and report an estimator for $\elpd$, denoted as $\lpdhat$ on the test set to evaluate predictive performance. Denote the draws from the posterior distribution by $\theta^{(s)}$ where $s=1,...,S$, then the $\elpd$ can be estimated by
%--------
\[  \widehat{\text{elpd}}= \sum_{i=1}^N \log \left(\frac{1}{S} \sum_{s=1}^S p(y_i|\theta^{(s)})  \right) .       \]
%--------
We proceed to discuss how out-of-sample predictive performance (as measured via elpd) is intrinsically related to the amount of shrinkage done by the prior. We also discuss how prediction can be improved by imposing a higher amount of shrinkage via the hyperparameters. As a reference, the $\elpd$ values on the training set are usually close to $-400 \pm 50$ for all cases, even as the number of covariates increases. Considering the test data data, Table \ref{tab:predtab1} shows that in the comparably simple case of $p=10$, all hyperparameter setups are providing similar point estimates for the $\elpd$ on test data and that it is similar to the $\elpd$ on training data. Figure \ref{fig:lpd_test_1_ridges} shows the distributions  of the point estimator of $\elpd$ pooled across the different $T=40$ trials arranged by setup and number of covariates along with $95\%$ confidence intervals and the median. When $p=10$, the distributions have similar behavior, indicating that under this relatively simple case (few predictor terms), we are able to properly generalize under the different hyperparameter setups.  
%-----------
%-----------

When $p=100$, Table \ref{tab:predtab1} shows that there is a noticeable variability between the results of the point estimates of the $\elpd$ in the test set. Figure \ref{fig:lpd_test_1_ridges} shows that the distributions have different concentration around their median and in the width of their $95\%$ confidence intervals. The fact that $\elpd$ decreases (while holding the number of test observations constant) is an indicator that the problem becomes more complex, to which the prior responds by increasing the amount of shrinkage it exerts. 

%-----------
 \begin{figure}[t!]%
	\centering
	\includegraphics[keepaspectratio, width=0.99\textwidth, height=0.20\textheight]{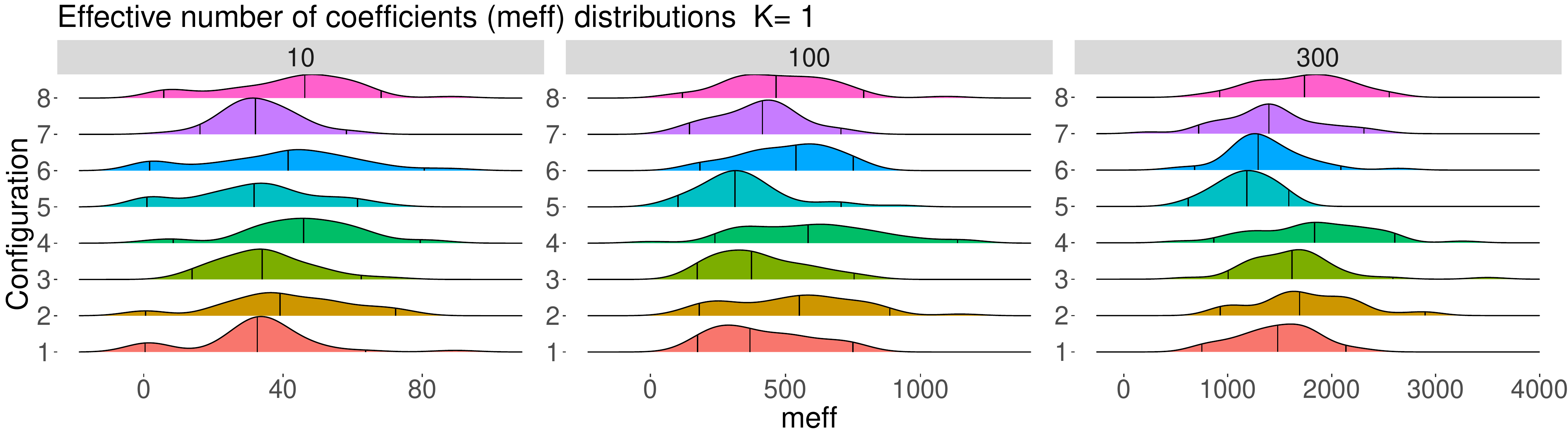}
	\caption{Estimated densities of the posterior median of the effective number of coefficients per hyperparameter setup and number of covariates when $K=1$. The variation depicted is across $T=40$ simulation trials. The vertical lines inside each density represent the $5\%, 50\%, 95\%$ quantiles.}
	\label{fig:post_meff_1}
\end{figure}
%-----------  

 Figure \ref{fig:post_meff_1} shows the distributions of the posterior medians of $\meff$ across the different $T$ trials arranged by hyperparameter setup and number of covariates, along with $95\%$ confidence intervals and the median. Indeed, setup (5) produces the most shrinkage, since it sacrifices detection of large signals by having a light tail and shifting mass near the origin, thus shrinking more. It is also expected that this hyperparameter setup performs well with respect to {\elpd} since the current simulation setup has high levels of sparsity present (around 95\% of the coefficients are truly zero). A somewhat better balanced condition, which presents a trade-off between predictive performance, detection of large signals and shrinkage, would be to consider conditions similar to setup (1) where $(\mu_{R^2}, \varphi_{R^2})=(0.1,0.5)$, which still allows for heavy tails. 
%--------------------
\begin{figure}[b!]%
	\centering
	\includegraphics[keepaspectratio, width=0.99\textwidth]{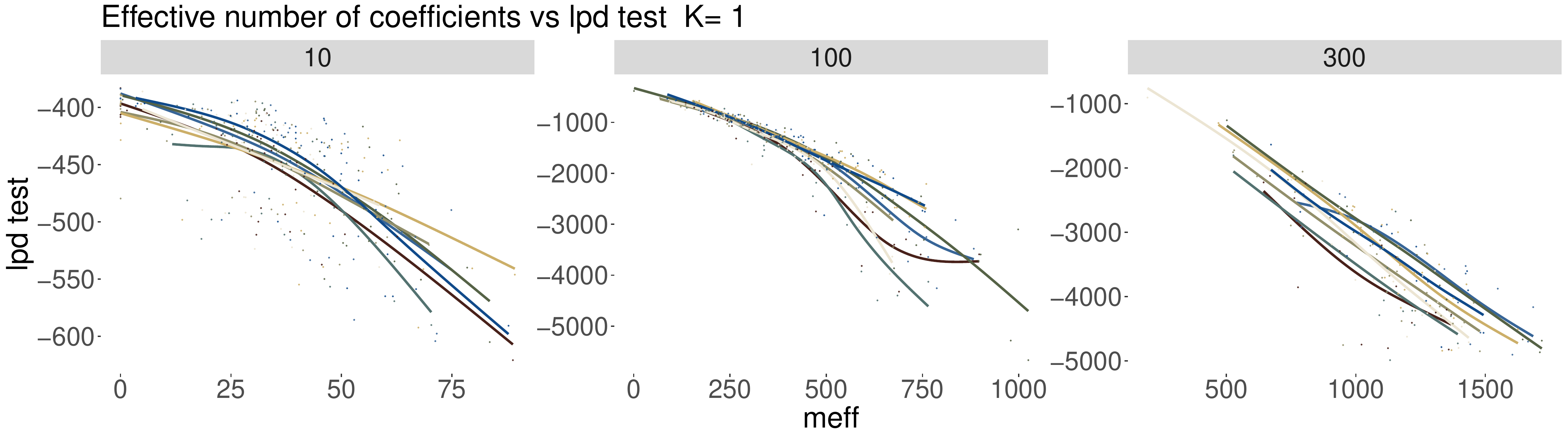}
	\caption{Relationship between $\elpd$ and shrinkage by hyperparameter setup when $K=1$. The non-linear relationships described by the colored lines in general shows a decreasing behavior as $\meff$ increases. Curves were estimated via thin-plate splines as implemented in the R package \texttt{mgcv} \citep{Wood2011mgcv}.  }
	\label{fig:meff-vs-lpd-grouped_1}
\end{figure}
%----------------------
Figure \ref{fig:meff-vs-lpd-grouped_1} shows the relationship between $\meff$ and $\elpd$ when estimating it over the different setups for a fixed value of $p$ across the different $T$ trials. Similar results are obtained when averaging over different hyperparameter setups but are not shown here. Importantly, this behavior indicates that over-shrinking has not yet occurred in any of the different configurations, as we would see an increasing and then decreasing behavior of elpd as $\meff$ increases. Overall, a general trend emerges in these results: As $\meff$ decreases, $\elpd$ increases and out-of-sample predictive performance thus improves. Therefore, to improve the latter, the user can impose hyperparameters $(\mu_{R^2}, \varphi_{R^2})$ that serve the purpose of regularizing instead of necessarily reflecting prior knowledge directly (as the latter might suggest a higher $R^2$ than is sensible for regularization).

\paragraph{Credible Interval Coverage}\mbox{}\\

We evaluated the coverage properties of the R2D2M2 model by inspecting marginal credibility intervals of the regression coefficients of different size $1-\alpha$. Table \ref{tab:errortab1} reports Type I errors, Type II errors and False Discovery Rates (FDR) on the set of all coefficients regardless of the type and on the set of overall coefficients in parenthesis. To carry out the hypothesis $H_0: \theta= \theta_0$ vs $H_1: \theta \neq \theta_0$ at level $\alpha$, we form the credibility interval of size $1-\alpha$ denoted by $I_\alpha$. We reject $H_0$ at level $\alpha$ if $\theta_0 \not\in I_\alpha$. To estimate the Type I and Type II error we consider those coefficients that are truly zero and truly non-zero, respectively. Denote by $V$ and $R$ the total number of hypothesis that are rejected but are true (also known as ''False Positives") and the total number of hypothesis respectively (namely, ''discoveries") and let $Q=V/R$, then the FDR is defined as $\text{FDR}=\mathbb{E}(Q)$ \citep{FDRBenjamini}. The FDR attempts to conceptualize Type I error in the multiple comparison setting and to identify as many discoveries, while incurring in a relatively low proportion of false positives. To estimate the FDR, we consider the truly zero coefficients and make use of the same test as before, while recording $Q$ for each simulation. When interpreting these results, keep in mind that we should not necessarily expect correct frequentist coverage from our models, since they were not built to satisfy that goal. 

%---------------- Error table
%--- K=1
\begin{table}[t!]
\caption{Error Table when $K=1$ and $\alpha=0.05$.  }
\label{tab:errortab1}
\centering
\resizebox{\textwidth}{!}{
\begin{tabular}{llllllllll}
\hline
\multicolumn{10}{c}{Scenario $K=1, \rho=0,  p_i=0.95, R_0^2=0.75, N=200 , \alpha=0.05$} \\
  \hline
p & Metric & 1 & 2 & 3 & 4 & 5 & 6 & 7 & 8 \\ 
  \hline
10 & Type I error & 0.00 (0.00) & 0.00 (0.00) & 0.00 (0.00) & 0.00 (0.00) & 0.00 (0.00) & 0.00 (0.00) & 0.00 (0.00) & 0.00 (0.01) \\ 
   & Type II error & 0.24 (0.20) & 0.24 (0.17) & 0.15 (0.09) & 0.17 (0.12) & 0.23 (0.18) & 0.28 (0.23) & 0.09 (0.08) & 0.25 (0.17) \\ 
   & FDR & 0.04 (0.04) & 0.01 (0.01) & 0.03 (0.03) & 0.01 (0.02) & 0.00 (0.01) & 0.02 (0.05) & 0.00 (0.00) & 0.09 (0.07) \\ 
  100 & Type I error & 0.00 (0.00) & 0.00 (0.00) & 0.00 (0.00) & 0.00 (0.00) & 0.00 (0.00) & 0.00 (0.00) & 0.00 (0.00) & 0.00 (0.01) \\ 
   & Type II error & 0.43 (0.26) & 0.40 (0.29) & 0.35 (0.22) & 0.45 (0.27) & 0.44 (0.24) & 0.47 (0.27) & 0.41 (0.22) & 0.36 (0.23) \\ 
   & FDR & 0.01 (0.01) & 0.02 (0.02) & 0.01 (0.01) & 0.08 (0.08) & 0.01 (0.01) & 0.05 (0.05) & 0.04 (0.04) & 0.01 (0.01) \\ 
  300 & Type I error &  0.00 (0.00) & 0.00 (0.00) & 0.00 (0.00) & 0.00 (0.00) & 0.00 (0.00) & 0.00 (0.00) & 0.00 (0.00) & 0.00 (0.01) \\ 
   & Type II error & 0.72 (0.58) & 0.68 (0.55) & 0.73 (0.57) & 0.73 (0.59) & 0.72 (0.59) & 0.75 (0.64) & 0.69 (0.58) & 0.76 (0.61) \\ 
   & FDR & 0.00 (0.00) & 0.00 (0.00) & 0.00 (0.00) & 0.00 (0.00) & 0.00 (0.00) & 0.00 (0.00) & 0.00 (0.00) & 0.00 (0.01) \\  
   \hline
\end{tabular}
}
\begin{tablenotes}
     \small 
     \item We show the results for all the coefficients as well as results for the overall coefficients inside the parenthesis.
\end{tablenotes}
\end{table}
%----------------

For a pre-specified level of $\alpha=0.05$, Table \ref{tab:errortab1} shows that Type I errors are controlled for (at level $\alpha$) in both the set of all coefficients and the set of overall coefficients (presented in parenthesis). The FDR is also controlled at this level for both types of terms. On the other hand, Type II errors increase monotonically across the hyperparameter combinations with increasing number of overall coefficients in the model. This is expected since credibility intervals contain zero even for the majority of true non-zero coefficients due to the shrinkage the prior imposes when responding to the complexity of the problem at hand. What is more, Type II errors are worse for the set of all coefficients than for the set of overall, which is the result of marginal posterior credibility intervals of varying coefficients being wider than their overall counterparts. This is bound to happen, since we have less information about the varying coefficients than we do about the overall coefficients and, as mentioned before in Section \ref{subsection:shrinkagefactors}, the former are thus subject to higher shrinkage.

The size of $\alpha$ determines the width of the credibility intervals and therefore has a direct influence on the error rates. Since the R2D2M2 prior produces Type I errors that are sufficiently low (and much lower than $\alpha$ itself), we can adjust $\alpha$ to decrease the Type II error, sacrificing Type I error control in exchange for statistical power. Type I errors are usually known as False Positive Rates (FPR). In this context, it is usual to study the complement of the Type II error known as the True Positive Rate (TPR). Likewise, modifying $\alpha$ will lead to a trade-off between the FPR and TPR and allows us to study ROC curves, which are usually presented when evaluating the efficiency of a classification algorithms \citep{roccurves}.     

\begin{figure}[t!]%
	\centering
	\includegraphics[keepaspectratio, width=0.99\textwidth, height=0.25\textheight]{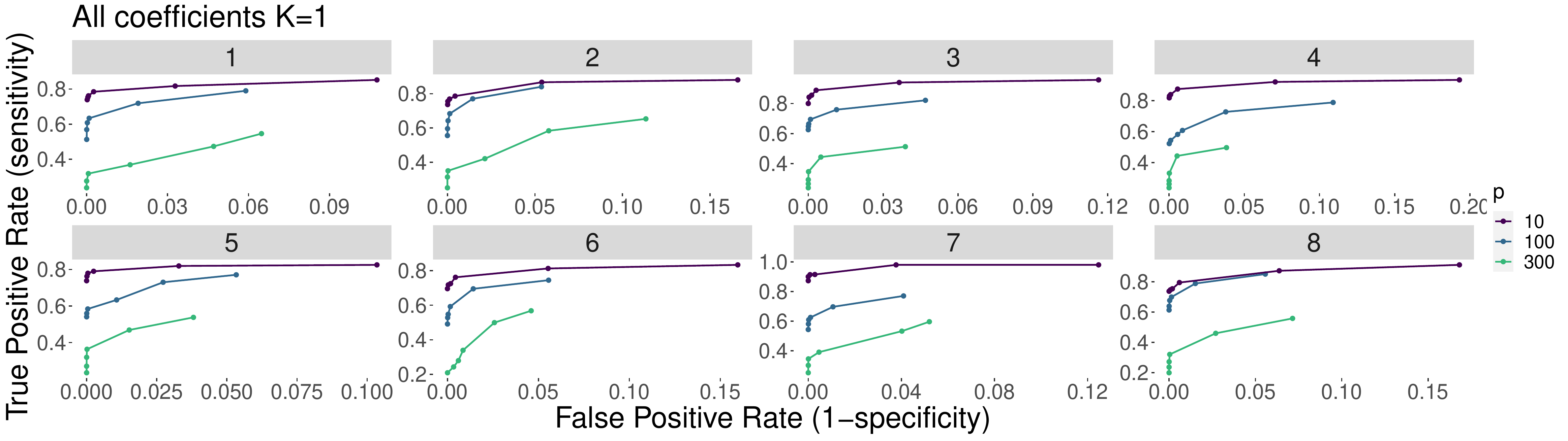}
	\caption{ROC curves for all coefficients arranged by hyperparameter setups when $K=1$. Points are calculated by changing $\alpha$ when computing credibility intervals. All ROC curves shown are above the identity line. }
	\label{fig:roc_ALL_1}
\end{figure}

\begin{figure}[t!]%
	\centering
	\includegraphics[width=0.99\textwidth, height=0.2\textheight]{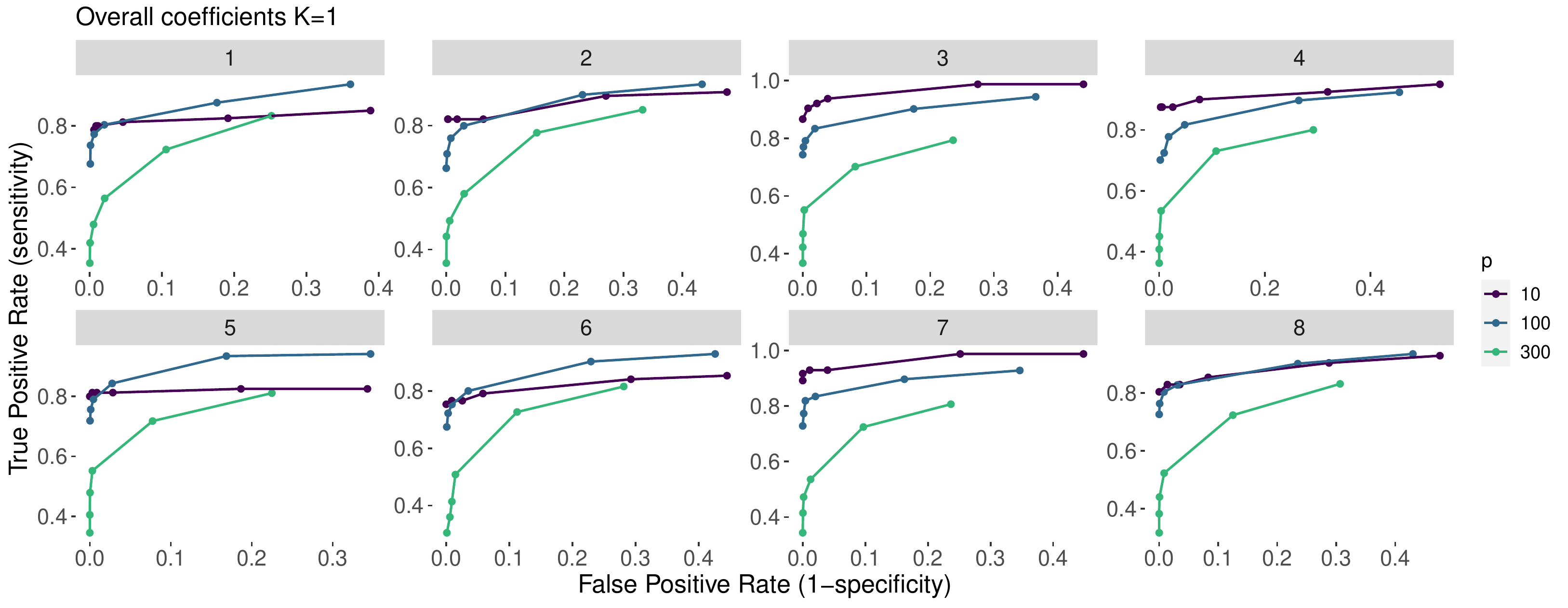}
	\caption{ROC curves for the overall coefficients when $K=1$. Points are calculated by changing $\alpha$ when computing credibility intervals. All ROC curves shown are above the identity line.}
	\label{fig:roc_OC_1}
\end{figure}

Figure \ref{fig:roc_ALL_1} shows how we can improve the TPR while increasing the FPR as we vary $\alpha \in \{ 0.02, 0.05, 0.10,0.20,0.50, 0.66 \}$ when forming the credibility interval $I_\alpha$ and considering all of the coefficients. Likewise, the same behavior is observed when taking into account only overall coefficients in Figure \ref{fig:roc_OC_1}. To illustrate how we can make better trade-offs between TPR and FPR, consider the most complex case when $p=300$ and a value of  $\alpha = 0.2$. This results in FPR $\approx 0.2$ and TPR $\approx 0.8$ for the set of overall coefficients. Figures \ref{fig:roc_ALL_1} and \ref{fig:roc_OC_1} also indicate that our procedure is substantially better than random classification even in the most complex cases. This occurs even without performing correction for multiple comparisons and although the prior was not built for variable selection per se. Nonetheless, we recommend to separate inference and selection procedures as argued by \cite{PiironenProjInf}. After performing inference via the R2D2M2 prior, projection prediction methods could be applied in a second step \cite{PiironenProjInf, catalina2020projection}, but studying this is out of scope of the present paper.

%----- Coverage tables
%--- K=1
\begin{table}[t!]
\caption{Coverage table when $K=1$. }
\label{tab:covtab1}
\centering
\scalebox{0.65}{
\begin{tabular}{lllllllllll}
\hline
\multicolumn{10}{c}{Scenario $K=1, \rho=0,  p_i=0.95, R_0^2=0.75, N=200 , \alpha=0.05$} \\
  \hline
p & Description & Metric & 1 & 2 & 3 & 4 & 5 & 6 & 7 & 8 \\ 
  \hline
10 & All & Coverage & 0.99 (0.91) & 0.99 (0.89) & 0.99 (0.90) & 0.99 (0.89) & 0.99 (0.89) & 0.99 (0.81) & 0.99 (0.91) & 0.99 (0.92) \\ 
   &  & Width & 0.86 (2.65) & 1.02 (2.56) & 0.96 (2.35) & 1.03 (2.21) & 0.85 (2.42) & 0.95 (2.27) & 0.91 (1.63) & 0.98 (2.30) \\ 
   & Overall & Coverage & 0.97 (0.83) & 0.98 (0.86) & 0.98 (0.92) & 0.98 (0.89) & 0.98 (0.91) & 0.96 (0.78) & 0.98 (0.90) & 0.96 (0.83) \\ 
   &  & Width & 0.44 (0.65) & 0.52 (0.74) & 0.50 (0.73) & 0.52 (0.69) & 0.44 (0.66) & 0.50 (0.70) & 0.47 (0.67) & 0.50 (0.66) \\ 
  100 & All & Coverage & 0.99 (0.71) & 0.99 (0.68) & 0.99 (0.66) & 0.99 (0.62) & 0.99 (0.68) & 0.99 (0.65) & 0.99 (0.70) & 0.99 (0.67) \\ 
   &  & Width & 1.40 (2.32) & 1.43 (1.98) & 1.33 (2.25) & 1.39 (1.94) & 1.33 (2.35) & 1.54 (2.18) & 1.3 (2.22) & 1.34 (1.99) \\ 
   & Overall & Coverage & 0.98 (0.81) & 0.98 (0.80) & 0.98 (0.76) & 0.97 (0.77) & 0.98 (0.82) & 0.98 (0.8) & 0.99 (0.81) & 0.98 (0.72) \\ 
   &  & Width & 0.88 (1.35) & 0.94 (1.36) & 0.84 (1.35) & 0.90 (1.33) & 0.85 (1.40) & 1.01 (1.45) & 0.8 (1.26) & 0.88 (1.33) \\ 
  300 & All & Coverage & 0.99 (0.53) & 0.99 (0.45) & 0.99 (0.54) & 0.99 (0.44) & 0.99 (0.51) & 0.99 (0.42) & 0.99 (0.54) & 0.99 (0.44) \\ 
   &  & Width & 1.75 (2.16) & 1.66 (1.96) & 1.79 (2.20) & 1.68 (2.01) & 1.73 (2.14) & 1.78 (2.07) & 1.67 (2.05) & 1.76 (2.06) \\ 
   & Overall & Coverage & 0.98 (0.67) & 0.97 (0.52) & 0.98 (0.69) & 0.97 (0.53) & 0.98 (0.66) & 0.96 (0.47) & 0.98 (0.64) & 0.97 (0.49) \\ 
   &  & Width & 1.41 (2.21) & 1.38 (2.08) & 1.44 (2.28) & 1.39 (2.10) & 1.4 (2.24) & 1.49 (2.17) & 1.34 (2.12) & 1.47 (2.17) \\ 
   \hline
\end{tabular}
}
\begin{tablenotes}
     \small 
     \item Credibility intervals were formed at a $\alpha=0.05$ level. We show results for the set of all coefficients mentioned in the description and for the non-zero corresponding cases inside parenthesis.
\end{tablenotes}
\end{table}
%----------------
Table \ref{tab:covtab1} reports coverage properties of $95\%$ credibility intervals for the set of all coefficients and for the set of all coefficients. The properties related to the subset of non-zero coefficients are reported in parenthesis. Two main points stand out: 1) Under a sparse regime, coverage of non-zero coefficients is in general a harder problem and the proportion of coverage is less for non-zero coefficients than when considering the set of all coefficients at once. Due to the high number of truly zero coefficients ($95\%$ in this scenario) the prior attempts to adapt to the sparsity at the expense of the non-zero coefficients.  2) On average, credibility intervals are wider for the non-zero coefficients. Since the prior adapts itself to the sparsity, it behaves conservatively when it detects a signal, performing shrinkage if the true value of the coefficient is not sufficiently high enough. Additionally, on average, the width of the credibility intervals increases with $p$ (or with $q$ for that matter) since there remains more epistemic uncertainty as model complexity increases given constant data set size.  
%---------------- ROC plots
%-----------
%-----------  

%-----------
%-----------  

\subsection{Application to riboflavin production data}

To illustrate the performance and usefulness of our developed models on real life challenges, we consider the \texttt{riboflavinGrouped} dataset provided by DSM (Kaiseraugst, Switzerland) and made publicly available by \cite{BuehlmannBio}. This dataset has been used to study the associations between gene expressions and riboflavin (vitamin $B_2$) \citep{BuehlmannBio, LinLina}. The response variable $y$ to be predicted is the logarithm of the riboflavin production rate. The data contains the expression levels of $p=4088$ candidate genes that might influence the riboflavin production. There are $N=111$ observations arranged into $L=28$ groups. Each group represents a strain (specimen) of a genetically engineered \textit{Bacillus subtilis} from which observations were taken. Different groups correspond to different strains of \textit{Bacillus subtilis}, each having from 2 to 6 observations at different time points where the logarithm of the riboflavin production rate and the expression levels of the $p=4088$ candidate genes were recorded. 

The size of the data $N=111$ is very small when compared to the number of candidate genes. \cite{BuehlmannBio} implemented a model with $p=4088$ overall coefficients of which $q=2$ were considered as varying among the different groups, however it is not specified in their work which are the ones they have considered to vary and why they have considered them.  \cite{LinLina} proposed modeling the association between the gene expressions and the production of riboflavin with a varying intercept model, i.e., $p=4088$ overall terms and $q=1$ varying term, however they do not take into account varying slopes nor models of different sizes. As an important result, both author groups reported the expression level of a specific gene, \textit{YXLE-at}, as significantly associated with riboflavin production.

Our main goal in this case study is to test under which circumstances the R2D2M2 model is able to  detect the gene \textit{YXLE-at} as influential on the riboflavin production rate as well, even when considering much more complicated models than in earlier analysis of this dataset. To do so we built models of increasing size in terms of number of regression coefficients that will always include this main covariate of interest.  A model with $p$ overall coefficients is formed in the following way: we first include the expression level of gene \textit{YXLE-at} as a covariate and then include the $p-1$ covariates most correlated with \textit{YXLE-at}; excluding those 5 covariates that are correlated with \textit{YXLE-at} above $|r| = 0.98$, which makes them practically indistinguishable from the latter. Thus the maximal model in terms of overall coefficients is of size $p=4083$. Once a covariate is included in the model, its varying coefficients counterpart is also included along with a varying intercept term. Including overall and varying coefficients the size of the model in terms of regression coefficients ranges from $P = 57$ when $p=1$ to $P = 118406$ when $p=4083$.  The hyperparameters of the R2D2M2 prior are set to $(\mu_{R^2}, \varphi_{R^2} , a_\pi )= (0.1,1,0.25)$ to encode that we expect a fair amount of noise and to allow it to shrink most of the covariates to zero. This makes sense here, since only very few genes are expected to influence riboflavin production rate.
%-----------
 \begin{figure}[t!]%
	\centering
	\includegraphics[keepaspectratio, width=0.99\textwidth, ]{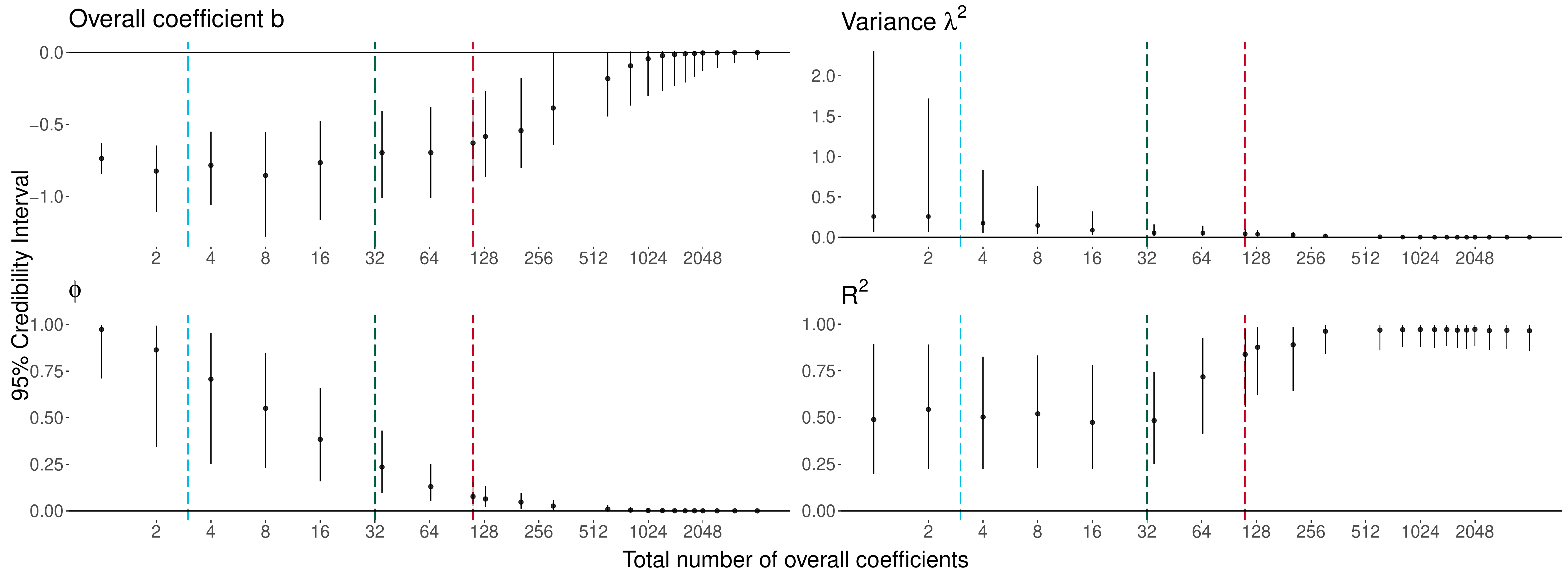}
	\caption{Plots of posterior means and $95\%$ credibility intervals related to model size in terms of the total number of overall coefficients in the model. Results for the overall coefficient and corresponding variance component of the overall coefficient of the label \textit{YXLE-at} are shown in the top left and top right plots respectively. The bottom left and bottom right plot show results for the proportion of allocated variance of \textit{YXLE-at} and $R^2$ respectively. The blue, green and red vertical lines indicate the point where the total number of coefficients, total number of terms and total number of overall coefficient exceed the number of observations $N=111$. }
	\label{fig:ribo-grid}
\end{figure}
%-----------

For the gene expression level of \textit{YXLE-at}, Figure \ref{fig:ribo-grid} shows plots of posterior means and $95\%$ credibility intervals of the corresponding overall coefficient $b$, variance $\lambda^2$ of the overall coefficient, and allocation component $\phi$ (we drop the indices since we will only discuss the case of this gene) as well as $R^2$ related to the size of the model in terms of overall coefficients. The point in which the amount of observations is surpassed by the total number of coefficients $p+(q+1)L$, number of terms $p+q+1$ (i.e., number of components in $\phi$) and number of covariates $p$ is indicated by blue, green and red vertical lines, respectively. The gene \textit{YXLE-at} is deemed as influential up to a certain point of $p$, when $p=613$ to be precise, as the credibility intervals do not include zero; however, detection becomes more complicated as the size of the model increases, since the model responds to the complexity of the problem by shrinking coefficients stronger towards zero until this point, even the coefficients of a gene we now know to be truly relevant. This is in line with what we have shown in our simulations in Subsection~\ref{subsection:GeneralSimMLM} and demonstrates that the theoretically expected behavior of the prior can also be found when analysing real data. We reiterate here that shrinkage priors are not made to directly select variables but rather to provided sensible inference. If selection was really the goal, two step procedures would be recommended, of which fitting an R2D2M2 prior model would only be the first step \citep{PiironenProjInf, catalina2020projection}.

\section{Discussion}
\label{sec:discussion}

In this work, we have studied how joint regularization of all regression coefficients in Bayesian linear multilevel models can be carried out. To do so, we have have proposed the R2D2M2 prior that imposes a prior on the global coefficient of determination $R^2$ and propagates the resulting regularization to  the individual regression coefficients via a Dirichlet Decomposition. The chosen parameterization of the R2D2M2 prior offers intuitive and interpretable hyperparameters that are easy to understand in practice and offers analysts the ability to readily incorporate prior beliefs into their multilevel model. Together with the fact that only very few hyperparameters have to be specified, this greatly simplifies prior specification for these models. 

We have also derived shrinkage factors and demonstrated how to compute the effective number of non-zero coefficients of the model. This allows the analyst to have a direct understanding of the implications that the chosen hyperparameters have on the amount of shrinkage imposed by the prior.  We have shown, theoretically and via simulations, that the R2D2M2 prior enables both local and global shrinkage of the regression coefficients and that it possesses vital properties that are required when working on high-dimensional regression problems, such as sufficient concentration of mass near the origin as well as heavy tails.  Finally we have demonstrated, via intensive simulations and by analysing real life data, that our prior implementation is not only well calibrated, but also offers reliable estimation even in the most complex cases that we investigated. 

Previously, established priors for regression coefficients in the context of multilevel modelling have only been able to either jointly regularize the overall coefficients or the varying coefficients but never both at the same time. In the case of the latter, it has been done either in the standard way of using multivariate normal distribution with fixed hyperparameters \citep{gelman_hill_2006} or by constructing a joint prior over their variance parameters \citep{Fulgstad2019}. Additionally, dealing with high dimensionality in the multilevel context has been fairly restrictive, since it has been concerned with the scenario of a larger number of overall coefficients, i.e., $p >N$, while including only very few varying terms, i.e $q \ll N$. The R2D2M2 prior allows for a joint regularization of both the overall and varying coefficients and provides the opportunity to study high dimensional scenarios in which both $p >N$ and $q>N$. 

Even though it is natural to wonder how the R2D2M2 prior model compares to frequentist methods to analyze multilevel models, we stress that we are focused on very high dimensional scenarios ($p,q>N$), where we need to regularize strongly due to the complexity of the problem. Existing frequentist methods and corresponding implementations for multilevel models that we are aware of are however not able to handle a high number of varying terms ($q>N$) and also struggle with a high number of overall terms ($p > N$) due to the resulting rank-deficient design matrices. Accordingly, comparing our proposed Bayesian model with frequentist methods would not be sensible at this point.

The present work can be extended in several ways. For instance, we could consider studying regularized versions of the R2D2M2 prior as \cite{PiironenHorseshoe} did for the original Horseshoe prior \citep{Horseshoe}. We have shown in simulations that as the complexity of the problem increases in terms of number of coefficients included the model, the prior reacts by increasing the amount of exerted shrinkage. However, coefficients that have a sufficiently high magnitudes will basically remain unaltered. Even though this is considered a key strength of shrinkage priors by some accounts \citep{BayesPenalizedRegSara}, it may not always be desirable and can be harmful when the parameters are only weakly identified by the data \citep{PiironenHorseshoe}.  Alternatively, deriving joint priors for non-normal likelihoods is an important area for future research. For instance \cite{bayesianworkflow} demonstrate for logistic regression that, as the number of covariates increases, quasi-complete separability becomes more likely, thus reducing the stability of the model if no substantial regularization is imposed.

Prior specification is a vital step in the Bayesian Workflow \citep{bayesianworkflow} and directly influences the performance of the models under consideration. Specifying joint priors allows us to account for the increasing complexity implied by increasing the amount of terms and parameters in the model, for example, in the shape of multilevel structure. What is more, they allow us to avoid some of the undesirable consequences of using independent, weakly informative priors in high dimensional setting. Ideally, joint priors should be able to express the user's prior knowledge while at the same time being able to perform efficient regularization, both globally and locally. These ideas together with the new developments presented here lay out a new avenue of research in the field of prior specification and elicitation.

%
% ** Acknowledgements **
\begin{acks}[Acknowledgments]
Funded by Deutsche Forschungsgemeinschaft (DFG, German Research Foundation) under Germany's Excellence Strategy EXC 2075-390740016 and DFG Project 500663361. We acknowledge the support by the Stuttgart Center for Simulation Science (SimTech).
\end{acks}

%% ** The bibliograhy **

\bibliographystyle{ba}
\bibliography{./bib/refs}

%% ** The appendix 
\appendix
\section{Proofs}
\label{appendixA}

%--------------------- Marginal distribution of b and u

\subsection*{Proof of proposition \ref{prop:marginalpriors}}

We provide a proof for the overall coefficients $b_i, i=1,...,p.$. The case of the varying coefficients $u_{ig_j}$ is handled similarly. We begin by providing an alternative parametrization of the R2D2M2 prior which is used in the proofs of propositions  \ref{prop:marginalpriors},\ref{prop:originprior}, and \ref{prop:tailprior}. The case of the varying coefficients is done in a similar way. \\

The prior for $b_i$ is  given by
    %---
	\begin{align}
	\tau^2 \sim \betaprime (a_1, a_2),  \phi \sim \dirichlet (\alpha), b_i| \phi, \tau^2,\sigma^2 \sim \normal \left(0, \frac{\sigma^2}{\sigma_{x_i}^2} \phi_i \tau^2  \right),
	\end{align}
	%---
	where $a_1=\mu_{R^2}\varphi_{R^2}, a_2=(1-\mu_{R^2})\varphi_{R^2}$. An observation from $\tau^2 \sim \betaprime (a_1, a_2)$ can be simulated by the chain (\cite{r2d2zhang})
	
	\begin{align*}
		\tau^2 | \xi \sim \gammadist (a_1, \xi), \ \  
		\xi \sim \gammadist(a_2, 1) ,
	\end{align*}
	%---
	where $\gammadist(a,b)$ denotes a Gamma distribution with shape $a$ and rate $b$. This provides an alternative representation of the R2D2M2 prior given by
	
	\begin{equation}
	\label{eq:r2d2altparam1}
	\begin{aligned}
	\xi \sim \gammadist(a_2, 1) , \tau^2 | \xi \sim \gammadist (a_1, \xi),  \phi \sim \dirichlet (\alpha) \\ 
	b_i| \phi, \tau^2,\sigma^2 \sim \normal \left(0, \frac{\sigma^2}{\sigma_{x_i}^2} \phi_i \tau^2  \right)
	\end{aligned}
	\end{equation}
	
	Let $\alpha$ consist of a single repeating element $a_\pi$, where $a_\pi= \frac{a_1}{\text{dim}(\alpha)}=\frac{\mu_{R^2}\varphi_{R^2}}{\text{dim}(\alpha)}$. This is an automatic way of specifying $a_\pi$ given that the user has provided a prior mean $\mu_{R^2}$ and a prior variance $\varphi_{R^2}$ for $R^2$. Then it follows that $\phi_{i} \tau^2 | \xi \sim \gammadist( a_\pi, \xi )$. By definition, $\lambda_{i}^2= \phi_{i}\tau^2$ and we can write the prior for $b _{i}$ as
	%----
	\begin{align}
	\label{eq:r2d2altparam3}
		\xi \sim \gammadist (a_2, 1) , \lambda_{i}^2| \xi \sim \gammadist \left( a_\pi, \xi \right) ,
		 b_{i} | \sigma^2, \lambda_{i}^2 \sim \normal \left( 0, \frac{\sigma^2}{\sigma_{x_i}^2 } \lambda_i^2   \right),
	\end{align}
	%----
	or as
	%----
	\begin{align*}
		\lambda_{i}^2 \sim \betaprime \left( a_\pi, a_2 \right),
		b_{i} |  \sigma^2, \lambda_{i}^2 \sim \normal \left( 0, \frac{\sigma^2}{\sigma_{x_i}^2 } \lambda_i^2\right).
	\end{align*}
	
	Let $q_i^2=\frac{\sigma^2}{\sigma_{x_i}^2}$, the prior marginal density of $b_i$ is given by
	
	\begin{align*}
		p( b_i | \sigma) &= \int_0^\infty \frac{1}{\sqrt{2\pi q_i^2 \lambda_{i}}^2 }\exp \left\lbrace  -\frac{ b_i^2  }{ 2 q_i^2\lambda_{i}^2}    \right\rbrace \frac{1}{\text{B}(a_\pi, a_2) } (\lambda_{i}^2)^{a_\pi-1} \left( \lambda_{i}^2+1
		\right)^{-a_\pi -a_2} d\lambda_{i}^2 \\
		&= \frac{1}{\sqrt{2\pi q_i^2} \text{B}(a_\pi, a_2) } \int_0^\infty \exp \left\lbrace  -\frac{ b_{i}^2  }{ 2q_i^2} t_i    \right\rbrace t_i^{\eta-1} (t_i+1)^{ \nu-\eta-1  } dt_i\\
		&= \frac{1}{ \sqrt{2\pi q_i^2} \text{B}(a_\pi, a_2) } \Gamma(\eta) U( \eta, \nu, z_i ), 
	\end{align*}
	
	where $t_i=\frac{1}{\lambda_i^2}$, $\eta=a_2+1/2$ and $\nu= 3/2-a_\pi$. $U(\eta, \nu, z_i)$ represents the confluent hypergeometric function of the second kind (see \cite{Zwillinger}) and $z_i=\frac{|b_i|^2}{2q_i^2}$. $U(\eta,\nu, z_i)$ is defined as long as the real part of $\eta$ and $z_i$ are positive, which is always the case since $a_2>0, q_i^2>0$ and $|b_i|>0$. \\

%-------------------- Behavior around the origin
	
\subsection*{Proof of proposition \ref{prop:originprior}}

The confluent hypergeometric function of the second kind $U(\eta, \nu, z)$ satisfies the following equation (see equation 13.2.40 \cite{nisthandbook})

\begin{align}
\label{eq:Utransformation}
   U(\eta,\nu, z)=z^{1-\nu} U(\eta-\nu+1, 2-\nu, z). 
\end{align}

We drop the index in $z$ since all cases are handled equally. Substituting $\eta=a_2+1/2, \nu= 3/2-a_\pi$ and $z= \frac{t^2}{2q^2}$ results in

\[ U\left(a_2+1/2, 3/2-a_\pi, \frac{t^2}{2q^2} \right)= (2q^2)^{1/2-a_\pi} t^{2a_\pi-1} U\left(a_2+a_\pi, a_\pi+1/2,  \frac{t^2}{2q^2}\right).\]

The right hand side has a singularity at $t=0$ when $a_\pi<1/2$.  \\ 

The limiting form $U$ has as $|z|\to 0$ follows equation 13.2.18 in \cite{nisthandbook} given by

\[  U(\eta, \nu, z)= \frac{\Gamma(\nu-1)}{\Gamma(\eta)} z^{1-
\nu}+\frac{\Gamma(1-\nu)}{\Gamma(\eta-\nu+1)} +\mathcal{O}\left( z^{2-\nu}\right) , \ \ 1< \nu <2 .\]

Substituting $\eta=a_2+1/2, \nu= 3/2-a_\pi$ and $z= \frac{t^2}{2q^2}$ results in
\begin{align*}
U\left( a_2+1/2, 3/2-a_\pi, \frac{t^2}{2q^2}\right)= C t^{2a_\pi-1} + \frac{\Gamma(a_\pi+1/2)}{\Gamma(a_2+1/2)} +\mathcal{O}\left( |t|^{1+2a_\pi} \right) , \ \ 0< a_\pi <1/2 ,    
\end{align*}

where $C= (2q^2)^{1/2-a_\pi} \frac{\Gamma(1/2-a_\pi)}{\Gamma(a_2+1/2)}$. The last expression is unboounded at $t=0$ when $a_\pi<1/2$. The case when $a_\pi=1/2$ follows equation 13.2.19 given by 

\[  U(\eta, \nu, z)= -\frac{1}{\Gamma(\nu)}\left( \ln(z)+\psi(\eta)+2\gamma \right)  +\mathcal{O}\left(  z \ln (z)  \right) , \ \ 1< \nu <2 ,\]

where $\psi(\cdot), \gamma$ represent the digamma function and the Euler-Mascheroni constant respectively. Substituting $\eta=a_2+1/2, \nu=3/2-a_\pi$ and $z=\frac{t^2}{2q^2}$ gives

\begin{align*}
U\left( a_2+1/2, 3/2-a_\pi, \frac{t^2}{2q^2}\right)&= -\frac{1}{\Gamma(a_2+1/2)}\left( \ln\left( \frac{t^2}{2q^2} \right)+\psi(a_2+1/2)+2\gamma \right) \\&\qquad +\mathcal{O}\left(  t^2 \ln \left(\frac{t^2}{2q^2} \right) \right).
\end{align*}

The last expression is not defined at $t=0$ and as $|t| \to 0$ it goes to infinity. To see the marginal priors are bounded when $a_\pi>1/2$ it is sufficient to consider the cases given by equations 13.2.20-13.2.22 in  \cite{nisthandbook} and making the proper substitutions of $\eta=a_2+1/2, \nu= 3/2-a_\pi$ and $z=\frac{t^2}{2q^2} $. In terms of the values  $a_\pi$ takes we have

\begin{align*}
U(\eta, \nu, t)&\sim
\begin{cases}
	 \mathcal{O} \left( t^{2a_\pi-1}   \right), \ \ \  &1/2<a_\pi<3/2 \\ 
	 \mathcal{O} \left( t^2 \ln(t^2/2)   \right), \ \ \  &a_\pi=3/2 \\ 
	 \mathcal{O} \left( t^2  \right), \ \ \  &a_\pi>3/2 \\ 
\end{cases}
\end{align*}

The derivative of $U(\eta, \nu, z)$ is given by equation 13.3.22 in \cite{nisthandbook} as

\[ \frac{d}{dz}  U(\eta, \nu, z)= - \eta U(\eta+1, \nu+1, z).   \]    

Combining this with Equation \eqref{eq:Utransformation} we have

\[  \frac{d}{dz}  U(\eta, \nu, z) = -\eta z^{-\nu} U(\eta-\nu  +1, 1- \nu, z ).    \]

Substituting $\eta=a_2+1/2, \nu= 3/2-a_\pi$ and $z= \frac{t^2}{2q^2}$ results in

\begin{align}
\label{eq:dudt}
\frac{d}{dt}  U\left( a_2+1/2 , 3/2-a_\pi , \frac{t^2}{2q^2} \right) = C t^{2a_\pi-2} U\left( a_2+a_\pi, a_\pi-1/2, \frac{t^2}{2q^2} \right),    
\end{align}

where $C=-(a_2+1/2)2^{3/2-a_\pi}(q^2)^{1/2-a_\pi}$. Equation \eqref{eq:dudt} is undefined at $t=0$ when $a_\pi<1$.

Making $t=b_i$ or $t=u_{ig_j}$ proves the proposition. \\

%-------------------- Asymptotic behavior of prior marginal distribution of coefficients b_i, u_{i g_j}

\subsection*{Proof of proposition \ref{prop:tailprior}}

To prove the proposition we make use of Watson's lemma as found in \cite{appliedasymptotic}.

\begin{lemma}[Watson's lemma]
Let $0\leq T \leq \infty$ be fixed. Assume $f(t)= t^{\lambda} g(t)$, where $g(t)$ has an infinite number of derivatives in the neighborhood of $t=0$, with $g(0)\neq 0$, and $\lambda > -1$. Suppose, in addition, either that $ |f(t)| < K e^{ct}$ for any $t>0$, where $K$ and $c$ are independent of $t$. Then it is true that for all positive $x$ that 
\[  \left| \int_0^T e^{-xt} f(t) dt \right| <\infty  \]
and that the following asymptotic equivalence holds:

\[	\int_0^T e^{-xt} f(t) dt  \sim \sum_{n=0}^\infty \frac{ g^{(n)}(0) \Gamma \left( \lambda+n+1 \right) }{n! x^{\lambda+n+1 }}	,\]
for $x>0$ as $x\to \infty$. \\
\end{lemma}

The marginal distribution of an overall coefficient $b$ is given by

\begin{align*}
    p(b|\sigma)&= \frac{1}{\sqrt{2\pi q^2} \text{B}(a_\pi, a_2) } \int_{0}^\infty \exp \left\lbrace  -\frac{|b|^2}{2q^2} t \right\rbrace t^{\eta-1} (t+1)^{\nu-\eta-1}  dt \\   &= \int_0^\infty \exp \left\lbrace  -z t  \right\rbrace f(t) dt , 
\end{align*}

where $ z=\frac{|b|^2}{2q^2},f(t)= C \, t^{\eta-1} (t+1)^{\nu-\eta-1} = t^{\eta-1} g(t), C=  \left(\sqrt{2\pi q^2} \text{B}(a_\pi, a_2) \right)^{-1},$ and $g(t)=C \,(t+1)^{\nu-\eta-1}$.  If we make $\lambda=\eta-1$, the hypothesis $\lambda >-1$ is satisfied since $a_2-1/2 > -1$ for $a_2>0$. $g(t)$ is infinitely differentiable around $t=0$ and $g(0)=0$. By Watson's Lemma, since $|f(t)|< K e^{ct}$ for all $t>0$ where $K$ and $c$ are independent of $t$, then as $|b| \to \infty,$

\[  p(z|\sigma)= \sum_{n=0}^\infty  \frac{g^{(n)}(0) \Gamma(\lambda+n+1)}{n! z^{\lambda+n+1}}.         \]

Truncating the sum at $n=2$ gives

\begin{align*}
    p(z|\sigma)&= C\left\lbrace \frac{ \Gamma(a_2+1/2)}{z^{a_2+1/2}} -
    \frac{ (a_\pi+a_2) \Gamma(a_2+3/2)}{z^{a_2+3/2}}+
    \frac{ (a_\pi+a_2) (a_\pi+a_2+1) \Gamma(a_2+5/2)}{z^{a_2+5/2}}  \right\rbrace  \\ 
    &\qquad + \mathcal{O}\left(  \frac{1}{ z^{a_2+7/2}}\right) \\
    &\sim \mathcal{O} \left(  \frac{1}{ z^{a_2+1/2}}\right).
\end{align*}

Therefore $p\left( |b| \,|\,\sigma \right)\sim \mathcal{O} \left(  \frac{1}{ |b|^{2a_2+1}}\right)$. When $a_2< 1/2$ as $|b|\to \infty$ and comparing with $\frac{1}{b^2}$ we have that 

\[ \frac{p(b|\sigma)}{ \frac{1}{b^2} } \sim \mathcal{O} \left( \frac{ 1 }{ b^{2a_2-1}  } \right)      \to \infty  . \]

\subsection*{Proof of proposition \ref{prop:r2d2bounded}}

We will first show that the R2D2M2 prior can be represented as Horseshoe type prior \citep{Horseshoe}. We then proceed to make use of Theorem 2 and Theorem 3 in \cite{Horseshoe} to show that the R2D2M2 prior with normal base distributions for the coefficients is of bounded influence. 

Consider representation \eqref{eq:r2d2altparam3} of the R2D2M2 prior given by 

\begin{align*}
    b_i | \sigma, \lambda_i^2 \sim \normal(0, \sigma^2 \lambda_i^2), \ \ 
    \lambda_i^2 | \xi \sim \gammadist(a_\pi, \xi), \ \  \xi \sim \gammadist(a_2,1).
\end{align*}

It can be shown that this representation is equivalent to 

	\begin{align*}
		b_{i} |  \sigma^2, \lambda_{i}^2 \sim \normal \left( 0, \sigma^2 \lambda_i^2 \right), \ \ \lambda_{i}^2 \sim \betaprime \left( a_\pi, a_2 \right).
	\end{align*}

When $a_\pi=1/2$ and $a_2=1/2$, then $\lambda_i \sim \text{Cauchy}^+(0,1)$ and we have

\begin{align*}
        b_i | \sigma, \lambda_i^2 \sim \normal(0, \sigma^2 \lambda_i^2), \ \ 
        \lambda_i \sim \text{Cauchy}^+(0,1).
\end{align*}

The Horseshoe prior \citep{HorseshoeProceedings} has the following representation:
%-------------
\begin{align*}
    b_i | \lambda_i, w &\sim \normal(0, w^2 \lambda_i^2)\\
    \lambda_i &\sim \text{Cauchy}^+(0,1),
\end{align*}
%-------------
where $\lambda_i$ are the local shrinkage parameters and $w$ is the global shrinkage parameter. Therefore the R2D2M2 prior can be represented as a Horseshoe type prior where the global scale is fixed and $w=1$. In the following consider $\sigma=1$ and $y|b \sim  \normal(b,1)$, since by hypothesis there are no varying coefficients in the model. Theorem 2 in \cite{Horseshoe} shows that conditioned on one sample $y^*$, we can write 
%-------------
\begin{align}
\label{eq:postmeannormalmeans}
    \mathbb{E}(b|y^*)= y^*+ \frac{d}{dy^*}\log m(y^*),
\end{align}
%-------------
where $m(y^*)$ is the marginal density for $y^*$ given by $m(y^*)=\int p(y^*|b) p(b) db$. Theorem 3 of \cite{Horseshoe} shows that for the Horseshoe prior as $|y^*|\to \infty$ then $$\lim\limits_{|y^*|\to \infty} \frac{d}{dy^*} \log m(y^*)=0,$$ 
%-------------
implying that as $|y^*|\to \infty$ then $\mathbb{E}(b|y^*)\approx y^*$ and thus showing that the R2D2M2 prior is of bounded influence. The exact form of $m(y^*)$ for the R2D2 prior is given by 

\begin{align*}
    m(y^*)= \frac{1}{(2\pi^3)^{1/2}} \int_0^\infty \exp \left( - \frac{y^{*2}/2}{1+\lambda^2}   \right) \frac{1}{(1+\lambda^2)^{3/2}} d\lambda.
\end{align*}
Making $z=\frac{1}{\lambda^2+1}$ results in

\begin{align*}
    m(y^*)&= \frac{1}{(2\pi^3)^{1/2}} \int_0^1 \exp \left(  -1/2 y^{*2} z   \right) z^{-1/2} dz \\
    &= \frac{1}{\pi } \frac{ \text{erf} \left(y^* / \sqrt{2} \right)}{y^*},
\end{align*}

where $\text{erf}(\cdot)$ denotes the error function given by $\text{erf}(x)= \frac{2}{\sqrt{\pi}} \int_0^x e^{t^2} dt$. Hence, $\frac{d}{dy^*}\log m(y^*)$ is given by 

\begin{align*}
    \frac{d}{dy^*}\log m(y^*)&= \frac{ \sqrt{\frac{2}{\pi}} e^{- y^{*2}/2 }} { \text{erf}\left(y^* / \sqrt{2} \right)} -\frac{1}{y^*},
\end{align*}
and $\lim_{|y^*|\to \infty} \frac{d}{dy^*}\log m(y^*)=0$. Using this and \eqref{eq:postmeannormalmeans} we have that as $|y^*|\to \infty $ then $\mathbb{E}(b|y^*) \to y^*$. 
\section{Posterior computation}
\label{section:appendixB}

\subsection{Stan}

To obtain draws from the posterior distribution we have implemented our model in Stan  \citep{StanJSS, stan2022}, which uses a substantially extended implementation of the No-U-Turn Sampler (NUTS) from \cite{nuts}. This sampler is an adaptive form of Hamiltonian Monte Carlo \citep{handbookmcmc}. The following code is fully functional however we also maintain a current version in \myosfresults .

This implementation considers $K$ grouping factors with multiple levels $L$ each and $D$ overall coefficients, including the overall intercept. All groups have the same number of levels $L_g$ as well as the same number of varying coefficients $D_g$. $D_g$ is the number of group level effects and is including the varying intercepts as well, i.e Group $K$ level $l$ has $D_g$ varying coefficients. This implementation is able to work with data where $D, K, L_g$ can vary with no need to manually modify the Stan code. However, the code can be easily modified to handle the case when groups have different amount of levels and terms per group. It is also possible to briefly modify this code in order to work with $q \leq p$ varying terms. The order of the varying terms should consider the relationship with the $i$th covariate since we scale by $\sigma_{x_i}$. 

\begin{small}
\begin{verbatim}
// Stan code for the R2D2M2 prior. 
functions {
  
  vector R2D2(vector z, vector sds_X, vector phi, real tau2) {
    /* Efficient computation of the R2D2 prior
    * Args:
    *   z: standardized population-level coefficients
    *   phi: local weight parameters
    *   tau2: global scale parameter (sigma is inside tau2)
    * Returns:
      *   population-level coefficients following the R2D2 prior
    */
      return  z .* sqrt(phi * tau2) ./ sds_X ;
  }
  
}
data {
  int<lower=1> N;  // total number of observations
  vector[N] Y;  // response variable
  int<lower=1> D;  // number of population-level effects
  matrix[N, D] X;  // population-level design matrix 
                   // including column of 1s
  int<lower=0> K; // number of groups
  vector[D-1] sds_X; // column sd of X before centering. 
                    //Estimate or real values.
  
  //---- data for group-level effects 
  
  int<lower=1> Lg;  // number of  levels per group (constant)
  int<lower=1> Dg; // number of coefficients per level per group 
                   //(D_g constant per group)
  int<lower=1> J[N,K]; // grouping indicator matrix 
                      // per observation per group K
  
  
  //---- group-level predictor values 
  matrix[Dg,N] Z[K]; 
  
  //---- data for the R2D2 prior
  vector<lower=0>[ (D-1)+K+(Dg-1)*K] R2D2_alpha; 
  real<lower=0> R2D2_mean_R2;  // mean of the R2 prior
  real<lower=0> R2D2_prec_R2;  // precision of the R2 prior
  int prior_only;  // should the likelihood be ignored?
}

transformed data {
  int Dc = D - 1; // 
  matrix[N, Dc] Xc; //centered version of X without an intercept
  vector[Dc] means_X;//column means of X before centering
  vector[Dc] var_X;
  vector[N] Yc;
  real Ymean;
  for (i in 2:D) {
    means_X[i - 1] = mean(X[, i]);
    var_X[i-1]= sds_X[i-1]^2;
    Xc[, i - 1] = (X[, i] - means_X[i - 1]) ; 
  }
  
  Ymean= mean(Y);
  for (i in 1:N) {
    Yc[i]= Y[i]-mean(Y);
  }
  
}

parameters {
  real Intercept;  // temporary intercept for centered predictors
  vector[Dc] zb; // standardized population-level effects
  matrix[Dg,Lg] z[K]; // standardized group-level effects 
  real<lower=0> sigma;  // standard deviation of response
  
  // local parameters for the R2D2M2 prior
  simplex[Dc+K+(Dg-1)*K] R2D2_phi; 
  // R2D2 shrinkage parameters 
  /*Convention of indexing: First Dc for overall effects,
                             group of K for varying intercepts,
                            Batches of Dc for each grouping factor */
                            
  real<lower=0,upper=1> R2D2_R2;  // R2 parameter
  
}

transformed parameters {
  
  vector[Dc] b;  // population-level effects
  
  matrix[Dg,Lg] r[K]; // actual group-level effects 
  
  real R2D2_tau2;  // global R2D2 scale parameter
  R2D2_tau2 =  R2D2_R2 / (1 - R2D2_R2);
  
  // compute actual overall regression coefficients
  b = R2D2(zb, sds_X, R2D2_phi[1:Dc], (sigma^2) * R2D2_tau2);  
  
  for(k in 1:K){
    // varying intercepts
    // Dc+k is the kth varying intercept
    r[k,1,] = (sigma * sqrt(R2D2_tau2 * R2D2_phi[Dc+k ])
              * (z[k,1,]));  
    for(d in 2: Dg){
      // group level effects
      // (k-1)Dc indexes the beginning of the kth batch of scales
      r[k,d,]= sigma /(sds_X[(d-1)]) * 
                sqrt(R2D2_tau2 *
                R2D2_phi[Dc+K+ (k-1)*(Dg-1) +(d-1) ]) * (z[k,d,]);  
      
    }
  }
}

model {
  // likelihood including constants
  
  if (!prior_only) {
    // initialize linear predictor term
    vector[N] mu = Intercept + rep_vector(0.0, N);
    for (n in 1:N) {
      // add more terms to the linear predictor 
      
      for(k in 1:K){
        mu[n]+=dot_product(r[k,,J[n,k]], Z[k,,n]) ;
        }
      
    }
    // mu+ Xc*b
    target += normal_id_glm_lpdf(Yc | Xc, mu, b, sigma); 
  }

  
  target += beta_lpdf(R2D2_R2 | R2D2_mean_R2 * R2D2_prec_R2,
            (1 - R2D2_mean_R2) * R2D2_prec_R2); // R^2 
  target += dirichlet_lpdf(R2D2_phi | R2D2_alpha); //phi 
  
  target += normal_lpdf(Intercept | 0, 10);  // Intercept
  target += std_normal_lpdf(zb); //zb: overall effects
  
  for(k in 1:K){
    for(d in 1: Dg){
      target += std_normal_lpdf(z[k,d,]); // z
    }
  }
  
  target += student_t_lpdf(sigma | 3, 0, sd(Yc));  // 
  
}
generated quantities {
  //---actual population-level intercept
  real b_Intercept = Ymean+Intercept - dot_product(means_X, b);
  
  //---y_tilde quantities of interest
  
  vector[N] log_lik; 
  real y_tilde[N];
  vector[N] mu_tilde = rep_vector(0.0, N)+Ymean+Intercept +Xc*b;
  vector<lower=0>[(D-1)+K+(Dg-1)*K] lambdas; 
  
  //---y_tilde calc

  for (n in 1:N) {
    for(k in 1:K){
      mu_tilde[n]+=dot_product(r[k,,J[n,k]], Z[k,,n]) ;
      }
    log_lik[n] =normal_lpdf( Y[n] | mu_tilde[n], sigma); 
    y_tilde[n]=normal_rng(mu_tilde[n], sigma);
  }
  
  //--- lambdas 
  
  //overall variances
  lambdas[1:Dc]= sigma^2*R2D2_phi[1:Dc]./ var_X *R2D2_tau2 ; 
  //varying int variances
  lambdas[(Dc+1):(Dc+K)]= sigma^2*R2D2_phi[(Dc+1):(Dc+K)]
                            *R2D2_tau2; 
  
  for(k in 1:K){
      // group level variances
      // (k-1)(Dg-1) indexes the start of the kth batch of scales
      lambdas[(Dc+K+(k-1)*(Dg-1)+1):(Dc+K+(k-1)*(Dg-1)+Dg-1)]= 
      sigma^2*R2D2_phi[(Dc+K+(k-1)*
                    (Dg-1)+1):(Dc+K+(k-1)*(Dg-1)+Dg-1)]
                    ./ var_X*R2D2_tau2;
  }

}
\end{verbatim}
        
\end{small}

\subsection{Gibbs sampling}

	To illustrate a Gibbs sampling approach to obtain posterior draws we consider, without loss of generality,  $K=1$ grouping factors and consider that the first $q$ terms (i.e covariates $x_1,...,x_q$ with $q\leq p$) are all varying over $L=l$ levels, where we will consider that $\sigma_{xi}=1 \forall i=1,...,p$.  Therefore $\phi$ is a simplex of dimension $r=p+1+q$ (due to the inclusion of the varying intercept term). We consider a symmetric Dirichlet distribution for $\phi$ with concentration vector $\alpha = (a_\pi,..., a_\pi)'$. Our procedure can be generalized for arbitrary concentration vectors $\alpha$. For ease of notation and to be able to develop a blocked Gibbs sampling we will write model \eqref{r2d2m2model} in matrix form. This will drastically improve performance over moving one regression term at a time . 
	
	Denote by $b=(b_1,...,b_p)'$, $u_j=(u_{0j},u_{1j},...,u_{qj} )'$ the varying coefficients in level $j=1,...,l$ and by $u=(u_1,...,u_l)' \in \mathbb{R}^{l(q+1)}$ the vector containing all varying coefficients. Notice that we have suppressed the notation that indicates the group since we only have one grouping factor $K$. Let $y=(y_1,...,y_N)'$ denote a vector containing the $N$ observations $y_i, i=1,...,N$, $X$ denote the standardized design matrix of dimension $N \times p$ and $Z$ denote the varying effects matrix   \citep[see][for details on how to construct $Z$]{lme4}. Since we only have one grouping factor, $Z$ is a block matrix, where each block contains the observations that vary in each level. In the following, we denote the linear predictor as $\mu= Xb + Zu$.
	
    Let $\Sigma_b=\sigma^2 \Gamma_b$ denote the covariance matrix of the vector of overall coefficients $b$ where $\Gamma_b=\text{diag}\left\lbrace \phi_1  \tau^2 ,..., \phi_p  \tau^2    \right\rbrace$. Let  $\Sigma_u \in \mathbb{R}^{l (q+1)\times l (q+1)}$ denote the block diagonal covariance matrix of $u$ which is given by  $\Sigma_u= \sigma^2 \Gamma_u $ where $\Gamma_u= \text{diag} \left \lbrace \gamma_{1},..., \gamma_{l} \right \rbrace$, and $\gamma_j$ is the covariance matrix associated to $u_j$ for $j=1,...,l$. In our context, $\gamma_j= \text{diag} \left\lbrace \phi_{p+1}\tau^2, \phi_{p+2} \tau^2, ..., \phi_{p+1+q} \tau^2  \right\rbrace $ for all $j=1,...,l$. Finally,  let $I_N$ represent the identity matrix of order $N$.  With the notation that has been introduced and using the representation \eqref{eq:r2d2altparam1}, we can write the R2D2M2 model \eqref{r2d2m2model} in matrix form as 
    \begin{equation}
    \label{eq:r2d2altparam2}
    \begin{aligned}
		y &\,|\, \mu, \sigma^2 \sim \normal(\mu, \sigma^2 I_N), \\  
		b&|\sigma, \tau^2, \phi \sim \normal \left(0, \Sigma_b \right), \ \ 
		u|\sigma, \tau^2, \phi \sim \normal \left(0,  \Sigma_u \right) \\
		\phi & \sim \dirichlet(\alpha) \\
	    \tau^2 | \xi &\sim \gammadist (a_1, \xi), \ \ 
		\xi \sim \gammadist(a_2, 1), \ \  \sigma \sim p(\sigma). 
	\end{aligned}
    \end{equation}

In the following we denote by $Z \sim \gigdist \left(  \chi, \rho, \nu  \right) $, the Generalized Inverse Gaussian distribution  \citep{RobertGIG} with parameters $\chi>0, \rho>0$ and $\nu \in \mathbb{R}$  if 
	\[ p(z) \propto  z^{\nu-1} \exp\left\lbrace   -\left(  \rho z+ \chi/z  \right)/2    \right\rbrace. \] 
	
	We denote by $\invgammadist(c,d)$ the Inverse Gamma distribution with shape $c$ and scale $d$ and consider that a priori  $\sigma^2 \sim \invgammadist(c,d)$. Consider representation \eqref{eq:r2d2altparam2} of the R2D2M2 prior, then the Gibbs sampling procedure is the following: 
	
	\begin{enumerate}
		\item Set initial values for $b, u, \sigma, \phi, \tau^2, \xi$. 
		\item Sample $b | u, \phi, \tau, \sigma, y  \sim \normal \left(  \bar{b}  , S_{b} \right)$, where 
		
		\begin{align*}
		    \bar{b}&= \left(X'X+ \Gamma_b^{-1} \right)^{-1} X'(y-Zu) \\
		    S_b&= \sigma^2 \left(X'X+ \Gamma_b^{-1} \right)^{-1}. \\
 		\end{align*}
		\item Sample $u | b, \phi, \tau, \sigma, y \sim \normal \left(  \bar{u}  , S_{u} \right)$, where 
		\begin{align*}
		    \bar{u}&= \left(Z'Z+ \Gamma_u^{-1} \right)^{-1} Z'(y-Xb) \\
		    S_u&= \sigma^2 \left(Z'Z+ \Gamma_u^{-1} \right)^{-1}. \\
 		\end{align*}
		
		\item Sample $\sigma^2 |  b, u, \phi, \tau^2  \sim \invgammadist \left(  \dot{c}, \dot{d}   \right)$, where 
		
		\begin{align*}
		  \dot{c}&=c+\frac{1}{2}\left( N+p+l(q+1) \right) \\
		  \dot{d}&=  d+   \frac{1}{2} \left(  ||  y-Xb-Zu||_2^2+  b' \Gamma_{b}^{-1} b + u' \Gamma_{u}^{-1} u \right).
		\end{align*}

		\item Sample  $\tau^2|  b, u, \sigma, \xi \sim \gigdist  \left( \chi, \rho, \nu \right)$, where
		
		\begin{align*}
		\chi=  \frac{1}{\sigma^2} \left(   b'   T_b^{-1}b  +  u' T_u^{-1} u    \right), \ \ 
		\rho= 2\xi , \ \ \nu= a_1- 1/2 \left(p+ l(q+1) \right),  \\
		\end{align*}
        
        where $T_b$ and $T_u$ are such that $\Gamma_b=T_b \tau^2 $ and $\Gamma_u=T_u \tau^2 $ respectively. Sampling from the Generalized Inverse Gaussian is not straightforward, but the reader can refer to \cite{GenInvGaussian} for efficient methods.\\
	
		\item Sample $\xi | \tau^2   \sim \gammadist \left( a_1+a_2 , 1+\tau^2 \right) $.\\
		\item Sample $\phi \,|\, b, u, \sigma, \xi $. \\ 
		
		To sample $\phi \,|\, b, u, \sigma $ we make use of the following proposition
		\begin{prop}
		    The joint posterior of $\phi \,|\,  b, u, \sigma, \xi  $ has the same distribution as $\left(T_1/T,..., T_r/T \right)$ where $T_j$ are independently drawn according to  
		\begin{align*}
		   T_j  &\sim \gigdist \left(  \frac{b_j^2}{2\sigma^2} , 2\xi, a_\pi-1/2\right) \ \  j=1,...,p \\
		  T_j  &\sim \gigdist \left( \frac{  \sum_{j=1}^l u_{ij}^2   }{2\sigma^2}  , 2\xi, a_\pi - l/2 \right)  \ \   j=p+1,..., q+p+1,
		\end{align*}
		where $r= p+1+q,$ and $T= \sum_{i=1}^r T_i$.
		\end{prop}
	
		The main idea is to integrate out $\tau^2$ and after doing so set $\phi_i= T_i/T$. To see this works, consider the joint posterior of $\phi  | b, u, \sigma, \xi$ that results from integrating $\tau^2$, 
		\begin{small}
		\begin{equation}
		    	\begin{aligned}
			\label{eq:gibbs1}
				p(\phi | b, u, \sigma,\xi  ) &\propto \prod_{i=1}^p \phi_i^{a_\pi-3/2} \prod_{i=p+1}^r \phi_i^{a_\pi-l/2-1} \\
				&\times \int_{0}^\infty (\tau^2)^{\nu-1} \exp \left\lbrace -\frac{1}{2} \left[  \rho \tau^2 +\chi/\tau^2    \right]  \right\rbrace d\tau^2,
			\end{aligned}
		\end{equation}
		\end{small}
		
		where $ \nu= a_1-1/2(p+l(q+1)), \rho= 2\xi, \chi= \frac{1}{\sigma^2} \left( \sum\limits_{i=1}^p \frac{b_i^2}{\phi_i}+ \sum\limits_{i=p+1}^r \sum\limits_{j=1}^l \frac{u_{ij}^2}{\phi_i} \right)$.

		Now consider the following proposition \citep[see][Annotation (36) for the proof]{SimplexKruijer}.
		
		\begin{prop}
			Suppose $T_1,..., T_r$ are independent random variables, with $T_i$ having a density $f_i$ on $(0,\infty)$. Let $\phi_i = T_i / T$ with $T= \sum_i T_i$, then the joint density $f$ of $(\phi_1,..., \phi_{r-1})$ supported on the simplex $\mathcal{S}^{r-1}$ has the form
		%-----------
		\begin{align}
			\label{eq:imp}
			p(\phi_1,...,\phi_{r-1})= \int_{0}^\infty t^{r-1} \prod_{i=1}^{n} f_i \left( \phi_i t \right) dt,
		\end{align}
		%-----------		
		where $\phi_r= 1- \sum_{i=1}^{r-1} \phi_j$.
		\end{prop}
		
		Setting $f_i(x)$ as 
		
		\begin{align*}
			f_i(x) &\propto x^{-\delta_1} \exp\left \lbrace  \frac{  b_i^2}{2\sigma^2} \frac{1}{x} \right \rbrace  \exp\left\lbrace  -\xi x\right \rbrace, \ \ i=1,...,p \\
			f_i(x) &\propto x^{-\delta_2} \exp \left \lbrace -\frac{ \sum_{j=1}^l u_{ij}^2}{ 2\sigma^2} \frac{1}{x} \right \rbrace \exp\left\lbrace  -\xi x\right \rbrace, \ \ i=p+1,...,p+q+1, \\
		\end{align*}

	    substituting $f_i(\phi_i \tau^2)$ into (\ref{eq:imp}) and simplifying we arrive to
		
		\begin{small}
	    \begin{equation}
	        \begin{aligned}
	        \label{eq:gibbs2}
			p(\phi_1,...,\phi_{r-1}) &=  \prod_{j=1}^p \phi_j^{ -\delta_1} \prod_{j=p+1}^{p+q+1} \phi_j^{ -\delta_2} \int_{0}^\infty  (\tau^2)^{r- \delta_1 p -\delta_2 (q+1 )-1 } \\
			&\times 
			\exp \left\{   -\frac{1}{2}  \left[ \frac{1}{\sigma^2} \left(  \sum_{i=1}^p b_i^2/\phi_i  + \sum_{i=p+1}^{r} \sum_{j=1}^l u_{ij}^2/\phi_i  \right)/\tau^2 + 2\xi \tau^2 \right]  \right\} d\tau^2. \\
		\end{aligned}
	    \end{equation}
		\end{small}
		
		Comparing equations \eqref{eq:gibbs1} and \eqref{eq:gibbs2} we have $\delta_1= 3/2-a_\pi , \delta_2= (l+2)/2-a_\pi$. The proof is completed by noticing that $f_j$ are Generalized Inverse Gaussian distributions. \\
		
		\item[h)] Repeat until convergence

	\end{enumerate}

\newpage

\section{Examples}
\label{section:appendixC}

\subsection{Simulation Based Calibration}

\begin{table}[ht]
\centering
\begin{tabular}{l|l|l}
Description & Hyperparameter  & {Values}                                      \\ \hline
Number of simulations & $T$ & 100 \\
Number of iterations  & $S$ & 3000 \\
Groups  & $K$             & $\{0,1\}$                                         \\
Levels & $L$             & 20                                                \\
Covariates & $p$             & $\{10,100,300\}$                                  \\
Prior mean of $R^2$ &$\mu_{R^2}$     & $\{0.1,0.5\}$                                 \\
Prior precision of $R^2$& $\varphi_{R^2}$ & $\{0.5,1\}$                                     \\
Concentration parameter & $a_\pi$         & \textbf{$\{0.5,1\}$}                          \\
Covariance matrix of $x$ & $\Sigma_x$      & $\{I_p, AR(\rho) \}$ where $\rho \in \{0.5\}$ \\
\end{tabular}
\caption{Values assigned to the hyperparameters in the Simulation Based Calibration experiment \ref{subsection:SBC}.}
\label{tab:sbchyperparams}
\end{table}

\subsection{Simulations from sparse multilevel models} 
 
\begin{table}[ht]
\centering
\begin{tabular}{l|l|l}
Description & Hyperparameter  & {Values}                                      \\ \hline
True value of $R^2$ & $R^2_0$      & $\{0.25,0.75\}$                                     \\
Groups  & $K$             & $\{1,3\}$                                         \\
Levels & $L$             & 20                                                \\
Covariates & $p$             & $\{10,100,300\}$                                  \\
Prior mean of $R^2$ &$\mu_{R^2}$     & $\{0.1,0.5\}$                                 \\
Prior precision of $R^2$& $\varphi_{R^2}$ & $\{0.5,1\}$                                     \\
Concentration parameter & $a_\pi$         & \textbf{$\{0.5,1\}$}                          \\
Covariance matrix of $x$ & $\Sigma_x$      & $\{I_p, AR(\rho) \}$ where $\rho \in \{0.5\}$ \\
Level of induced sparsity &$v, z$      & $\{0.5,0.95\}$                 
\end{tabular}
\caption{Values assigned to the hyperparameters in the data generating process in experiment \ref{subsection:GeneralSimMLM}.  }
\label{tab:sim2datahyperparams}
\end{table}
%----- 

\subsubsection{Additional tables and figures}

%--- Pred table K=3
\begin{table}[H]
\centering
\caption{Predictive Table results with $K=3$.}
\resizebox{\textwidth}{!}{
\begin{tabular}{llrrrrrrrr}
\hline
\multicolumn{10}{c}{Scenario $K=3, \rho=0,  p_i=0.95, R_0^2=0.75, N=200$} \\
 \hline
p & Metric & 1 & 2 & 3 & 4 & 5 & 6 & 7 & 8 \\ 
  \hline
10 & RMSE & 0.05 & 0.09 & 0.08 & 0.09 & 0.04 & 0.06 & 0.09 & 0.03 \\ 
   & elpd & -477 & -491 & -491 & -461 & -470 & -478 & -497 & -499 \\ 
   & $\meff$ & 71 & 93 & 90 & 80 & 71 & 115 & 84 & 88 \\ 
  100 & RMSE & 0.01 & 0.01 & 0.01 & 0.02 & 0.01 & 0.02 & 0.02 & 0.02 \\ 
   & elpd & -2700 & -3049 & -3806 & -3440 & -2899 & -2920 & -2744 & -3681 \\ 
   & $\meff$  & 1109 & 1402 & 1388 & 1764 & 1101 & 1213 & 1260 & 1629 \\ 
  300 & RMSE & 0.02 & 0.02 & 0.01 & 0.02 & 0.01 & 0.01 & 0.02 & 0.01 \\ 
   & elpd & -6210 & -4413 & -7064 & -6285 & -4480 & -3413 & -6471 & -5501 \\ 
   & $\meff$ & 2532  & 2104 & 2461 & 2722 & 2025 & 1669 & 2642 & 2405 \\ 
   \hline
\end{tabular}
}
\label{tab:predtab3}
 \begin{tablenotes}
      \small
      \item  The table shows results for the Root Mean Squared Error (RMSE), expected logpointwise predictive density (elpd) on the test set and the effective number of nonzero coefficients $\meff$. .
\end{tablenotes}
\end{table}

%--- Error table K=3
\begin{table}[H]
\centering
\caption{Error Table $K=3$}
\resizebox{\textwidth}{!}{
\begin{tabular}{llllllllll}
\hline
\multicolumn{9}{c}{Scenario $K=3, \rho=0,  p_i=0.95, R_0^2=0.75, N=200, \alpha=0.05$ } \\
  \hline
p & Metric & 1 & 2 & 3 & 4 & 5 & 6 & 7 & 8 \\ 
  \hline
10 & Type I error & 0.00 (0.00) & 0.00 (0.00) & 0.00 (0.00) & 0.00 (0.00) & 0.00 (0.00) & 0.00 (0.00) & 0.00 (0.00) & 0.00 (0.00) \\ 
   & Type II error & 0.34 (0.32) & 0.32 (0.24) & 0.23 (0.13) & 0.31 (0.18) & 0.23 (0.21) & 0.21 (0.13) & 0.17 (0.17) & 0.22 (0.20) \\ 
   & FDR & 0.12 (0.00) & 0.11 (0.00) & 0.11 (0.00) & 0.11 (0.00) & 0.03 (0.00) & 0.08 (0.00) & 0.15 (0.01) & 0.02 (0.00) \\ 
   & TNDR & 0.99 (0.95) & 0.98 (0.96) & 0.98 (0.98) & 0.98 (0.97) & 0.99 (0.96) & 0.99 (0.98) & 0.98 (0.97) & 0.99 (0.97) \\ 
  100 & Type I error & 0.00 (0.00) & 0.00 (0.00) & 0.00 (0.00) & 0.00 (0.00) & 0.00 (0.00) & 0.00 (0.00) & 0.00 (0.00) & 0.00 (0.00) \\
   & Type II error & 0.50 (0.37) & 0.48 (0.39) & 0.40 (0.33) & 0.52 (0.35) & 0.35 (0.27) & 0.47 (0.35) & 0.49 (0.37) & 0.51 (0.30) \\ 
   & FDR & 0.00 (0.00) & 0.00 (0.00) & 0.00 (0.00) & 0.00 (0.00) & 0.00 (0.00) & 0.00 (0.00) & 0.02 (0.02) & 0.01 (0.00) \\ 
   & TNDR & 0.99 (0.97) & 0.99 (0.98) & 0.99 (0.98) & 0.99 (0.97) & 0.99 (0.98) & 0.99 (0.97) & 0.99 (0.97) & 0.99 (0.98) \\ 
  300 & Type I error & 0.01 (0.01) & 0.01 (0.01) & 0.00 (0.00) & 0.00 (0.00) & 0.00 (0.00) & 0.00 (0.00) & 0.00 (0.00) & 0.00 (0.00) \\
   & Type II error & 0.83 (0.72) & 0.89 (0.80) & 0.85 (0.76) & 0.90 (0.78) & 0.82 (0.73) & 0.85 (0.81) & 0.87 (0.75) & 0.87 (0.80) \\ 
   & FDR & 0.02 (0.02) & 0.02 (0.02) & 0.00 (0.00) & 0.00 (0.00) & 0.00 (0.00) & 0.00 (0.00) & 0.00 (0.00) & 0.00 (0.00) \\ 
   & TNDR & 0.99 (0.96) & 0.99 (0.95) & 0.99 (0.96) & 0.99 (0.96) & 0.99 (0.96) & 0.99 (0.95) & 0.99 (0.95) & 0.99 (0.95) \\ 
   \hline
\end{tabular}
}
\begin{tablenotes}
     \small 
     \item We show the results for all the coefficients as well as results for the overall coefficients inside the parenthesis. The numbers 1-8 indicate the hyperparameter setup used.
\end{tablenotes}
\label{tab:errortab2}
\end{table}

\begin{table}[H]
\centering
\caption{Coverage Table $K=3$}
\resizebox{\textwidth}{!}{
\begin{tabular}{lllllllllll}
\hline
\multicolumn{10}{c}{Scenario $K=3, \rho=0,  p_i=0.95, R_0^2=0.75, N=200 , \alpha=0.05$} \\
  \hline
p & Description & Metric & 1 & 2 & 3 & 4 & 5 & 6 & 7 & 8 \\ 
  \hline
10 & All & Coverage & 0.99 (0.42) & 0.99 (0.46) & 0.98 (0.46) & 0.98 (0.43) & 0.99 (0.47) & 0.99 (0.51) & 0.98 (0.41) & 0.99 (0.60) \\ 
   &  & Width & 0.79 (1.50) & 0.94 (2.16) & 1.00 (2.06) & 0.96 (2.04) & 0.80 (2.02) & 0.94 (2.14) & 1.01 (1.87) & 0.84 (1.84) \\ 
   & Overall & Coverage & 0.97 (0.80) & 0.97 (0.77) & 0.98 (0.90) & 0.96 (0.74) & 0.97 (0.80) & 0.97 (0.80) & 0.98 (0.88) & 0.98 (0.86) \\ 
   &  & Width & 0.46 (0.75) & 0.55 (0.82) & 0.58 (0.89) & 0.57 (0.80) & 0.48 (0.73) & 0.55 (0.81) & 0.57 (0.84) & 0.51 (0.72) \\ 
  100 & All & Coverage & 0.99 (0.39) & 0.99 (0.33) & 0.99 (0.37) & 0.99 (0.33) & 0.99 (0.40) & 0.99 (0.37) & 0.99 (0.34) & 0.99 (0.36) \\ 
   &  & Width & 1.07 (1.42) & 0.95 (1.54) & 0.96 (1.09) & 1.16 (1.59) & 0.96 (1.35) & 1.06 (1.52) & 1.01 (1.38) & 1.20 (1.67) \\ 
   & Overall & Coverage & 0.98 (0.73) & 0.97 (0.55) & 0.98 (0.68) & 0.97 (0.53) & 0.98 (0.71) & 0.97 (0.54) & 0.98 (0.69) & 0.97 (0.59) \\ 
   &  & Width & 0.82 (1.42) & 0.75 (1.43) & 0.74 (1.34) & 0.91 (1.50) & 0.73 (1.33) & 0.85 (1.42) & 0.77 (1.37) & 0.95 (1.57) \\ 
  300 & All & Coverage & 0.98 (0.29) & 0.98 (0.26) & 0.99 (0.29) & 0.99 (0.28) & 0.99 (0.30) & 0.99 (0.22) & 0.99 (0.28) & 0.99 (0.29) \\ 
   &  & Width & 1.32 (1.55) & 1.29 (1.51) & 1.27 (1.55) & 1.37 (1.53) & 1.26 (1.58) & 1.23 (1.37) & 1.39 (1.61) & 1.36 (1.62) \\ 
   & Overall & Coverage & 0.95 (0.34) & 0.94 (0.23) & 0.96 (0.35) & 0.96 (0.24) & 0.96 (0.33) & 0.96 (0.25) & 0.96 (0.35) & 0.96 (0.24) \\ 
   &  & Width & 1.21 (2.03) & 1.19 (1.79) & 1.15 (1.95) & 1.26 (1.91) & 1.16 (2.01) & 1.14 (1.73) & 1.27 (2.16) & 1.26 (1.92) \\ 
   \hline
\end{tabular}
}
\begin{tablenotes}
     \small 
     \item Credibility intervals were formed at a $\alpha=0.05$ level. We show results for the set of all coefficients mentioned in the description and for the non-zero corresponding cases inside parenthesis. The numbers 1-8 indicate the hyperparameter setup used.
\end{tablenotes}
\end{table}

%----------- Images for when K=3
%-----------
 \begin{figure}[H]%
	\centering
	\includegraphics[keepaspectratio, width=0.99\textwidth, height=0.23\textheight]{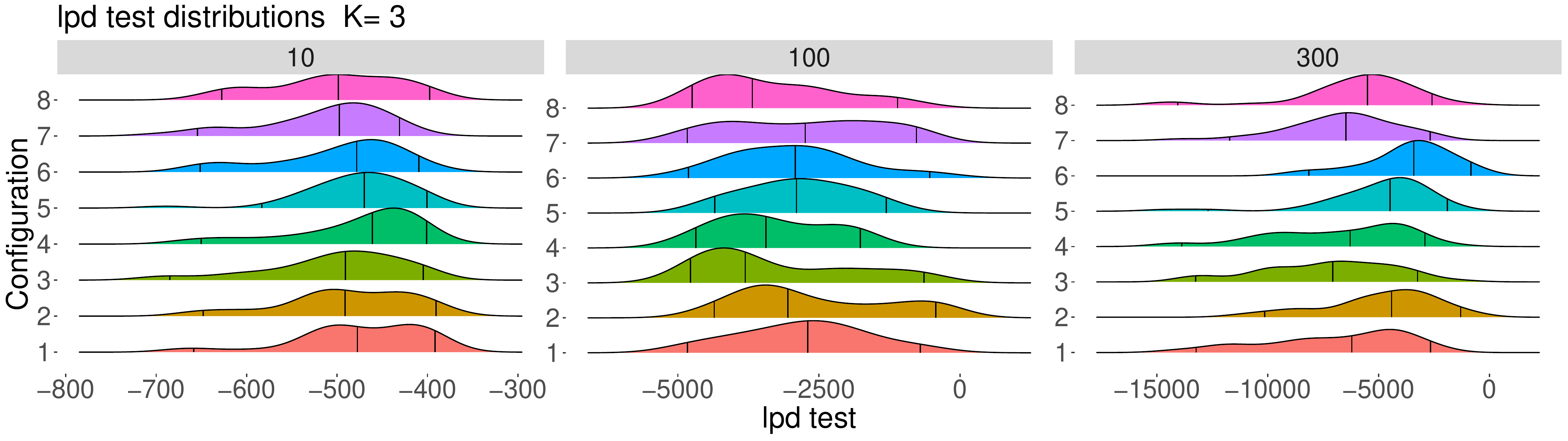}
	\caption{ Densities of the $\elpd$ estimators on the test set as $p$ increases and arranged by hyperparameter configurations when $K=3$. The vertical lines inside each density represent the $5\%, 50\%, 95\%$ quantiles from left to right respectively. This shows that, in average, we can expect similar results, however for single realizations differences can be observed due to the heavy tails present. Hence, proper hyperparameter specification should be done.  }
	\label{fig:lpd_test_3_ridges}
\end{figure}
%-----------
%-----------
 \begin{figure}[H]%
	\centering
	\includegraphics[keepaspectratio, width=0.99\textwidth, height=0.20\textheight]{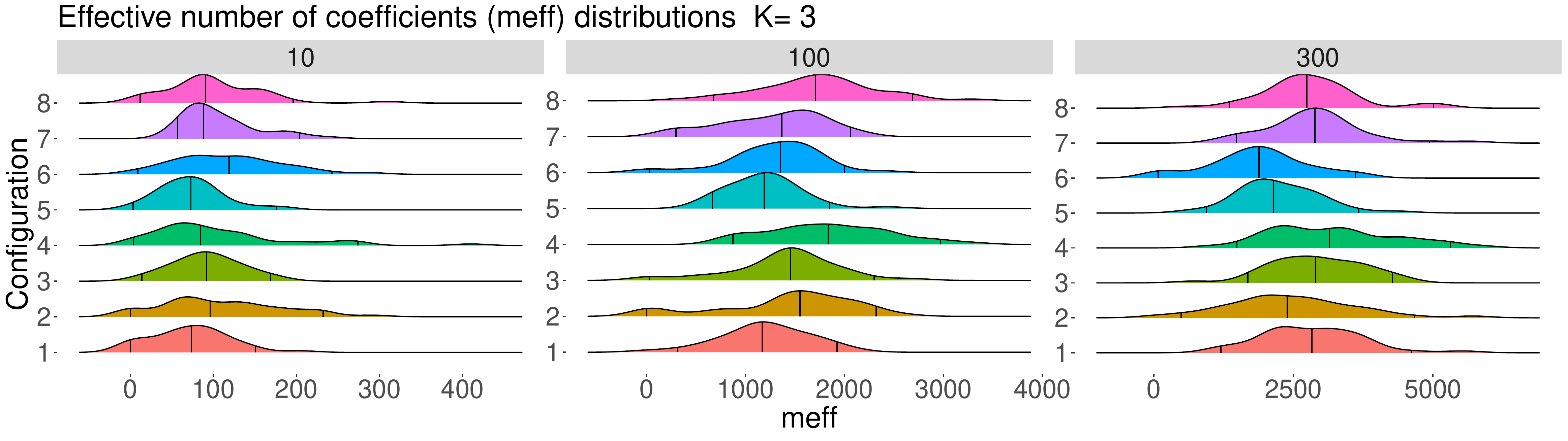}
	\caption{Densities of the posterior median of the effective number of coefficients per hyperparameter configuration and number of covariates when $K=3$. }
	\label{fig:post_meff_3}
\end{figure}
%-----------  

%--------------------
\begin{figure}[H]%
	\centering
	\includegraphics[keepaspectratio, width=0.95\textwidth, height=0.23\textheight]{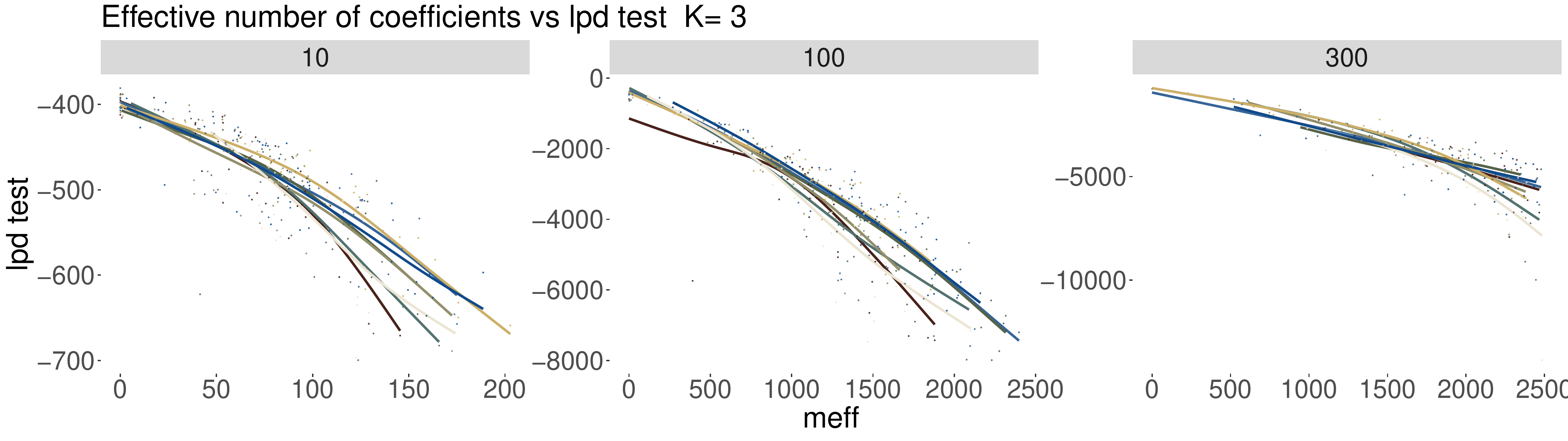}
	\caption{Relationship between $\elpd$ and shrinkage by hyperparameter configuration when $K=3$. The non-linear relationships described by the colored lines in general shows a decreasing behavior as $\meff$ increases. }
	\label{fig:meff-vs-lpd-grouped_3}
\end{figure}

\begin{figure}[H]%
	\centering
	\includegraphics[keepaspectratio, width=0.99\textwidth, height=0.33\textheight]{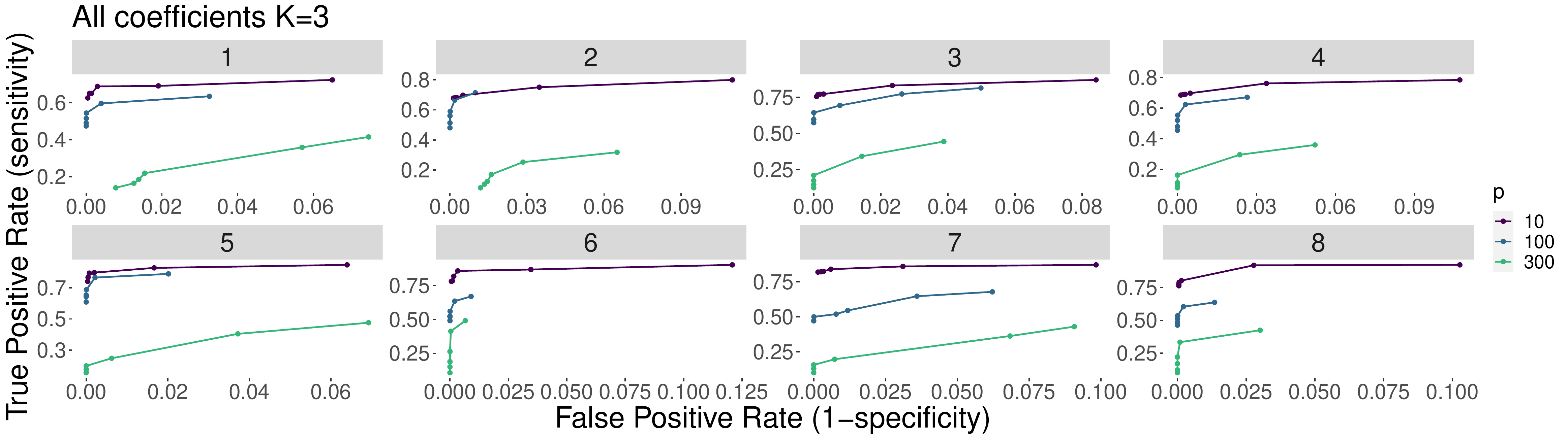}
	\caption{ROC curves for all the coefficients arranged by hyperparameter configurations when $K=3$. Points are calculated by moving $\alpha$ when forming credibility intervals. Notice that the shape of the curve is due to the fact that even for high values of $\alpha$, False Positive Rates don't increase. All ROC curves shown are above the identity line.}
	\label{fig:roc_ALL_3}
\end{figure}
%-----------  
 \begin{figure}[H]%
	\centering
	\includegraphics[keepaspectratio, width=0.99\textwidth, height=0.33\textheight]{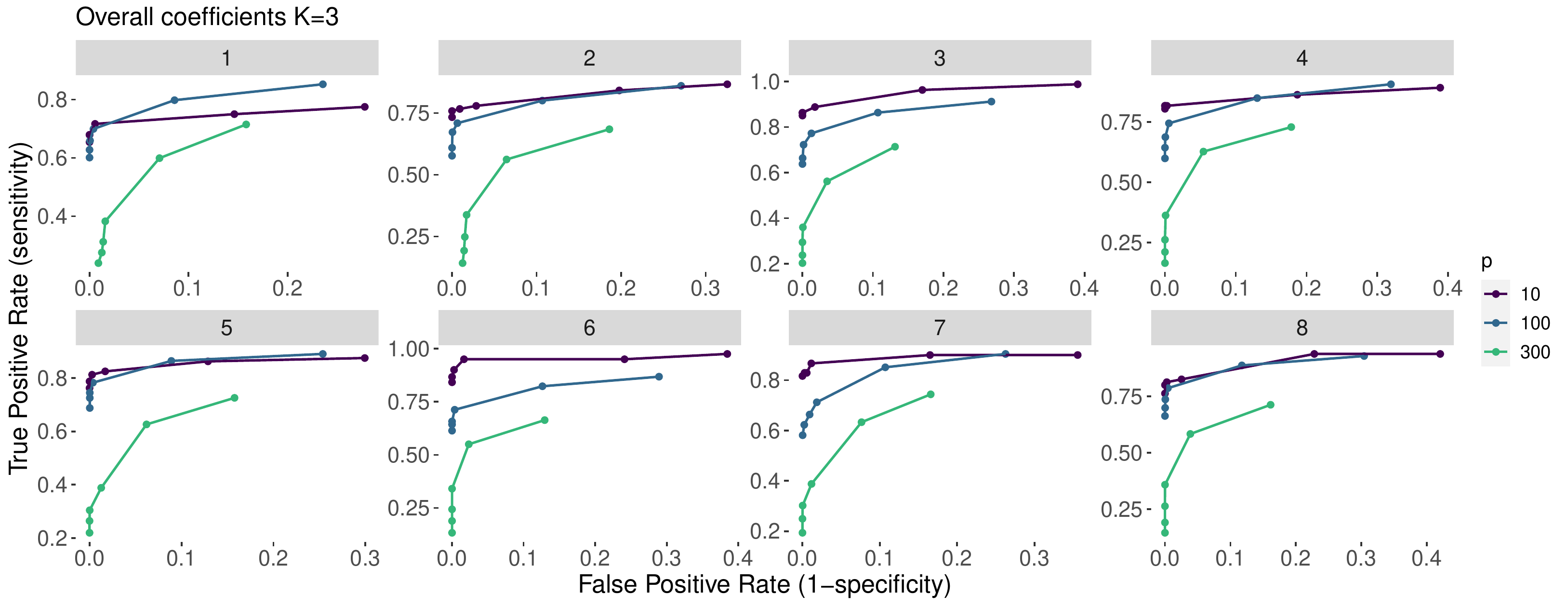}
	\caption{ROC curves for the overall coefficients when $K=3$. Points are calculated by moving $\alpha$ when forming credibility intervals. Notice that the upper bound on the curve is due to the fact that even for high values of $\alpha$, False Positive Rates don't increase. All ROC curves shown are above the identity line.}
	\label{fig:roc_OC_3}
\end{figure}
%-----------  

%--bib	

% 

\end{document}